\documentclass{siamart190516}

\usepackage{amsfonts,amssymb,amstext,latexsym,amsmath}

\usepackage{array}

\usepackage[caption=false]{subfig}

\usepackage{pgfplots}
\pgfplotsset{compat=1.13}

\usepackage{graphicx,epstopdf}

\Crefname{ALC@unique}{Line}{Lines}

\usepackage{xspace}

\usepackage{hyperref}

\usepackage{balance}

\usepackage{cleveref}

\usepackage[ruled,linesnumbered,vlined,algo2e]{algorithm2e}

\DontPrintSemicolon
\SetKwComment{tcp}{$\triangleright$ }{}
\SetVlineSkip{0cm}
\SetKwInOut{Param}{Param}
\SetKwInOut{Input}{Input}
\SetKwComment{kc}{$\triangleright$~}{}

\SetCommentSty{mycommfont}

\usepackage{graphicx}

\usepackage{colortbl}
\usepackage{multirow}

\usepackage{pifont}
\newcommand{\cmark}{\text{\ding{51}}}
\newcommand{\xmark}{\text{\ding{55}}}

\usepackage{xargs}
\usepackage[colorinlistoftodos,prependcaption,textsize=scriptsize]{todonotes}

\usepackage{mathtools}

\newcommandx{\mLI}{\ensuremath{\text {\sc mLI}}\xspace}
\newcommandx{\mNC}{\ensuremath{\text {\sc mNC}}\xspace}

\newcommandx{\nic}{\ensuremath{\text {\sc Nic}}\xspace}
\newcommandx{\uni}{\ensuremath{\text {\sc Uni}}\xspace}

\newcommandx{\rac}{\ensuremath{\text {\sc RaC}}\xspace}
\newcommandx{\pal}{\ensuremath{\text {\sc PaL}}\xspace}
\newcommandx{\opal}{\ensuremath{\text {\sc oPaL}}\xspace}

\newcommandx{\bac}{\ensuremath{\text {\sc BaC}}\xspace}
\newcommandx{\bal}{\ensuremath{\text {\sc BaL}}\xspace}

\newcommand{\vcp}{VC\xspace}
\newcommand{\srpp}{SRP\xspace}

\newcommand{\vect}[1]{\ensuremath{\left\langle #1 \right\rangle}}
\newcommand{\interval}[2]{\ensuremath{\left[#1\,..\,#2\right]}}

\title{On Symmetric Rectilinear Matrix Partitioning}

\author{
  Abdurrahman Ya\c{s}ar\thanks{School of Computational Science and Engineering, Georgia Institute
  of Technology, Atlanta, GA 30332
  ({\{ayasar, balin, anxiaojing, kaan, umit\}@gatech.edu})}
\and
  Muhammed Fat{\.i}h Balin\footnotemark[1]
\and
  Xiaojing An\footnotemark[1]
\and
  Kaan Sancak\footnotemark[1]
\and
  \"{U}mit V. \c{C}ataly\"{u}rek\footnotemark[1]
}

\begin{document}


\maketitle
\slugger{sisc}{xxxx}{xx}{x}{x--x}

\begin{abstract}
Even distribution of irregular workload to processing units
is crucial for efficient parallelization in many applications. In this work, we are concerned with a spatial
partitioning called rectilinear partitioning (also known as generalized block distribution)
of sparse matrices.
More specifically, in this work, we address the problem of symmetric rectilinear partitioning of
a square matrix. By symmetric, we mean the rows and columns of the matrix are identically partitioned
yielding a tiling where the diagonal tiles (blocks) will be squares.
We first show that the optimal solution to this problem is NP-hard, and we propose four heuristics to
solve two different variants of this problem. We present a thorough analysis of the computational
complexities of those proposed heuristics. To make the proposed techniques more
applicable in real life application scenarios, we further reduce their computational complexities by
utilizing effective sparsification strategies together with an efficient sparse
prefix-sum data structure.
We experimentally show the proposed algorithms are efficient and effective
on more than six hundred test matrices.
With sparsification, our methods take less than $3$ seconds in the Twitter graph
on a modern $24$ core system and output a solution whose load imbalance is no
worse than $1\%$.
\end{abstract}



\pagestyle{myheadings}
\thispagestyle{plain}

\markboth{Ya\c{s}ar, Bal{\i}n, An, Sancak and \c{C}ataly\"{u}rek}{On Symmetric Rectilinear Matrix Partitioning}

\section{Introduction}
\label{sec:intro}

After advances in social networks and the rise of web interactions,
we are witnessing an enormous growth in the volume of generated data.
A large portion of this data remains sparse and irregular and is stored as graphs
or sparse matrices. However,
analyzing data stored in those kinds of data structures is
challenging, especially for traditional architectures due to the
growing size and irregular data access pattern of these problems.
High-performance processing of this data is an important and a pervasive
research problem.
There have been many studies developing parallel sparse matrix~\cite{davis2016survey},
linear-algebra~\cite{Balay97-MSTSC,benzi2002preconditioning,Heroux05-TOMS,Bell09-HIPCS} and graph
algorithms~\cite{lu2014large, Quinn84-CSUR,Tarjan85-SIJC,Caceres97-ICALP,George12-SPRINGER}
for shared and distributed memory systems as well as GPUs and hybrid
systems~\cite{Garland08-PDAC,Buatois09-ICPEDS,Hong11-ICPACT,shi2018graph}.
In such platforms,
the balanced distribution of the computation and data
to the processors is crucial for achieving better efficiency.

In the literature, balanced partitioning techniques can be broadly divided into two
categories: connectivity-based
(e.g.,~\cite{Karypis98-SISC,Catalyurek99,Hendrickson00-PARCO,Catalyurek10-SISC})
and spatial/geometric
(e.g.,~\cite{Berger87-TC,Manne96-IWAPC,Ujaldon96-ICS,Pilkington96-TPDS,Saule12-JPDC-spart}).
Connectivity-based methods model the load balancing problem through
a graph or a hypergraph. In general, computation volumes are weighted
on the nodes and communication volumes are weighted on the edges or
hyperedges.
Connectivity-based techniques explicitly model the computation,
and the communication, hence, they are generally computationally more expensive.
This paper tackles the lightweight spatial (i.e., geometric) partitioning problem
of two-dimensional sparse matrices, which mainly focuses on load-balancing, and
communication is only implicitly minimized by localizing the data and {\em neighbors}
that need the data.

\begin{figure}[htb]
  \centering
  \subfloat[An example partition]{\includegraphics[width=.3\linewidth]{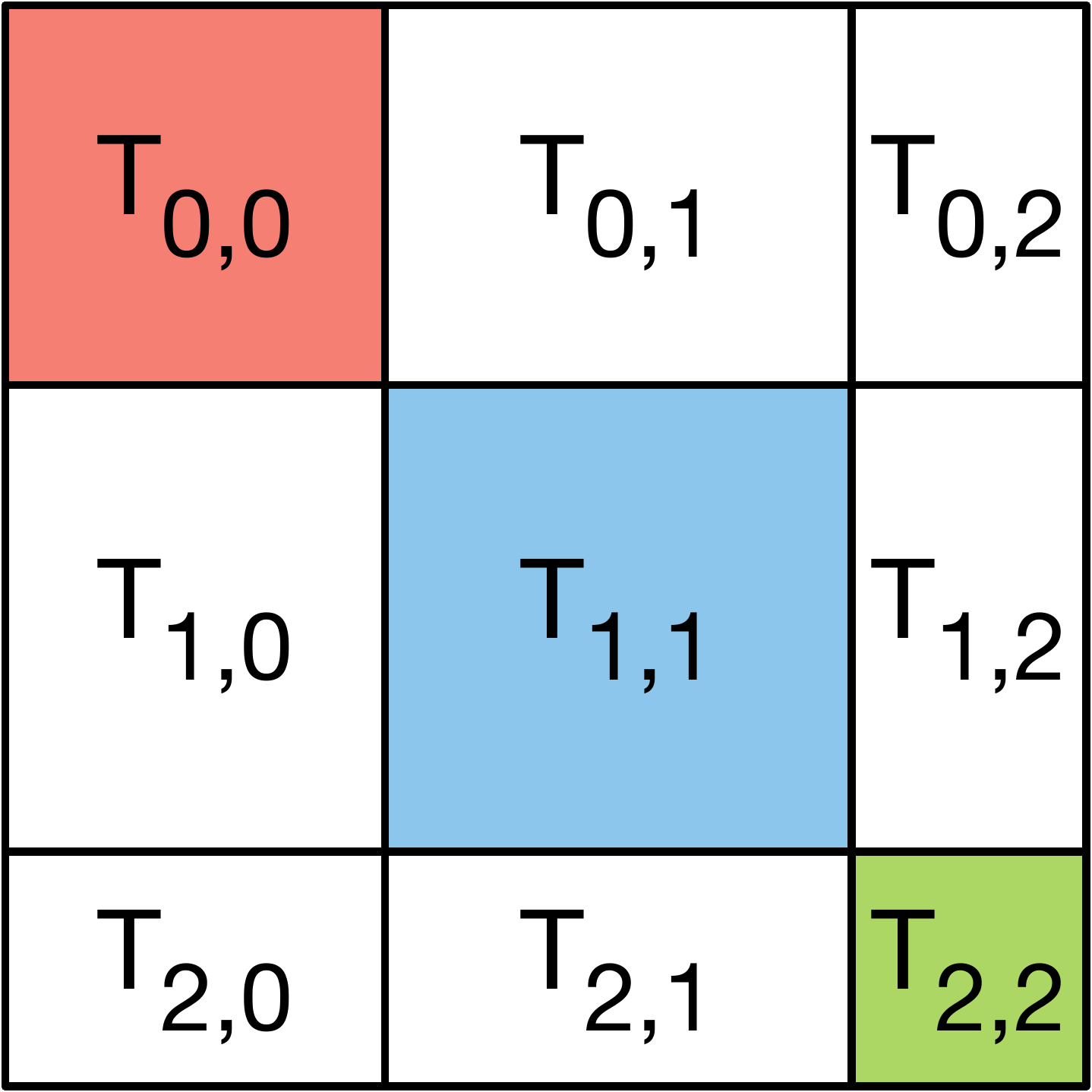}
  \label{fig:ex-symrec}}
  \hspace{3em}
  \subfloat[Gather/Scatter data]{\includegraphics[width=.3\linewidth]{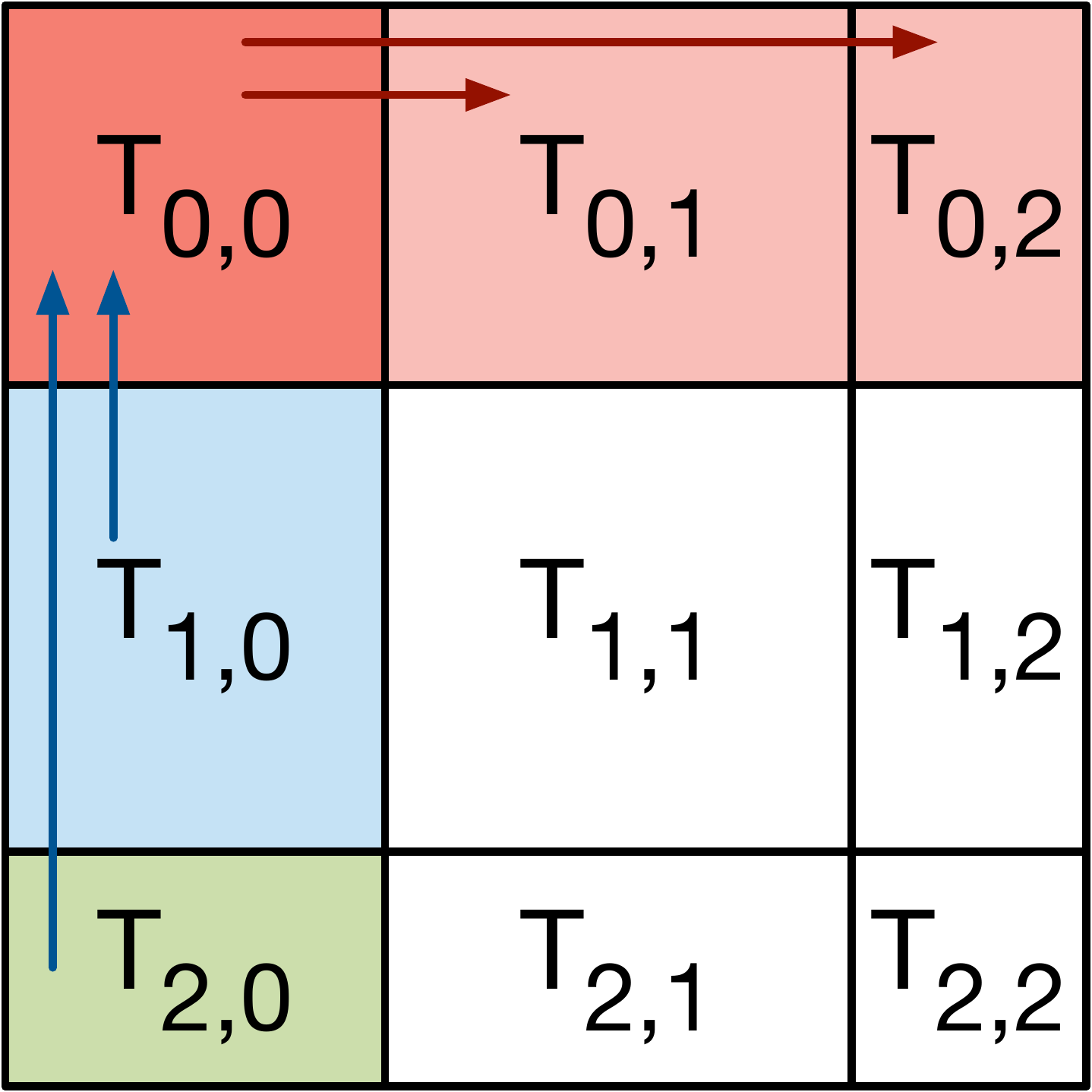}
  \label{fig:ex-symrec-info}}
  \caption{Symmetric rectilinear matrix partitioning.}
  \label{fig:intro-ex}
\end{figure}

A large class of work uses two-dimensional sparse matrices in their
design~\cite{Hendrickson950IJHSC,Im04-ICHPCA,Boman13-SC,Kepner16-HPEC}.
For these applications, spatial partitioning methods focus on dividing the load
using geometric properties of the workload. However, finding the optimal load
distribution among the partitions and also minimizing the imposed communication
(such as, the communication among {\em neighboring} parts)
is a difficult problem. For instance, uniform partitioning is useful to
regularize and {\em limit} the communication, but it ends up with highly
imbalanced partitions. Some of the most commonly used spatial partitioning
techniques like, Recursive Coordinate Bisection (RCB)~\cite{Berger87-TC},
jagged (or $m$-way jagged) partitioning~\cite{Ujaldon96-ICS,Saule12-JPDC-spart}
are useful to output balanced partitioning but may yield highly
irregular communication patterns. Rectilinear partitioning (i.e., generalized
block distribution)~\cite{Manne96-IWAPC,Grigni96-IWPAISP} tries to address these issues by aligning two different partition
vectors to rows and columns, respectively. In rectilinear partitioning, the
tiles are arranged in an orthogonal, but unevenly spaced grid. This partitioning
has three advantages; first, it limits the {\em number of neighbors} to 4
(or 8). Second, if communication along the {\em logical} rows and columns are needed, they will
also be bounded to a smaller number of processors (e.g., for $P=p\times p$ processor system,
it will be limited to $p$, i.e., $\sqrt{P}$). Third, more balanced blocks (in comparison
to uniform partitioning) can be generated.
Thus, the rectilinear partitioning gives a simple and well-structured communication
pattern if the problem has a local communication structure.

In many applications where the internal data is {\em square} matrices, such
as graph problems and iterative linear solvers for symmetric and non-symmetric
square systems, the sparse matrix (the adjacency matrix in graph algorithms),
represents the dependency of {\em input} elements to {\em output} elements.
In many cases, the next iteration's input elements are simply computed via
linear operations on previous input's output elements. For example, in graph algorithms,
the inputs and outputs are simply the same entity, vertices of the graph.
Hence, gathering information along the rows and then distributing the result
along the columns is an essential step, and generic rectilinear partitioning
would require additional communication for converting outputs of the previous
iteration to inputs of next iteration. One natural way to address
these issues is to use a conformal partitioning where diagonal tiles are squares.
This is a restricted case of rectilinear partitioning in which a partition vector
is aligned to rows and columns. We call this problem as {\em Symmetric Rectilinear
Partitioning Problem}, which is also known as, the symmetric generalized block
distribution~\cite{Grigni96-IWPAISP}.

This paper tackles the the {\em Symmetric Rectilinear Partitioning Problem},
finding an optimal rectilinear partitioning where diagonal tiles are squares
(see \Cref{fig:ex-symrec}).
Here, we assume that the given matrix is square and we partition that matrix into $p \times p$ tiles
such that by definition diagonal blocks will be squares.
In this type of partitioning diagonal tiles are the owners of matching input and
output elements. Hence, under this partitioning
scheme distributing/gathering information along the rows/columns becomes very
convenient  (see \Cref{fig:ex-symrec-info}).
Also, in the context of graphs, each tile can be visualized as sub-graphs where
diagonal tiles are the owners of the vertex meta-data and
any other tile represents the edges between two sub-graphs.
This type of partitioning becomes highly useful to reason about graph algorithms.
For instance, in a concurrent work, we have leveraged the symmetric rectilinear
partitioning for developing a block-based triangle counting formulation~\cite{Yasar19-HPEC}
that reduces data movement, during both sequential and parallel
execution, and is also naturally suitable for heterogeneous architectures.

The optimal rectilinear partitioning problem was shown to be NP-hard by Grigni
and Manne~\cite{Grigni96-IWPAISP}.
Symmetric rectilinear matrix partitioning is also a challenging problem even though
it appeared to be simpler than the rectilinear matrix partitioning, yet until our
work its complexity was unknown. In this work we show that the optimal symmetric rectilinear
partitioning problem is also NP-hard.
Here, we also define two variants of the symmetric rectilinear partitioning
problem and we propose refinement-based and probe-based partitioning heuristics
to solve these problems. Refinement-based
heuristics~\cite{Nicol94-JPDC,Manne96-IWAPC} apply a dimension reduction
technique to map the two-dimensional problem into one dimension and compute a partition
vector on one-dimensional data by running an optimal partitioning
algorithm~\cite{Nicol94-JPDC,Pinar04-JPDC}. Probe-based algorithms compute the partitioning
vector by seeking for the best cut for each point.
We combine
lightweight spatial partitioning techniques with simple heuristics. Contributions of
this work are as follows:

\begin{itemize}
\item Presenting two formulations for the symmetric rectilinear matrix partitioning
problem; {\sc minLoadImbal} (\mLI), {\sc minNumCuts} (\mNC), which are dual problems
of one another (\Cref{sec:pdef}).

\item Proving that optimal symmetric rectilinear partitioning is NP-hard (\Cref{ref:ssec:npc}).

\item Proposing efficient and effective heuristics for the symmetric rectilinear
partitioning problem (\Cref{sec:algs}).

\item Implementing an efficient sparse prefix-sum data structure
to reduce the computational complexity of the algorithms (\Cref{sec:spp}).

\item Evaluating the effectiveness of sparsification techniques on the
proposed algorithms (\Cref{ssec:sparsification}).

\item Extensively evaluating the performance of proposed algorithms in different
settings on more than six hundred real-world matrices (\Cref{sec:experimental}).
\end{itemize}

Our experimental results show that our proposed algorithms can
very efficiently find good symmetric rectilinear partitions and
output nearly the optimal solution on about $80$\% of the $375$ small graphs and
do not produce worse than $1.9$ times the optimal load imbalance.
We have also run our algorithms on more than 600 matrices and hence
experimentally validate the effectiveness of our proposed algorithms, as well as
our proposed sparsification techniques and efficient sparse prefix-sum
data structures.
Our algorithms take less than $10$ seconds in the Twitter graph that has
approximately $1.46$ billion edges on a modern $2\times12$ core system.
With sparsification, our algorithms can process Twitter graph in less than
$3$ seconds and output a solution whose load imbalance is no worse than $1\%$.

\section{Problem Definition and Notations}
\label{sec:pdef}

In this work, we are concerned with partitioning sparse matrices.
Let $A$ be a two-dimensional square matrix of size $n \times n$ that has
$m$ nonnegative nonzeros, representing the weights for spatial loads.
In the context of this work, we are also interested in partitioning the adjacency matrix
representation of graphs. A weighted directed graph $G=(V, E, w)$, consists of a
set of vertices $V$, a set of edges $E$, and a function mapping edges to weights, $w : E \to \mathbb{R}^+$.
A directed edge $e$ is referred to by $e = (u, v) \in E$, where $u, v \in V$, and $u$ is called
the source of the edge, $v$ is called the destination.
The neighbor list of a vertex $u \in V$ is defined as $N(u) = \{v\in V\mid(u, v)\in E\}$.
We use $n$ and $m$ for the number of vertices and edges, respectively,
i.e., $n=|V|$ and $m=|E|$. Let $A_G$ be the adjacency matrix representation
of the graph $G$, that is, $A_G$ is an $n\times n$ matrix,
where $\forall (u, v) \in E$, $A_G(u,v) = w(u, v)$, and everything else will be 0.
Without loss of generality, we will assume source vertices are represented
as rows, and destination vertices are represented as columns. In other words,
elements of $N(u)$ will correspond to column indices of nonzero elements
in row $u$. We will also simply refer to matrix $A_G$ as $A$ when $G$ is
clear in the context. \Cref{tab:notation} lists the notation used in
this paper.

\begin{table}[ht]
\centering
\caption{Notation used in this paper.}
\label{tab:notation}
  \begin{tabular}{r  l}
    \textbf{Symbol} & \textbf{Description}                \\
    \hline
    $G=(V, E, w)$   & A weighted directed graph $G$ with a vertex set, $V$, a edge\\
                    & \ set, $E$, and a nonnegative real-value weight function $w$\\
    $n=|V|$         & number of vertices \\
    $m=|E|$         & number of edges \\
    $A_G$           & $n\times n$ adjacency matrix of $G$, or simplified as $A$ \\
    $A(i, j)$       & the value at $i$th row and $j$th column in matrix $A$ \\
    $N(u)$          & Neighbor list of vertex $u$\\
    $\interval{a}{b}$        & Integer interval, all integers between $a$ and $b$ included\\
    $C$             & A partition vector; $C=\vect{c_0, c_1, \dots, c_{|C|-1}}$, $c_0<\dots<c_{|C|-1}$\\
    $C(i)$          & The $i$th lowest element in partition vector $C$, i.e., $c_i$.\\
    $C_r$, $C_c$    & Row and column partition vectors\\
    $T_{i,j}$       & Tile $i,j$ \\
    $\phi(T_{i,j})$           & Load of $T_{i,j}$, i.e., sum of the nonzeros in $T_{i,j}$\\
    $L_{max}(A, C_r, C_c)$    & Maximum load among all tiles\\
    $L_{avg}(A, C_r, C_c)$    & Average load of all tiles\\
    $\lambda(A, C_r, C_c)$    & Load imbalance for partition vectors, or simplified as $\lambda$\\
    $\lambda(A, C_r, C_c, k)$ & Load imbalance among $T_{i,j}$ st. $i,j \leq k$\\
    $s$             & Sparsification factor where $s\in [0,1]$ \\
    $\epsilon$      & Error tolerance for automatic sparsification factor selection\\
    \hline
  \end{tabular}
\end{table}

Given an integer $p$, $1 \leq p \leq n$,
let $C=\vect{c_0, c_1, \dots, c_p}$ be a partition vector
that consists of sequence of $p+1$ integers such that
$0=c_0<c_1<\dots<c_p=n$. Then $C$ defines a partition of $\interval{0}{n-1}$ into $p$ integer
intervals $\interval{c_i}{c_{i+1}-1}$ for $0 \leq i \leq p-1$.

\begin{definition}{Rectilinear Partitioning.}
Given $A$, and two integers, $p$ and $q$, a rectilinear partitioning
consists of a partition of $\interval{0}{n-1}$ into $p$ intervals
($C_r$, for rows) and into $q$ intervals ($C_c$, for columns) such that $A$ is partitioned into
non-overlapping $p \times q$ contiguous tiles.
\end{definition}

\begin{figure}[ht]
  \centering
  \subfloat[Non-Symmetric: $C_r=\{0,1,4,6\}$, $C_c=\{0,3,5,6\}$]{\includegraphics[width=.34\linewidth]{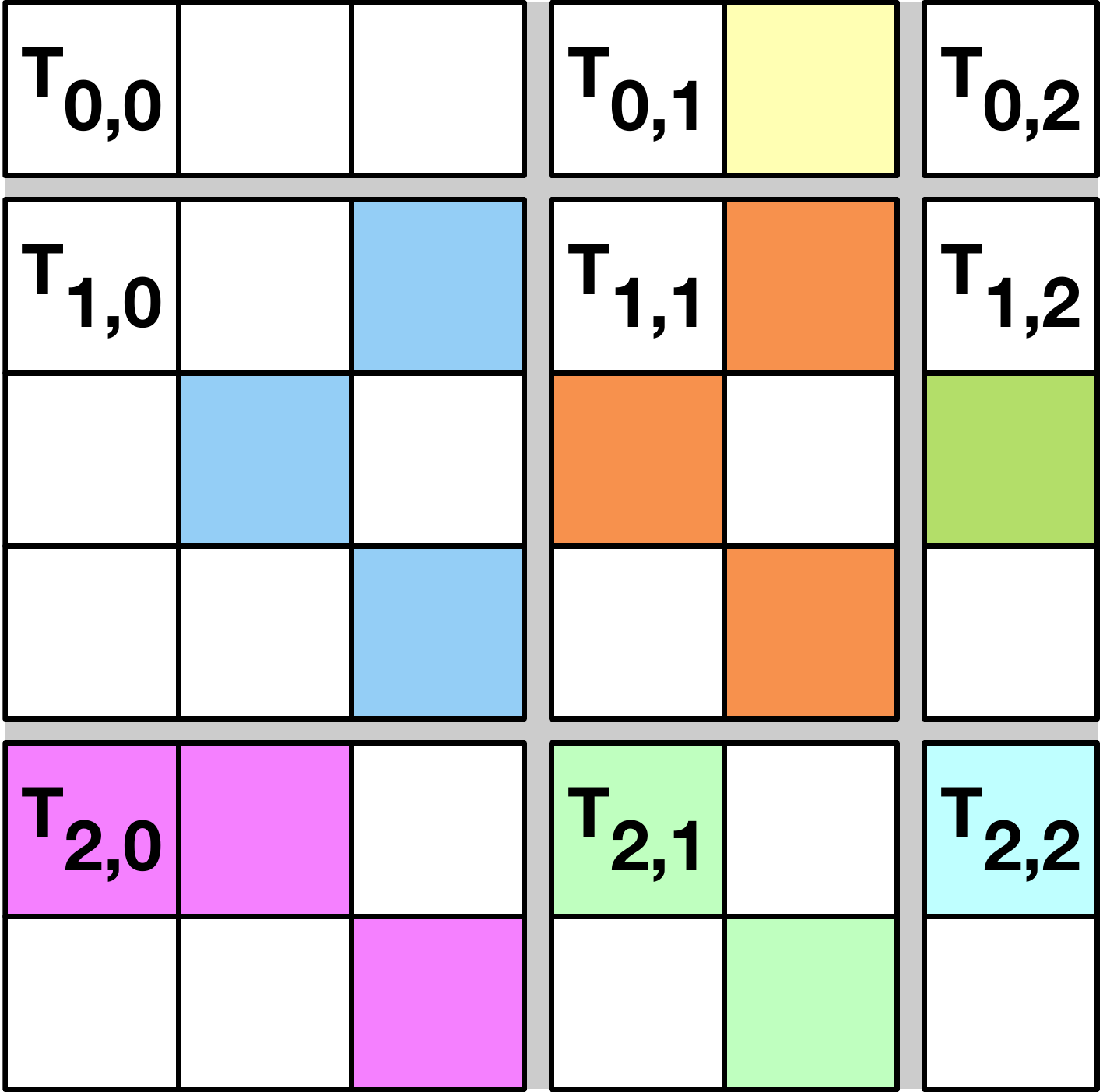}
  \label{fig:ex-nic}}
  \hspace{2em}
  \subfloat[Symmetric: $C = \{0,3,5,6\}$]{\includegraphics[width=.34\linewidth]{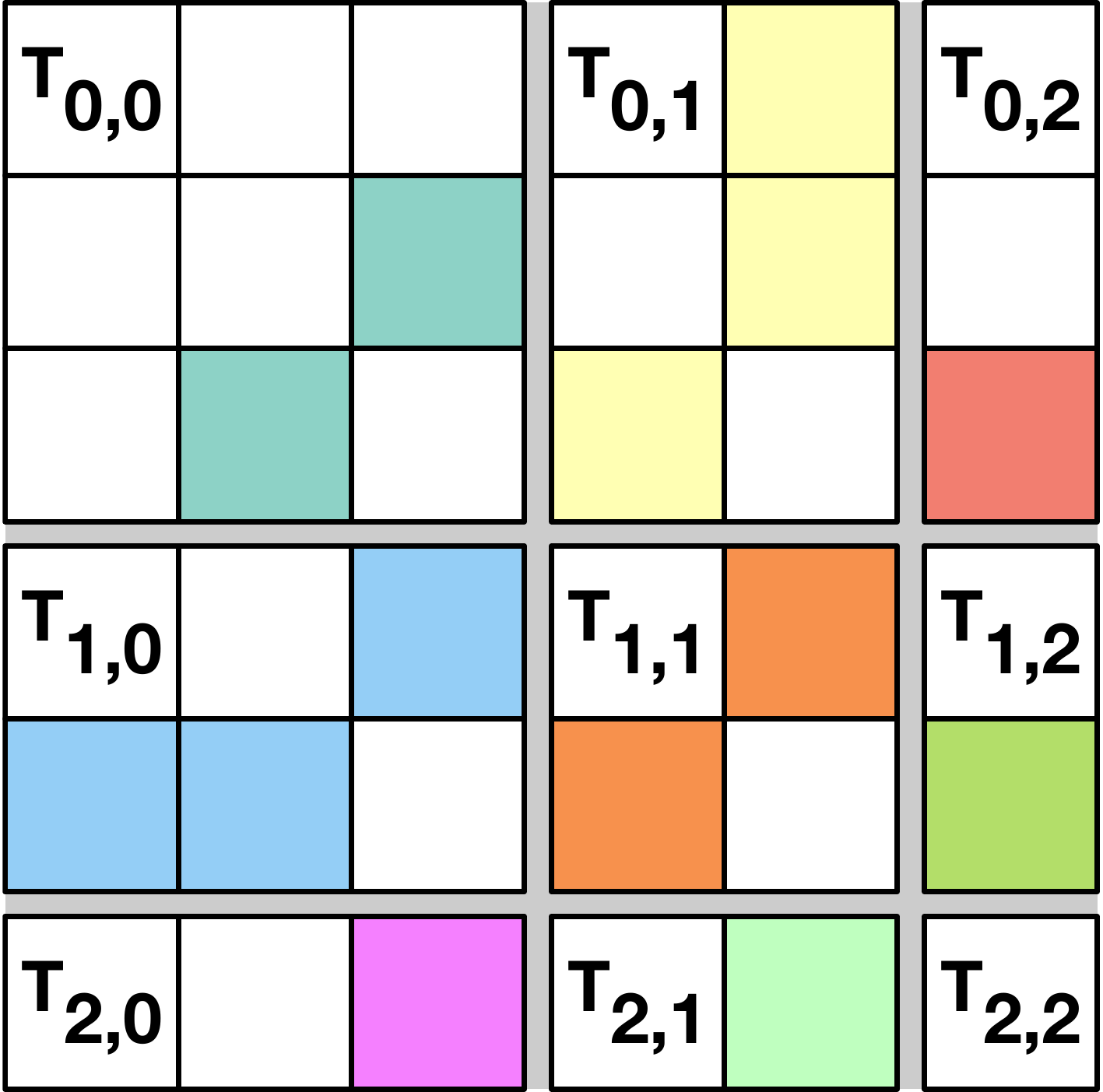}
  \label{fig:ex-ptc}}
  \caption{$3 \times 3$ Non-Symmetric/Symmetric rectilinear partitioning examples on
  the adjacency matrix representation of the toy graph in \Cref{fig:vpc-ex-graph}.}
  \label{fig:exsr}
\end{figure}

In rectilinear partitioning, a row partition vector, $C_r$, and a column partition vector, $C_c$,
together generate $p \times q$ tiles.
For $i \in [0..p-1]$ and $j\in[0..q-1]$, we denote $(i,j)$-th tile by $T_{i,j}$.
$\phi(T_{i,j})$ denotes the load of $T_{i,j}$, i.e., the sum of nonzero values in $T_{i,j}$.
Given partition vectors, quality of partitioning can be defined using
load imbalance, $\lambda$, among the tiles, which is computed as
$$\lambda(A, C_r, C_c) = \frac{L_{max}(A, C_r, C_c)}{L_{avg}(A, C_r, C_c)}$$
\noindent where
$$L_{max}(A, C_r, C_c) = \max_{i, j} \phi(T_{i,j}) $$ and
$$L_{avg}(A, C_r, C_c) =\frac{\sum_{i, j}\phi(T_{i,j})}{p \times q}.$$
A solution which is perfectly balanced achieves a load imbalance,
$\lambda$, of $1$. \Cref{fig:ex-nic} presents a toy example for
rectilinear partitioning where $C_r=\vect{0, 1, 4, 6}$, $C_c=\vect{0, 3, 5, 6}$
and $\lambda(A, C_r, C_c) = \frac{3}{(14/9)} \approx 1.9$ when we assume that
all nonzeros are equal to $1$.

\begin{definition}{Symmetric Rectilinear Partitioning.}
Given $A$ and $p$, symmetric rectilinear partitioning
can be defined as partitioning $\interval{0}{n-1}$ into $p$ intervals,
applying which to both rows and columns, such that
$A$ partitions into $p \times p$ non-overlapping contiguous tiles where
diagonal tiles are squares.
\end{definition}

In symmetric rectilinear partitioning, the same partition vector, $C_c= C =C_r$,
is used for row and column partitioning.
\Cref{fig:ex-ptc} presents a toy example for the symmetric rectilinear
partitioning  where $C = \vect{0,3,5,6}$
and $\lambda(A, C, C) = \frac{3}{(14/9)} \approx 1.9$.

In the context of this work, we
consider two symmetric rectilinear partitioning problems; {\sc minLoadImbal}
and {\sc minNumCuts}. These two problems are the dual of each other.

\begin{definition}{{\sc minLoadImbal} \textup{(}\mLI\!\textup{)} Problem.} \label{def::mli}
Given a matrix $A$ and an integer $p$, the \mLI problem consists in finding the
optimal partition vector, $C$, of size $p$
that minimizes the load imbalance:
\begin{align*}
\mLI( A, p ) &= \min_{C} \lambda(A, C, C)
\end{align*}
\end{definition}

\begin{definition}{{\sc minNumCuts} \textup{(}\mNC\!\textup{)} Problem.} \label{def::mnc}
Given a matrix $A$ and a maximum load limit $Z$, the \mNC problem consists of finding the minimum number of intervals $p$
that will partition the matrix $A$ so that the sum of nonzeros in all tiles are bounded by $Z$.

\begin{align*}
\mNC( A, Z ) = \min_{C} |C|,\\
\text{ s.t. } L_{max}(A, C, C) \leq Z
\end{align*}
\end{definition}

\section{Related Work}
\label{sec:related}

\begin{figure}[!ht]
\centering
\subfloat[Cartesian]{\includegraphics[width=0.21\linewidth]{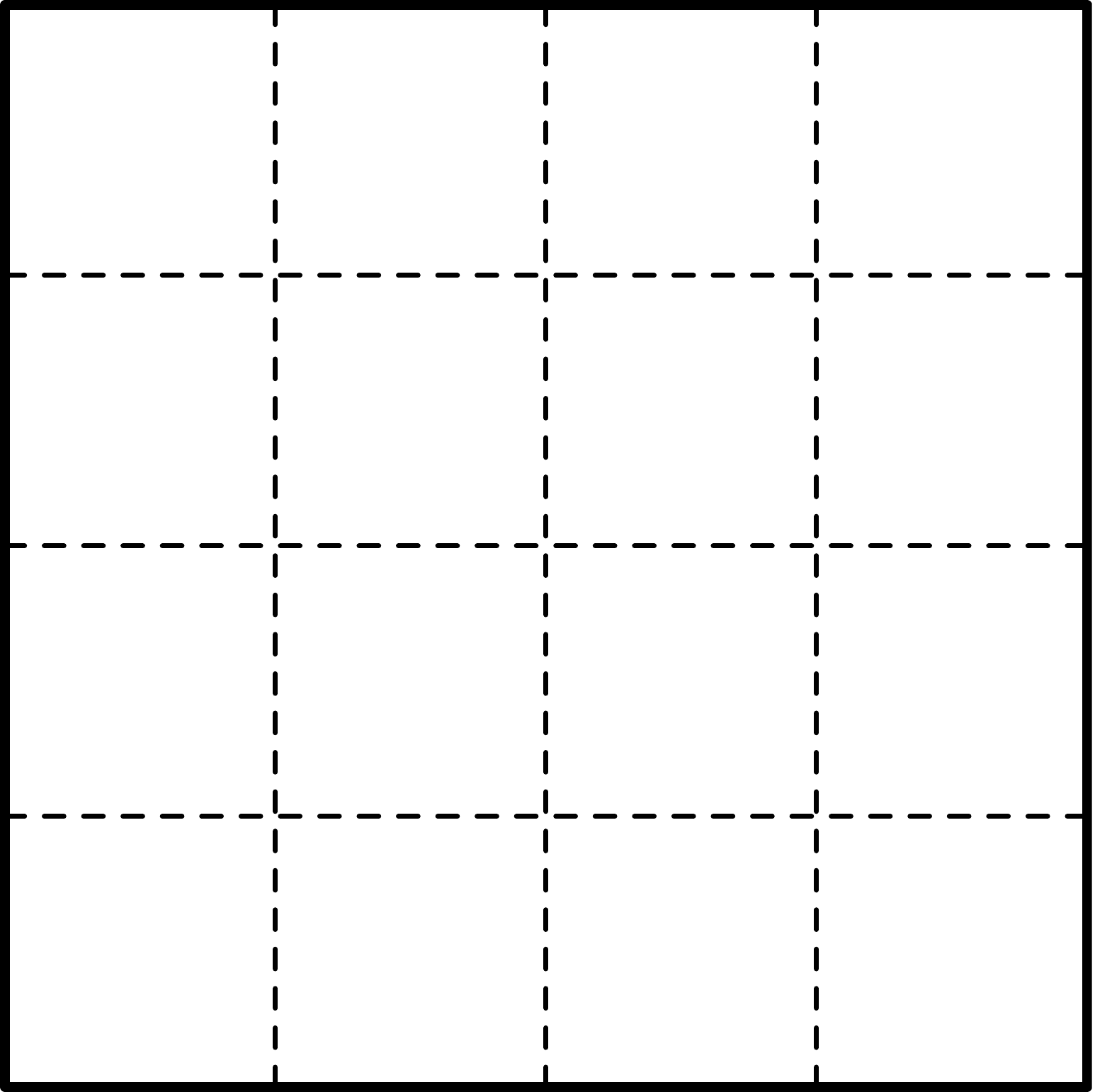}
\label{fig:uni2d}}
\hspace{1em}
\subfloat[Rectilinear]{\includegraphics[width=0.21\linewidth]{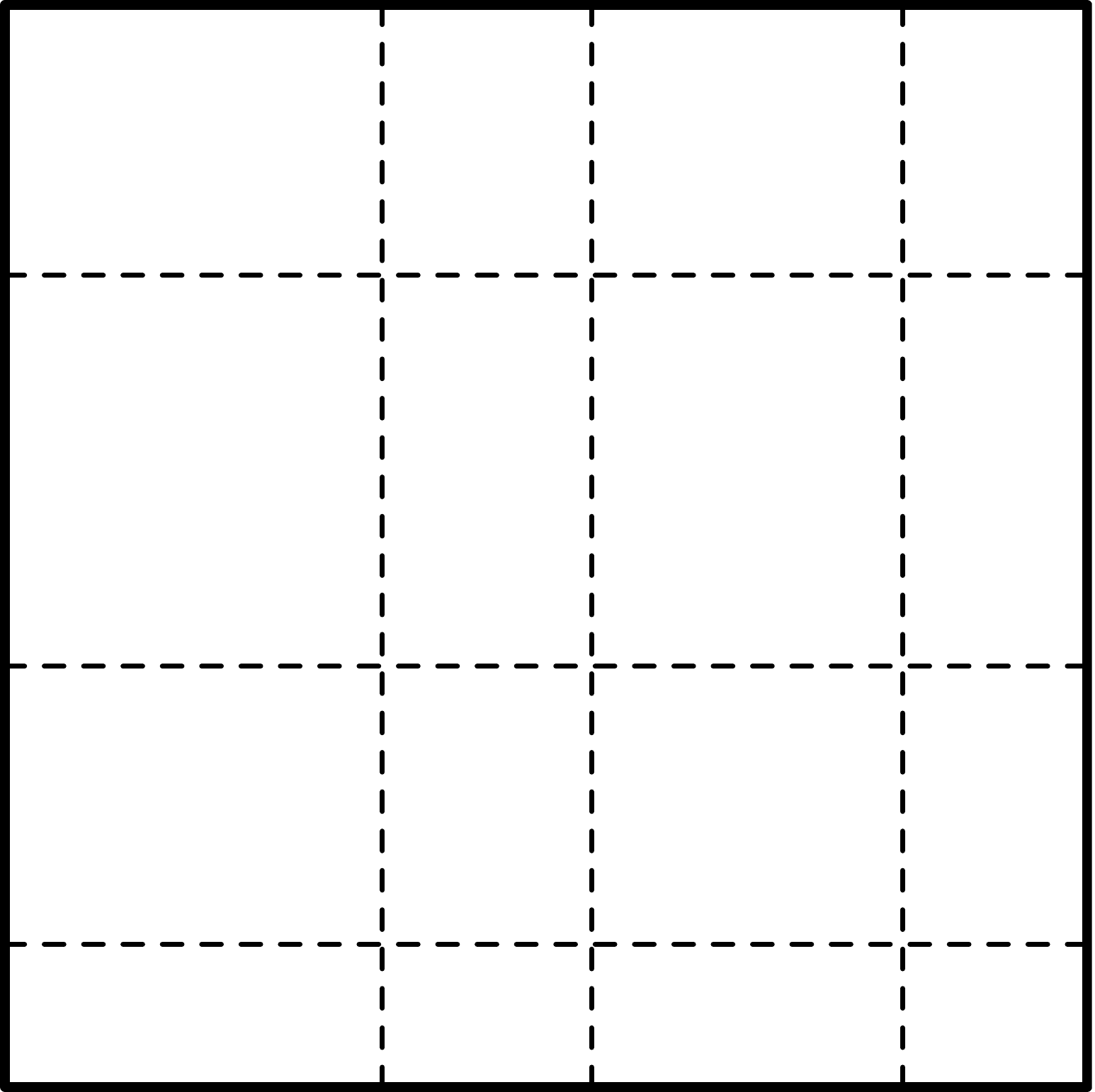}
\label{fig:recti}}
\hspace{1em}
\subfloat[RCB]{\includegraphics[width=0.21\linewidth]{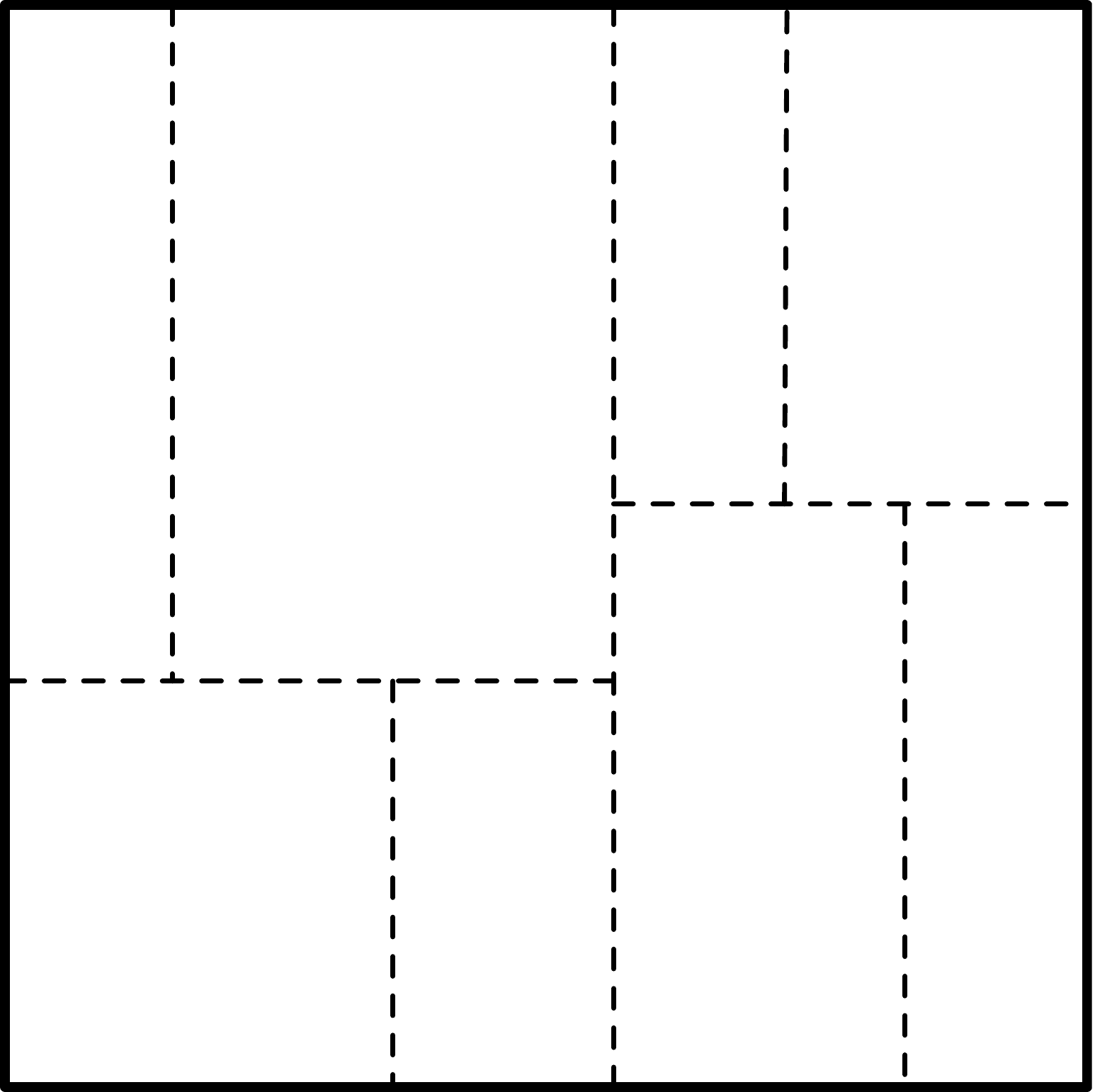}
\label{fig:rcb}}
\hspace{1em}
\subfloat[Jagged]{\includegraphics[width=0.21\linewidth]{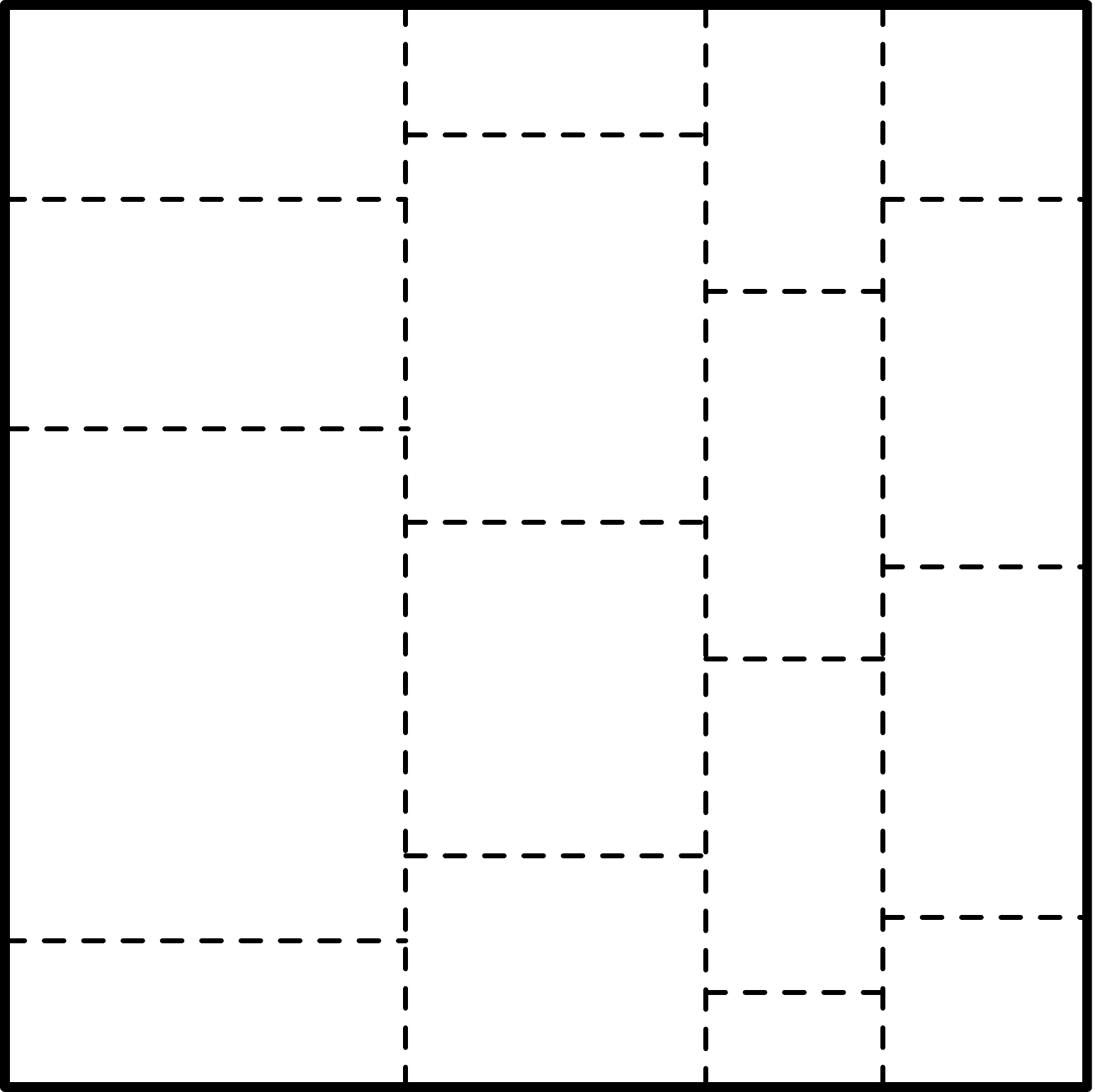}
\label{fig:jagged}}
\caption{Spatial partitioning examples.}
\label{fig:spatial}
\end{figure}

Two-dimensional matrix distributions have been widely used in dense
linear algebra~\cite{Hendrickson950IJHSC,Lewis94-HPCC}.
Cartesian~\cite{Hendrickson950IJHSC} distributions (see \Cref{fig:uni2d})
where the same partitioning vector is used to partition rows and columns are
widely used. This is due to the partitioned matrix naturally mapping onto a
two-dimensional mesh of processors. This kind of partitioning becomes highly
useful to limit the total number of messages on distributed settings.
Dense matrices or well structured sparse matrices can be easily partitioned
with cartesian partitioning. However, for sparse and irregular
problems finding a good vector that can be aligned with both dimensions
is a hard problem. Therefore, many non-cartesian two-dimensional matrix
partitioning methods have been
proposed~\cite{Berger87-TC,Meagher82-CGIP, Nicol94-JPDC, Manne96-IWAPC,Ujaldon96-ICS, Saule12-JPDC-spart} for
sparse and irregular problems.
As a class of shapes,
rectangles implicitly minimize communication,
allow many potential allocations, and
can be implemented efficiently with simple operations and data structures.
For these reasons, they are the main preferred shape.
For instance, recursive coordinate
bisection (RCB)~\cite{Berger87-TC}
is a widely used technique
that rely on a recursive decomposition of the domain (see \Cref{fig:rcb}).
Another widely used technique is called jagged partitioning~\cite{Ujaldon96-ICS,
Saule12-JPDC-spart}
which can be simply achieved by first partitioning the matrix into one-dimensional (1D) row-wise or
column-wise partitioning, then independently partitioning in each part (see \Cref{fig:jagged}).

One way to overcome the hardness of proposing one partition vector for rows and
columns is to propose different partition vectors for rows and columns.
This problem is named as rectilinear partitioning~\cite{Nicol94-JPDC}
(or generalized block distribution~\cite{Manne96-IWAPC}).
Independently, Nicol~\cite{Nicol94-JPDC}
and Manne and S{\o}revik~\cite{Manne96-IWAPC} proposed an algorithm to solve this
problem that is based on iteratively improving a given solution by alternating
between row and column partitioning.
These algorithms
transform two-dimensional (2D) rectilinear partitioning problem into
1D partitioning problem using a heuristic
and iteratively improves the solution,
in which the load of a interval is calculated as the maximum of loads in columns/rows in the interval of rows/columns.
This refinement
technique is presented in Algorithm~\ref{alg:rowparti}.
Here, {\sc optimal1DPartition}($P$)~\cite{Nicol94-JPDC} is a function that returns the optimal
partition on rows of $P$.
Hence, Algorithm~\ref{alg:rowparti} returns the optimal 1D row partition for
the given column partition $C_c$.

\begin{algorithm2e}[ht]
  \caption{\label{alg:rowparti} {\sc refinement($A, C_c, p$)}}
  \tcp*[l]{$P$ is a $n\times p$ matrix to store interval sums for each row}
  $P(i,j) \leftarrow 0$, for $i \in \interval{0}{n}$ and $j \in \interval{0}{p-1}$\;

  \tcp*[l]{for each row}
  \For{$i=0$ to $n-1$}{
    \For{{\bf each} $j$, where $A(i,j) \neq 0$}{
      $k \leftarrow 0$ \tcp*[l]{Interval index}
      \While{$j \geq C_c(k+1)$}{
        $k \leftarrow k+1$ \tcp*[l]{Find the interval}
      }
      $P(i+1,k) \leftarrow P(i+1,k)+A(i+1,j)$ \;
    }
  }

  \tcp*[l]{Compute $p$ prefix sums for each interval}
  \For{$j=0$ to $p-1$}{
      \For{$i=1$ to $n$}{
      $P(i,j) \leftarrow P(i,j)+P(i-1,j)$ \;
    }
  }

  \tcp*[l]{Return the optimal partitioning on rows of $P$}
  \Return \sc{optimal1DPartition($P$)}

\end{algorithm2e}

The optimal solution of the rectilinear partitioning was shown to be NP-hard by Grigni and
Manne~\cite{Grigni96-IWPAISP}. In fact, their proof shows that the
problem is NP-hard to approximate within any factor less than $2$. Khanna et
al.~\cite{Khanna97-ICALP} have shown the problem to be constant-factor
approximable.

Rectilinear partitioning may still cause high load-imbalance due to
generalization. Jagged partitions~\cite{Ujaldon96-ICS} (also called Semi Generalized Block
Distribution~\cite{Grigni96-IWPAISP}) tries to overcome this problem by
distinguishing between the main dimension and the auxiliary dimension (see \Cref{fig:jagged}).
The main dimension is split into $p$ intervals and each of these intervals
partition into
$q$ rectangles in the auxiliary dimension.
Each rectangle of the solution must have its main dimension matching one of
these intervals. The auxiliary dimension of each rectangle is arbitrary.
We refer readers to Saule et al.~\cite{Saule12-JPDC-spart} which presents multiple variants and
generalization of jagged partitioning, and also detailed comparisons of various
2D partitioning techniques.

Most of the algorithms we present require querying the load in a rectangular
tile and this problem is known as the dominance counting problem in
the literature~\cite{Joseph96}. Given a set of $d$-dimensional points $S$ and a $d$-dimensional
query point $x = \vect{x_1, \ldots ,x_d}$, dominance counting problem returns the number of
points $y = \vect{y_1, \ldots ,y_d}$, such that $y\in S$, and $y_i \leq x_i, \forall i\in \interval{1}{d}$.
\cite{Joseph96} presents an efficient data structure that can answer such
queries in $O\left(\frac{\log |S|}{\log \log |S|}\right)$ time with $O(|S|)$ space usage. However,
this data structure is very complex and hard to implement. Hence we propose another
data structure that we are going to cover in \Cref{sec:spp}.

\section{Symmetric Rectilinear Partitioning is NP-hard}
\label{ref:ssec:npc}

We first define the decision problem of the symmetric rectilinear partitioning for the proof.

\begin{definition}{Decision Problem of the Symmetric Rectilinear Partitioning.}
Given a matrix $A$, the number of intervals $p$, and a value $Z$,
decision problem of the symmetric rectilinear partitioning (\srpp)
seeks whether there is a partition vector of size $p+1$ such that the sum
of the nonzero values in each tiles are bounded by $Z$.
\end{definition}

It's clear \srpp is in NP.
We show that it is NP-complete by reducing a well-known NP-complete problem, vertex cover problem(\vcp),
to \srpp.

\begin{definition}{Vertex Cover Problem (\vcp).}
Given an undirected graph $G=(V, E)$ and an integer $K$, \vcp is to decide whether
there exist a subset $V'$ of the vertices of size $K$ such that at least one end point of every edge
is in $V'$, i.e., $\forall (u, v) \in E$, either $u \in V'$ or $v \in V'$.
\end{definition}

\begin{figure}[ht]
    \centering
    \subfloat[Graph]{\includegraphics[width=.35\linewidth]{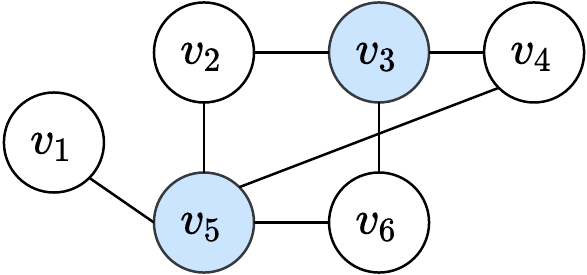}
    \label{fig:vpc-ex-graph}}
    \hspace{5em}
    \subfloat[Adjacency matrix]{\includegraphics[width=.45\linewidth]{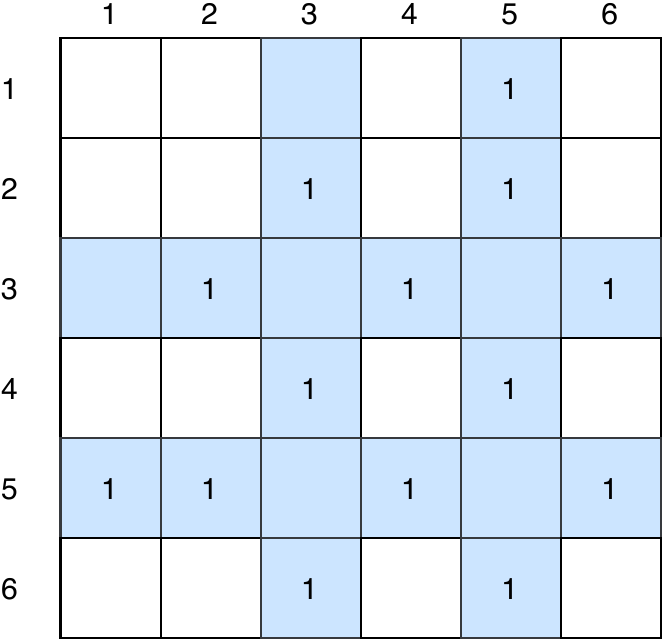}
    \label{fig:vpc-ex-mtx}}
    \vspace{1em}
    \subfloat[Transformation]{\includegraphics[width=.45\linewidth]{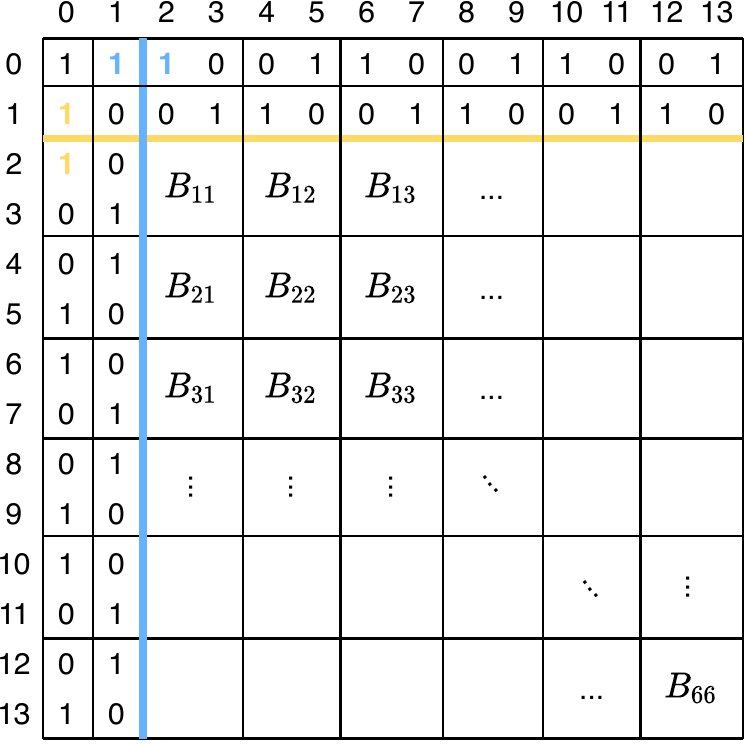}
    \label{fig:npc-ex1}}
    \hspace{1em}
    \subfloat[Tile matrix]{\includegraphics[width=.45\linewidth]{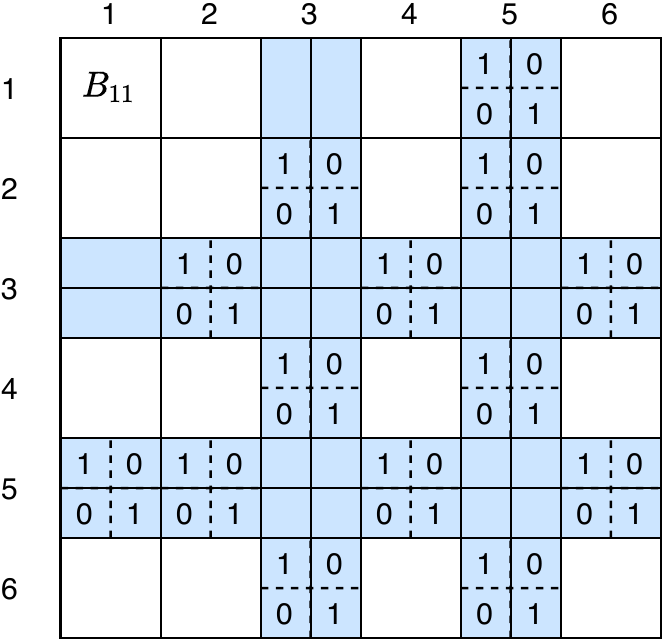}
    \label{fig:npc-ex2}}
    \caption{A toy example for \vcp $(K=2)$ and its equivalent \srpp instance $(Z=1, p=10)$}
    \label{fig:vpc-ex}

\end{figure}

\Cref{fig:vpc-ex-graph} illustrates a toy example for \vcp. In this example the graph
consists of $6$ vertices and $7$ edges. For $K=2$, $V'=\{v_3,v_5\}$ is a
solution.

We extend Grigni and Manne's~\cite{Grigni96-IWPAISP} input reduction
technique to reduce \vcp to \srpp.
Given a graph $G=(V, E)$, $V=\{v_1, v_2, \dots, v_n\}$, its adjacency matrix $A$ (see \Cref{fig:vpc-ex-mtx}), and an integer $K$ we apply six
transformation steps.
First, we create a new square binary matrix, $A'$, of size $(2n+2) \times (2n+2)$,
initialized with zeros (see \Cref{fig:npc-ex1}).
Second, we limit the sum of nonzeros in each result tile to
be at most one ($Z=1$).
This limitation allows us to enforce cuts by placing nonzeros at
adjacent positions in the matrix. For instance, in \Cref{fig:npc-ex1},
consequent nonzeros at $A'(0,1)$ and $A'(0,2)$ enforce a cut between the
column $1$ and the column $2$. Two sample cuts are highlighted in \Cref{fig:npc-ex1}.
As the third step, we initialize the first two rows
and two columns as follows: we set $A'(0,0)=1$, then in first row and column we
put two $1$s followed by two $0$s, until the end of row or column. Similarly,
starting $A'(1,1)$ position in second row and column, we now first put two $0$s
followed by two $1$s.
Fourth, the rest of the $A'$ matrix is tiled as $n\times n$
tiles of sizes $2\times 2$ (see \Cref{fig:npc-ex1}). Here, let
$B_{i,j}$ represent the $2\times 2$ tile located at
position $A'(2i, 2j)$ in the matrix, where $i,j \in \interval{1}{n}$.
Fifth, we initialize each $B_{i,j}$ with an identity matrix of size two
($I_2$) if $A(i,j)=1$ (see \Cref{fig:npc-ex2}).
Since we limit the number of nonzero elements in each result tile to be at most one,
an identity matrix has to be cut by at least one horizontal or vertical cut.
Last, we set the number of intervals, $p$, as $n+2+K$, thus, $n+3+K$ cuts will
be sought.
With the enforced $n+1$ cuts from the third step and the $2$ cuts at the beginning and the end,
possible cuts left are only in between of each row (and column)
of the $2 \times 2$ tiles and there are only $n$ of them.
Thus the problem becomes, choosing $K$ rows (columns) of $2 \times 2$
tiles to be cut, among $n$ possible cuts,
s.t., all $2 \times 2$ identity matrices are covered.

The equivalence between the constructed \srpp instance and the \vcp comes from both
about choosing rows and columns to cover the nonzero elements. They only differ in choosing
elements, the former is a $2 \times 2$ matrix, and the latter is $1 \times 1$ matrix,
as shown with the example in \Cref{fig:npc-ex2} and \Cref{fig:vpc-ex-mtx}.
The formal proof shows there is a solution for one instance if and only if there
is for the other, as follows.

\begin{proof}{NP-Completeness Proof of \srpp.}
Let $C$ denote the partition vector.
Let a set $S_0$ contains the trivial cuts in $C$, i.e., $\vect{0, 2n+2}$,
set $S_1$ contains the forced cuts, i.e.,
$\{1\}\cup \{2i \mid i\in [1, n]\}$, and $S_2$ contains the remainder cuts in $C$.

$\Rightarrow$ Suppose $V'$ is a solution to the \vcp instance. Then, let $S_2 = \{ 2i+1 | v_i \in V' \}$.
Since $|S_0 \cup S_1|=3+n$, and $|S_2|=K$, we have $|C| = n+3+K = p+1$.

The tiles in first two rows and columns all have a load of at most 1 after the forced cuts.
For the rest of the $2 \times 2$ nonzero tiles $B_{i, j}$, we have the following:
\begin{align*}
    B_{i,j} \text{ is an identity matrix} &\implies A(i, j) \neq 0\\
    &\implies v_i \in V' \text{~or~} v_j \in V' \\ &\implies 2i+1 \in S_2 \text{~or~} 2j+1 \in S_2
\end{align*}

All nonzero $B_{i,j}$ are cut by $S_2$ such that the load is at most 1 for tiles, showing $C$ is valid.

$\Leftarrow$ a similar logic can be applied in the reverse order to complete the proof. We are
omitting for the sake of brevity.
\end{proof}

\subsection{A mathematical model for the \srpp problem}
\label{ref:ssec:mm}

To the best of our knowledge, this is the first work that tackles the symmetric
rectilinear partitioning problem. Symmetric rectilinear partitioning is a
restricted problem, therefore comparing our algorithms with more relaxed
partitioning algorithms (such as jagged, rectilinear etc.) does not provide
enough information about the quality of the found partition vectors. Hence,
we implemented a mathematical model that finds the optimal solution and run
this model on small matrices to compare optimal solutions with the output of
our algorithms.

\begin{align}
  \text{\bf minimize} & \quad {L_{\text{max}}} \nonumber\\
  \text{\bf subject to} & \nonumber\\
  C         &= \vect{0=c_0<c_1\dots c_{p-1}<c_p=n} \label{mpl:cutvect}\\
  I(i,j)    &= 1 \iff c_i \leq j & \scriptstyle (i = 0,\dots, p) (j=0,\dots,n-1) \label{mpl:nnzpos} \\
    X(i,j,u, v)  &= 1 \iff 2 = I(c_i,u) - I(c_{i + 1},u)  & \scriptstyle (i,j=0,1,\dots,p-1) \label{mpl:nnztile} \\
            & \qquad \qquad + I(c_j,v) - I(c_{j + 1},v) &\scriptstyle (u, v = 0, \dots, n-1) \nonumber \\
    \phi(i,j)   &= \sum_{u,v=0,\dots,n-1} X(i,j,u,v)\times A(u, v) & \scriptstyle (i,j=0,1,\dots,p-1) \label{mpl:load}\\
  L_{\text{max}} & \geq \phi(i,j) &\scriptstyle (i,j=0,1,\dots,p-1) \label{mpl:maxload}
\end{align}

In the above model, the $C$ vector (\Cref{mpl:cutvect}) represents the monotonic
cut vector where each cut is an integer; $C \in \mathbb{Z}^{(p + 1)}$.
$I$ denotes
a $(p+1)\times n$ binary matrix, i.e., $I \in \{0, 1\}^{(p + 1) \times n}$,
where $I(i,j)=1$ if and only if the $i^{th}$ cut is to the left of $j^{th}$ column.
The $I$ matrix allows us to identify in which partition a row/column
appears, for example $I(c_i, j) - I(c_{i+1}, j) = 1$ if and only if $j$th row/column is in the $i$th partition.
$X$ is a binary $p\times p \times n \times n$ matrix. $X(i,j,u,v)=1$ if and only if
$A(u,v)$ is in tile $T_{i, j}$.
As shown in \Cref{mpl:nnztile} $X$ can be constructed using $I$.
Then, using $X$, we can represent tile loads as presented in \Cref{mpl:load}.
Finally, we define a variable, $L_{\text{max}}$ that stores the load of a
maximum loaded tile and the goal of the above model is to minimize
$L_{\text{max}}$.

\section{Algorithms for Symmetric Rectilinear Partitioning}
\label{sec:algs}

We propose two algorithms for the \mLI problem (Definition.~\ref{def::mli}) and two algorithms for the \mNC problem (Definition.~\ref{def::mnc}).
At a high level, those algorithms can be classified as refinement-based and probe-based.
In this section, we explain how these algorithms are designed.

\subsection{{\sc minLoadImbal} (\mLI) problem}

We propose two algorithms for the \mLI problem. One of those algorithms,
{\em Refine a cut} (\rac)
adopts previously defined refinement technique (see Algorithm~\ref{alg:rowparti})
into the symmetric rectilinear partitioning problem.
Note that the \rac algorithm has no convergence guarantee.
The second algorithm, {\em Bound a cut} (\bac),
implements a generic algorithm that takes an algorithm which solves the \mNC problem
as input and solves the \mLI problem.

\subsubsection{Refine a cut (\rac)}
\label{subsec:rac}

\rac algorithm first applies the refinement on rows, and then on columns independently.
Then it computes the load imbalances for the generated partition vectors.
The \rac algorithm chooses the direction (row or column) that gives a better load imbalance.
Then, iteratively applies the refinement algorithm only in this direction until it reaches
the iteration limit ($\tau$). This procedure is presented in the Algorithm~\ref{alg:rac}.

\begin{algorithm2e}[ht]
  \caption{\label{alg:rac} {\rac}($A, p$)}
  \tcp*[l]{Current ($C$) and previous ($C'$) partition vectors}
  $C(0) \leftarrow 0$; $C(j) \leftarrow n$, for $ 1 \leq j \leq p$ \;

  \tcp*[l]{Apply 1D partitioning refinement}
  $C_r \leftarrow$ {\sc refinement}($A, C, p$) \tcp*[r]{Row based}
  $C_c \leftarrow$ {\sc refinement}($A^T, C, p$) \tcp*[r]{Column based}

  \tcp*[l]{Aligning same partition vector for rows and columns}
  $L_{r} \leftarrow \lambda(A, C_r, C_r )$ \tcp*[r]{Row based imbalance}
  $L_{c} \leftarrow \lambda(A, C_c, C_c )$ \tcp*[r]{Column based imbalance}
  \eIf{ $L_{r}<L_{c}$ }{
      $C \leftarrow C_r$ \;
    }{
      $C \leftarrow C_c$ \;
      $A \leftarrow A^T$ \;
    }

  $i \leftarrow 0$ \;
  \While{ $i<\tau$ }{
    $C \leftarrow$ {\sc refinement}($A, C, p$) \;
    $i \leftarrow i+1$ \;
  }

  \Return $C$
\end{algorithm2e}

The primary advantage of this algorithm is its simplicity. This algorithm can be
easily parallelized as shown in~\cite{Manne96-IWAPC,Nicol94-JPDC}.
However, choosing a direction at the beginning may result in a missed opportunity to converge
to a better partition vector using the other direction.

\subsubsection{Bound a cut (\bac)}
\label{subsec:ptc}
\bac algorithm solves the \mLI problem given an algorithm that solves the \mNC problem.
Given a matrix $A$ and an integer $p$, the \bac algorithm seeks for the minimal load size,
$B$, such that \mNC algorithm returns a partition vector of size $p+1$.
In this approach, the \bac algorithm does a binary search over the range starting from $0$ to
the sum of nonzeros. In each iteration of the binary search, it runs \mNC algorithm
with the middle target load between lower and upper bounds, and halves the search
space. This procedure is presented in the Algorithm~\ref{alg:bac}.
Note that binary searching on the exponent first and then on the fraction can enable
efficient float value binary search in order to deal with the machine precision of real values.

\begin{algorithm2e}[ht]
  \caption{\label{alg:bac} {\bac}($A, p$)}
  \tcp*[l]{Initialize temporary partition vector}
  $C(0) \leftarrow 0$; $C(j) \leftarrow n$, for $ 1 \leq j \leq p+1$ \;

  $l \leftarrow 0$ \;
  $r \leftarrow \sum_{0\leq u,v <n} A(u,v)$ \;

  \tcp*[l]{Probe in binary search fashion}
  \While{$l<r$}{
    $B \leftarrow (l+r)/2$ \;
    $C \leftarrow $ \mNC($A, B$) \;
    \eIf{$|C| \leq p + 1$}{
      $r \leftarrow B$ \;
    }{
      $l \leftarrow B+1$ \;
    }
  }

  \Return \mNC($A, l$)
\end{algorithm2e}

\subsection{{\sc minNumCuts} (\mNC) problem}

Given a matrix, $A$, and an integer, $Z$, the \mNC problem
aims to output a partition vector, $C$, with the minimum number of intervals, $p$,
where the maximum load of a tile in the corresponding partitioning is less than $Z$, i.e.,
$\max_{0 \leq i,j \leq p}\phi(T_{i,j}) \leq Z$.

\begin{algorithm2e}[ht]
  \caption{\label{alg:pal} {\pal}($A, Z$)}

  \tcp*[l]{Initially we do not know partition vector's size}
  $C(0) \leftarrow 0$ \;
  $i \leftarrow 1$\;
  \While{$C(i-1)\neq n$}{
    $C(i) \leftarrow  \beta(A, C, i, Z) $\;
    $i \leftarrow i+1$ \;
  }

  \Return $C$

\end{algorithm2e}

\subsubsection{Probe a load (\pal)}
\label{subsec:pal}

Compared to our refinement based algorithm (\rac) which does not have any convergence guarantee,
the \pal algorithm guarantees outputting a partition vector at the local optimal in the sense
that removal or moving forward of any of the cuts will increase the maximum load.
That's why the \pal
algorithm is more stable and usually performs better than the \rac algorithm.
\pal is illustrated in Algorithm~\ref{alg:pal}.
The elements of $C$ are found through binary search, $\beta$, on the matrix.
In this algorithm, $\beta(A, C, i, Z)$, searches $A$ in the $\interval{C[i-1]}{n}$
to compute the largest $i^{th}$ cut point such that
$\max_{0\leq j,k \leq t}\{ \phi(T_{j,k}) \}\leq Z$.
Note that the \pal algorithm considers more cases in a two-dimensional fashion.

\subsubsection{Ordered probe a load (\opal)}
\label{subsec:opal}
The \opal algorithm tries to reduce computational complexity of the \pal algorithm by
applying a coordinate transformation technique to the input matrix,
presented in Algorithm~\ref{alg:trmtx}.
In Algorithm~\ref{alg:trmtx}, $A'$ is a three-dimensional matrix
where $A'(\max(i,j), \min(i,j),i>j)=v$, if $A(i,j)=v$.
To construct $A'$, for each nonzero, $A(i,j)\neq 0$,
in the $A$ matrix,
if $i>j$, we assign $A'(i,j,$ {\tt True}$)=v$, and otherwise $A'(i,j,$ {\tt False}$)=v$.
Note that, we visit each nonzero in $A$ following the row-major order and update $A'$
accordingly.
To avoid dynamic or dense memory allocations, we first, pre-calculate size of each row, i.e., $A'(i, :, :)$,
then allocate and insert nonzeros.

\begin{algorithm2e}[ht]
  \caption{\label{alg:trmtx} {\sc Transform}($A$)}

    \tcp*[l]{Initialize the three-dimensional matrix}
    $A'(i, j, b) \leftarrow 0$, for $0 \leq i,j \leq n-1$ and $b\in$ \{{\tt True}, {\tt False}\}\;

    \For{{\bf each} $i=0$ {\bf to} $n - 1$}{
      \For{{\bf each} $j=0$ {\bf to} $n - 1$}{
        \If{$A(i,j) \neq 0$}{
          $A'(\max(i, j), \min(i, j), i > j) \leftarrow A(i, j)$\;
        }
      }
    }

  \Return $A'$

\end{algorithm2e}

\begin{figure}[htb]
  \centering
  \subfloat[Transformed matrix, $A'$]{\includegraphics[width=.27\linewidth]{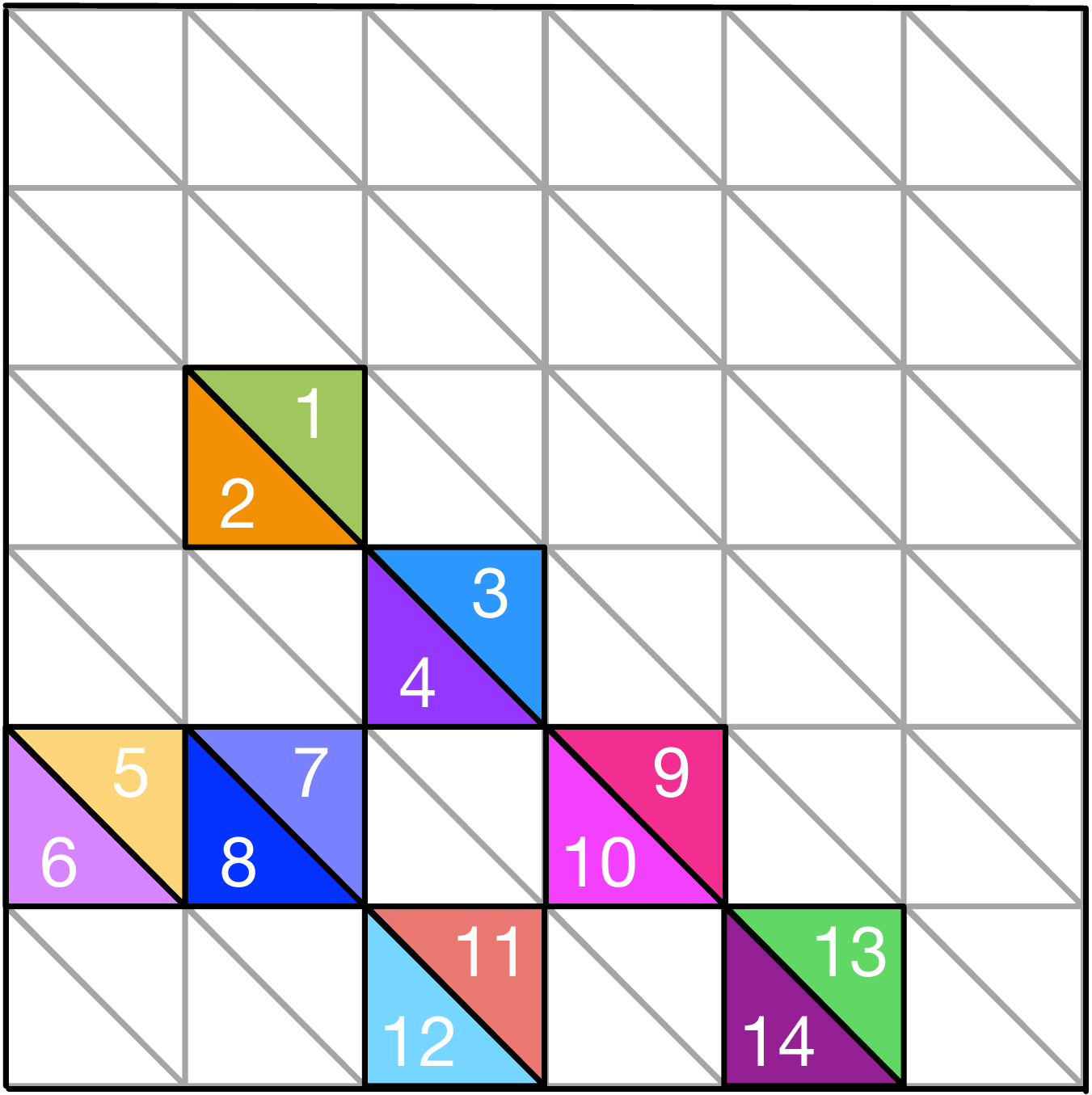}
  \label{fig:opal-transform}}
  \hspace{3em}
  \subfloat[Order of nonzeros]{\includegraphics[width=.27\linewidth]{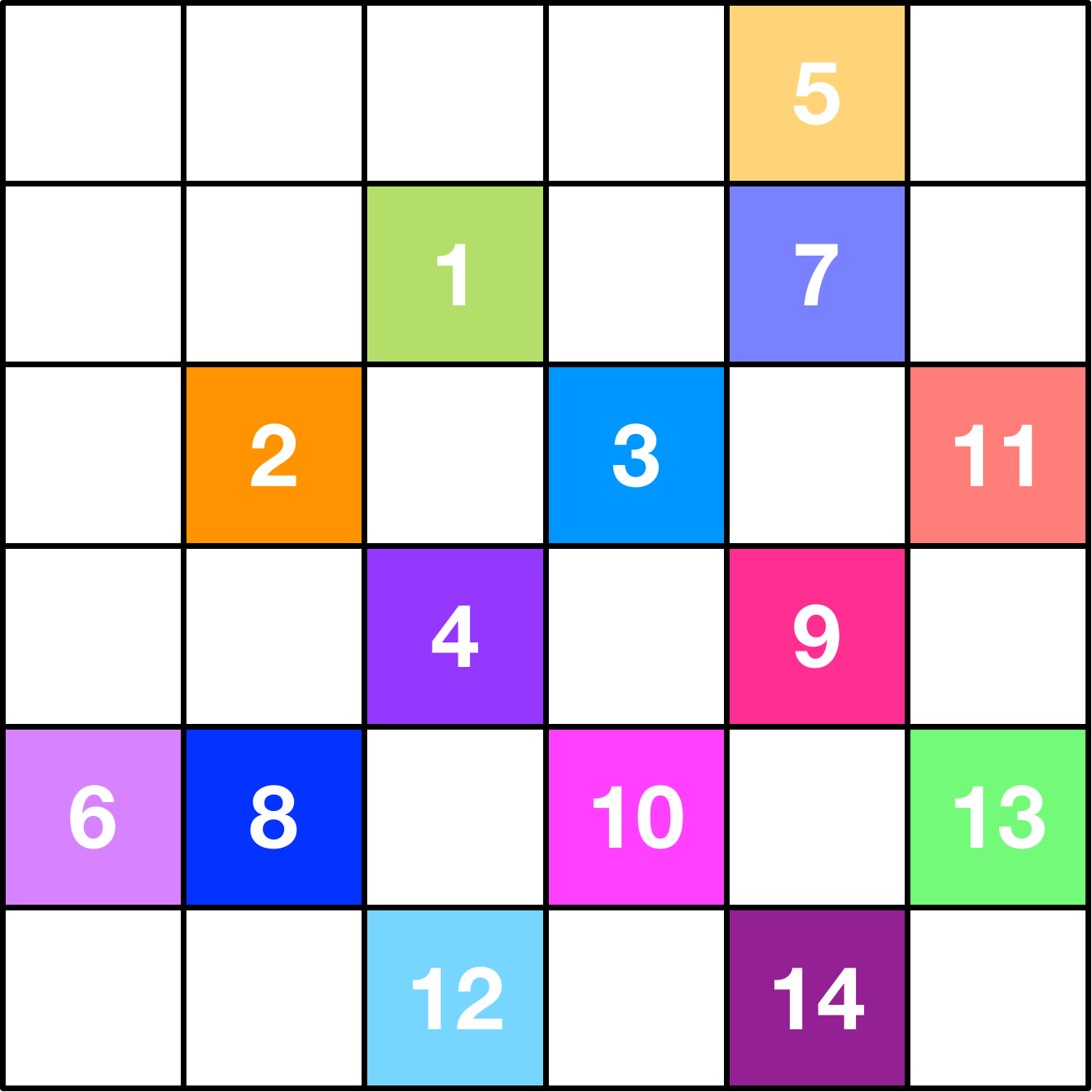}
  \label{fig:opal-traverse}}
  \caption{An example transformation for the toy graph in \Cref{fig:vpc-ex-graph}.
  \Cref{fig:opal-transform} $A'(\max(i,j), \min(i,j),i>j)=A(i,j)$.
  In \Cref{fig:opal-transform}, upper triangles represent $(i\leq j)$ and
  lower triangles represent $(i>j)$ for the third dimension in each cell.
  \Cref{fig:opal-traverse} illustrates the adjacency matrix where the order of each nonzero
  written in the corresponding cell when $A'$ is accessed in row-major order.}
  \label{fig:opal}
\end{figure}

After the transformation, going over the transformed matrix in row-major order becomes
equivalent to going over the original matrix in the diagonal-major order
(see \Cref{fig:opal-traverse}). Hence, we can make a single pass over the whole matrix to
compute the same partition vector as the \pal algorithm.

\opal is presented in Algorithm~\ref{alg:opal}.
When we are going over the transformed matrix in row-major order, we are trying to find the furthest
point from the previous cut so that if we put the next cut at that point, none of the newly created tiles
will exceed the load bound $Z$. When processing a single nonzero, we first find which of the newly created
tiles it will fall into and increment its load by the weight of the nonzero. Then we update the maximum
load of the newly created tiles, $l_{\text{max}}$. We stop when $l_{\text{max}}$ exceeds $Z$ and add the
index of the row we are
currently processing to the partition vector $C$. The algorithm terminates either when all the nonzeros have been processed
or when two of the same cuts are present in $C$ indicating infeasibility of partitioning with a load bound of $Z$.
Note that the \pal and the \opal algorithms outputs the same partition vector.
Our goal to propose the \opal algorithm is to decrease the computational complexity
of the \pal algorithm. The \pal algorithm is pleasingly parallel, hence the execution time can be decreased
significantly on a multi-core machine. However, on sequential execution the \pal algorithm's complexity
is worse than the \opal algorithm. Hence, the \opal algorithm is highly beneficial for the sequential
execution and the \pal algorithm is better to use on parallel settings.

\begin{algorithm2e}[ht]
  \caption{\label{alg:opal} {\opal}($A, Z$)}

  $A' \leftarrow$ {\sc Transform}($A$)\;
  $C(0) \leftarrow 0$ \tcp*[r]{Initially we do not know partition vector's size}
  $c \leftarrow 1$ \tcp*[r]{Track cut indices}
  $i \leftarrow 0$ \;
  \While{$i < n$}{
    $L_{1}(c) \leftarrow 0$, for $ 0 \leq c \leq |C|-1$ \tcp*[r]{Lower triangle and diagonal loads}
    $L_{2}(c) \leftarrow 0$, for $ 0 \leq c \leq |C|-1$ \tcp*[r]{Upper triangle loads}
    $l_{\text{max}} \leftarrow 0$\;
    \While(\tcp*[f]{To find the next cut}){$i < n$}{
      \For{{\bf each} $j=0$ {\bf to} $n - 1$ {\bf and} $b \in \{$ {\tt True}, {\tt False} $\}$}{
          \If{$A'(i, j, b) \neq 0$ }{
            $v \leftarrow A'(i,j,b)$\;
            $t \leftarrow \max_t \{t \mid C(t)\leq j\}$\;
            \If(\tcp*[f]{A lower or a diagonal tile}){ $b=${\tt True} {\bf or} $t = |C| - 1$} {
              $L_{1}(t) \leftarrow L_{1}(t) + v$\;
              $l_{\text{max}} \leftarrow \max(l_{\text{max}}, L_{1}(t))$\;
            }
            \Else{
              $L_{2}(t) \leftarrow L_{2}(t) + v$\;
              $l_{\text{max}} \leftarrow \max(l_{\text{max}}, L_{2}(t))$\;
            }
            \If{$l_{\text{max}}> Z$}{
              {\tt break}\;
            }
        }
      }
      \If{$l_{\text{max}} > Z$}{
        {\tt break}\;
      }
      $i \leftarrow i+1$\;
    }
    \If{$C(c-1) = i$}{
      {\tt break} \tcp*[r]{Infeasible $Z$}
    }

    $C(c) \leftarrow i$ \tcp*[r]{Append new cut to the partition vector}
    $c \leftarrow c+1$ \tcp*[r]{Increment index}
  }

  \Return $C$

\end{algorithm2e}

\subsubsection{Bound a load (\bal)}
\label{subsec:bal}
One can solve the \mNC problem using any algorithm that is proposed for solving the \mLI
problem, using binary searches over the possible number of cuts.
We call this procedure as {\em bounding a load} (\bal),
displayed in Algorithm~\ref{alg:bal}.
This approach can be improved in certain cases by bounding the search space of the candidate
number of cuts to decrease the number of iterations. For instance,
when the given matrix is binary, the search space can be initialized as $[1, \lceil \frac{n}{\lceil \sqrt{Z} \rceil} \rceil]$,
where the upper bound is derived by considering the dimension
$\lceil \sqrt{Z} \rceil$ of the smallest matrix that can contain $Z$ nonzeros.

\begin{algorithm2e}[ht]
  \caption{\label{alg:bal} {\bal}($A, Z$)}

  \tcp*[l]{$l$ and $u$ are the upper and lower bounds}
  $l \leftarrow 1$ \tcp*[l]{At least 1 cut}
  $u \leftarrow n$ \tcp*[l]{Number of rows}

  \While{$l<u$}{
    $p \leftarrow (l+u)/2$ \;
    $C \leftarrow $ \mLI($A, p$) \;
    $Z' \leftarrow L_{max}(A, C, C)$ \;
    \eIf{$Z'<Z$}{
      $u \leftarrow p$ \;
    }{
      $l \leftarrow p+1$ \;
    }
  }

  \Return \mLI($A, l$)
\end{algorithm2e}

\section{Sparse Prefix Sum data structure and Computational Complexity}
\label{sec:spp}

Given the partition vectors, querying numbers of nonzeros in each
tile is one of the computationally heavy steps of our proposed algorithms; a naive
approach requires iterating over all edges. We address this issue by proposing
a data structure to reduce the complexity of this query and thus
reduce the complexity of our algorithms.

\begin{algorithm2e}[ht]
  \caption{\label{alg:spsinit} {\sc SPSConstruction}($A$)}
  \tcp*[l]{Assuming that 1-based indexing is used for matrix $A$ and $A \geq 0$}

    \tcp*[l]{Initialize a zero $n\times n$ sparse matrix}
    $S(i, j) \leftarrow 0$ for $ 1 \leq i,j \leq n$ \;

    \For{{\bf each} $i=1$ {\bf to} $n$}{
      \For{{\bf each} $j=1$ {\bf to} $n$}{
        \If{$A(i,j)\neq 0$}{
          $v \leftarrow A(i, j)$\;
          $t \leftarrow j$\;
          \While{$t \leq n$}{
            $S(i, t) \leftarrow v + \max_{k \leq i} S(k, t)$\; \label{spp:persistent} 
            \tcp*[l]{Increment $t$ by adding its least significant bit to itself.}
            $t \leftarrow t + $ {\tt LSB}$(t)$\; \label{spp:insert}
          }
        }
      }
    }

  \Return $S$
\end{algorithm2e}

\subsection{Sparse prefix sum data structure}
\label{sec:ds}

\begin{figure}[ht]
  \centering
  \includegraphics[width=.95\linewidth]{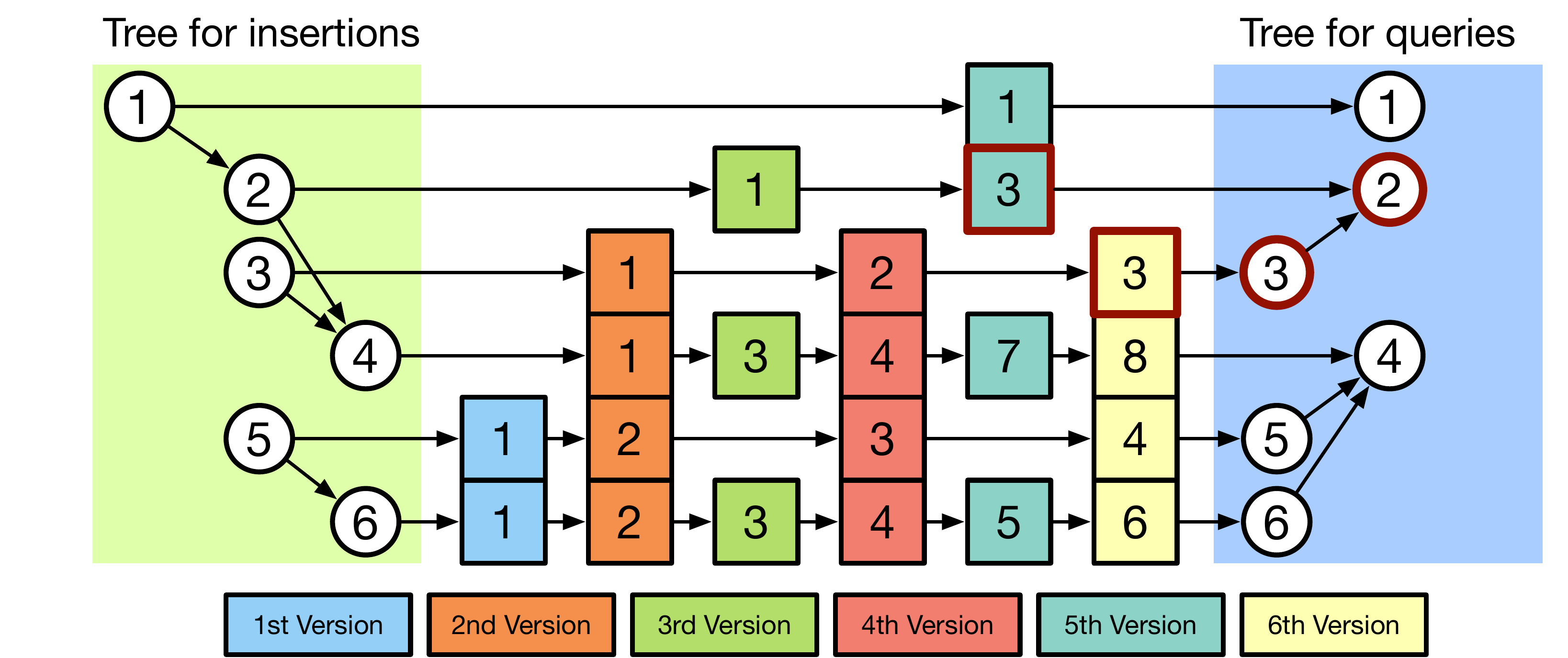}
  \caption{An illustration of our sparse prefix sum data structure on the toy
  graph presented in~\Cref{fig:vpc-ex-graph}. Sum of the highlighted cells
  gives us the result a query for the $3^{rd}$ index from version $6$ which
  is the load of the rectangle from $(1, 1)$ to $(6, 3)$ that is equal to 6.}
  \label{fig:spp-ds}
\end{figure}

One can query number nonzeros within
a rectangle using a two-dimensional cumulative sum matrix in constant time. However,
such a matrix requires $\Theta(n^2)$ space, which is infeasible for large problem
instances. Here, we propose an elastic sparse prefix sum
data structure which can query the load of a rectangle in $O(\log^2 n)$
time and requires $O(m \log n)$ memory space.
The data structure is essentially a persistent
Binary Indexed Tree (BIT)~\cite{Fenwick94-SPE}. We use fat node approach to transform BIT into
a persistent data structure as described in~\cite{Driscoll86-STC}. BIT is a data structure to query and maintain prefix sums
in a one-dimensional array of length $n$ using $O(n)$ space and $O(\log n)$ time.
Persistent data structures are dynamic data structures that let you query from a previous version
of it. The column indices of the nonzeros are inserted into the BIT in a row
major order with version number being the row index of the nonzeros.
Algorithm~\ref{alg:spsinit} presents the high-level algorithm and \Cref{fig:spp-ds}
illustrates an example representation of our data structure for the toy
graph presented in~\Cref{fig:vpc-ex-graph}. In \Cref{fig:spp-ds} tree
on the left is used to construct the data structure (see line~\ref{spp:insert} in Algorithm~\ref{alg:spsinit}).
For instance, based on that example
$A(i=1,j=5)\neq 0$, hence $i=1$ is the version number and we update
$5^{\text{th}}$ and $6^{\text{th}}$ indices
(see tree for insertions in \Cref{fig:spp-ds}) of the first version.
Also, $A(i=2,j=5)\neq 0$ hence $i=2$ is the version number and again
we need to update $5^{\text{th}}$ and $6^{\text{th}}$ indices
of the second version. Since we use the fat node approach to provide persistence;
in the second version we fetch the values of $5^{\text{th}}$ and $6^{\text{th}}$ indices
from the closest previous version and then update the second version (see line~\ref{spp:persistent} in
Algorithm~\ref{alg:spsinit}) by setting the values of those indices as $2$.

\begin{algorithm2e}[ht]
  \caption{\label{alg:spsquery} {\sc SPSQuery}($S, i, j$)}
  \tcp*[l]{Assuming that $i, j$ are 1-based indices and $S \leftarrow$ \sc SPSConstruction(A)}
    $r \leftarrow 0$\;
    \While{$j > 0$}{ \label{spp:qwhile}
      $r \leftarrow r + \max_{k \leq i} S(k,j)$\; \label{spp:queryclosest}
      \tcp*[l]{Decrement $j$ by clearing its least significant bit.}
      $j \leftarrow j - $ {\tt LSB}$(j)$\; \label{spp:queryindices}
    }

  \Return $r$

\end{algorithm2e}

When we finish the construction of the data structure, the number of nonzeros for a rectangle with
corners $(1, 1)$ and $(i, j)$ is found by making a query for the $j$th index from the BIT with
the version $i$. Algorithm~\ref{alg:spsquery} presents the high-level query algorithm.
Similar to initialization process (Algorithm~\ref{alg:spsinit}) in Algorithm~\ref{alg:spsquery}
while loop (line~\ref{spp:qwhile}) operates like a tree for queries. In \Cref{fig:spp-ds}
we illustrate an example query tree (right side) for the toy graph.
For instance, to compute the load of the rectangle from $(1, 1)$ to $(6, 3)$
we have to query the $6^{\text{th}}$ version and sum the values of the
$3^{\text{rd}}$ and $2^{\text{nd}}$ indices (see line~\ref{spp:queryindices} in Algorithm~\ref{alg:spsquery}).
The $3^{\text{rd}}$ index of the
$6^{\text{th}}$ version is $3$ and for the $2^{\text{nd}}$ index $6^{\text{th}}$ version
is empty hence we find the closest previous version to
the $6^{\text{th}}$ version (see line~\ref{spp:queryclosest} in Algorithm~\ref{alg:spsquery})
in which $2^{\text{nd}}$ index is not empty, which is the $5^{\text{th}}$ version whose value is $3$.
Therefore the load of the rectangle from $(1, 1)$ to $(6, 3)$ is $6$.

A query on a BIT takes $O(\log n)$ time, searching for the correct version
for each entry also takes $O(\log n)$ time. Thus, a single query to the persistent BIT data
structure takes $O(\log^2 n)$ time.
When updating a BIT, each update changes $1/2 \cdot \log n$ entries on average. In order to
have persistence, each changed field has to be stored. Thus, the construction time and space requirement of our
data structure is $O(n + m \log n)$, where the number of columns is $n$ and number of nonzeros is $m$.

Note that, to make the implementation more efficient and avoid multiple memory allocations,
number of versions that each entry of the BIT is going to have is precomputed. Finally,
Compressed Sparse Column (CSC) format is used to build and store the final persistent BIT
which is effectively a sparse matrix. For simplification and visualization purposes we
do not use CSC like representation in our example \Cref{fig:spp-ds}.

\subsection{Complexity analysis}
\label{subsec:complexity}

In addition to our proposed algorithms, we implemented Nicol's~\cite{Nicol94-JPDC}
two-dimensional rectilinear partitioning algorithm (\nic in short). Note that
\nic does not output symmetric partitions hence, we also use uniform partitioning
(\uni in short) as a baseline.  The \uni algorithm is the simplest
checkerboard partitioning, where each tile has an equal number of
rows and columns. The \uni algorithm runs in constant time.
\Cref{tab:algs} summarizes the high-level characteristics of the algorithms
that are covered in this work and \Cref{tab:complexity} displays the computational
complexities of those algorithms.

\begin{table}[!ht]
\caption{Algorithms covered in this work. Problem: the problem that an algorithm
tackles with. Approach: the approach used by an algorithm. Symmetric: is the output partitioning
is symmetric.}
\label{tab:algs}
\begin{center}
\begin{tabular}{l l l c}
\hline
\textbf{Algorithm} & \textbf{Problem} & \textbf{Approach} & \textbf{Symmetric}\\
\hline
Uniform (\uni)       & N/A                & N/A              & \cmark\\
Nicol's 2D (\nic)    & Rectilinear - \mLI & Refinement & \xmark\\
Refine a cut (\rac)  & Rectilinear - \mLI & Refinement & \cmark\\
Bound a cut (\bac)   & Rectilinear - \mLI & Generalized      & \cmark\\
Probe a load (\pal)  & Rectilinear - \mNC & Probe      & \cmark\\
Ordered probe a load (\opal)  & Rectilinear - \mNC & Probe      & \cmark\\
Bound a load (\bal)  & Rectilinear - \mNC & Generalized      & \cmark\\
\hline
\end{tabular}
\end{center}
\end{table}

\begin{table}[!ht]
\caption{Worst case complexities of algorithms with and without sparse-prefix-sum
data structure.}
\label{tab:complexity}
\begin{center}
\begin{tabular}{l l l}
\hline
\textbf{Algorithm} & \textbf{Without BIT} & \textbf{With BIT}\\
\hline
\bf \nic        & $O( \tau (m + n + p^3 \log^2 \frac{n}{p}))$       & $O( \tau p^2 \log\frac{n}{p} \log^2n)$\\
\bf \rac        & $O( m + n + \tau p^3 \log^2 \frac{n}{p} )$        & $O( p^2 \log^2n + \tau p^3 \log^2 \frac{n}{p} )$\\
\bf \bac(\pal)  & $O(p m \log m \log n \log p)$                     & $O(p^2 \log m \log^3n )$\\
\bf \pal        & $O(p m\log n \log p)$                             & $O( p^2 \log^3 n)$\\
\bf \bac(\opal) & $O(m \log p \log m)$                              & $O(p^2 \log m \log^3n)$\\
\bf \opal       & $O(m\log p)$                                      & $O( p^2 \log^3 n)$\\
\bf \bal(\rac)  & $O( \log n (m + n + \tau p^3 \log^2\frac{n}{p}))$ & $O( p^2 \log n(\log^2n+\tau p\log^2\frac{n}{p}))$\\
\hline
\end{tabular}
\end{center}
\end{table}

\nic's refinement algorithm~\cite{Manne96-IWAPC,Nicol94-JPDC}
(Algorithm~\ref{alg:rowparti}) has a worst-case complexity of
$O(m + n + qp^2\log^2 \frac{n}{p} + pq^2\log^2 \frac{n}{q})$~\cite{Saule12-JPDC-spart} for
non-symmetric rectilinear partitioning. The Algorithm is guaranteed to converge with
at most $n^2$ iterations when the matrix is square. However, as noted in those earlier work,
in our experiments, we observed that algorithm converges very quickly, and hence
for the sake of fairness we have decided to use the same limit on the number of iterations, $\tau$.
For the symmetric case, where $p=q$, refinement algorithm runs in
$O(m + n + p^3\log^2 \frac{n}{p})$, and this what we displayed in \Cref{tab:complexity}.

\rac algorithm first runs Algorithm~\ref{alg:rowparti} and then computes
the load imbalance. These operations can be computed in $O(p^3\log^2 \frac{n}{p})$ and
in $O(m + n)$ respectively. In the worst-case, Algorithm~\ref{alg:rowparti} is called $\tau$ times.
Hence, \rac algorithm runs in $O(m + n + \tau p^3(\log \frac{n}{p})^2)$.
Note that this is the naive computational complexity of the \rac algorithm. Using
our sparse prefix sum data structure we can compute load imbalance in $O(p^2\log^2(n))$ time.
Hence using our data structure computational complexity of the \rac algorithm can be
defined as $O( p^2 \log^2 n + \tau p^3 \log^2\frac{n}{p} )$.

The \bac algorithm in the worst-case calls $O(\log(m))$ times a
given \mNC algorithm (such as \pal). So, the \bac algorithm runs in $O(p m\log m \log n \log p)$
when \pal is used as the secondary algorithm.
Using sparse prefix sum data structure we can further improve this computational complexity
to $O(p^2 \log m \log^3n)$.

The \pal algorithm (Algorithm~\ref{alg:pal}) does $O(m \log n)$ computations in
the worst-case to find a cut point; $O(\log n)$ searches and $O(m \log p)$ for load imbalance
computation. Since there are
going to be $O(p)$ cut points the \pal algorithm runs in $O(p m\log n \log p)$. We reduce this
computational complexity to $O( p^2 \log^3 n)$ using our sparse prefix sum data structure.

The \opal algorithm (Algorithm~\ref{alg:opal}) transforms a matrix in $O(m + n)$ time and then
passes over the matrix to find the cut points. For each nonzero, a $O(\log p)$ binary search is done
to determine which tile the nonzero is in. In the same way as \pal, this complexity reduces to
$O( p^2 \log^3 n)$ using our sparse prefix sum data structure.

The \bal algorithm in the worst-case calls $O(\log m)$ times a given \mLI algorithm (such as \rac).
So, the \bal algorithm runs in $O(\log m (m + n + \tau p^3 \log^2 \frac{n}{p}))$
when \rac is used as the secondary algorithm.

\section{Matrix Sparsification}
\label{ssec:sparsification}

When there are many nonzeros in a matrix, we can use a fraction of the nonzeros
to approximately determine the load imbalance for a given partition vector.
We can sample the nonzeros by flipping a coin for each nonzero with a
keeping probability of $s$, which we will call sparsification factor.

\begin{figure}[ht]
  \centering
  \subfloat[$s=1$]{\includegraphics[width=.3\linewidth]{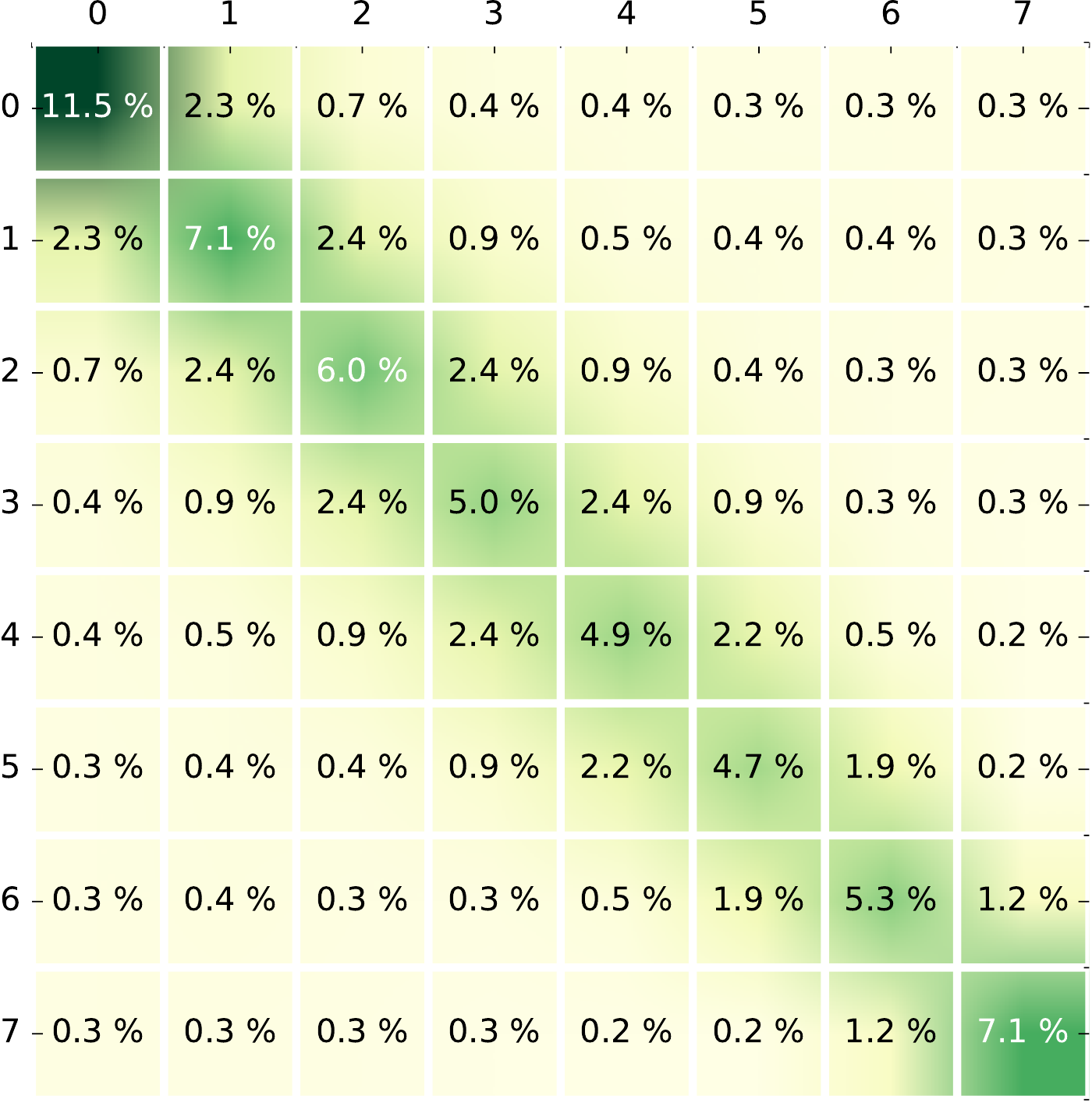}
  \label{fig:sp-amazon-1}}
  \hspace{1em}
  \subfloat[$s=0.1$]{\includegraphics[width=.3\linewidth]{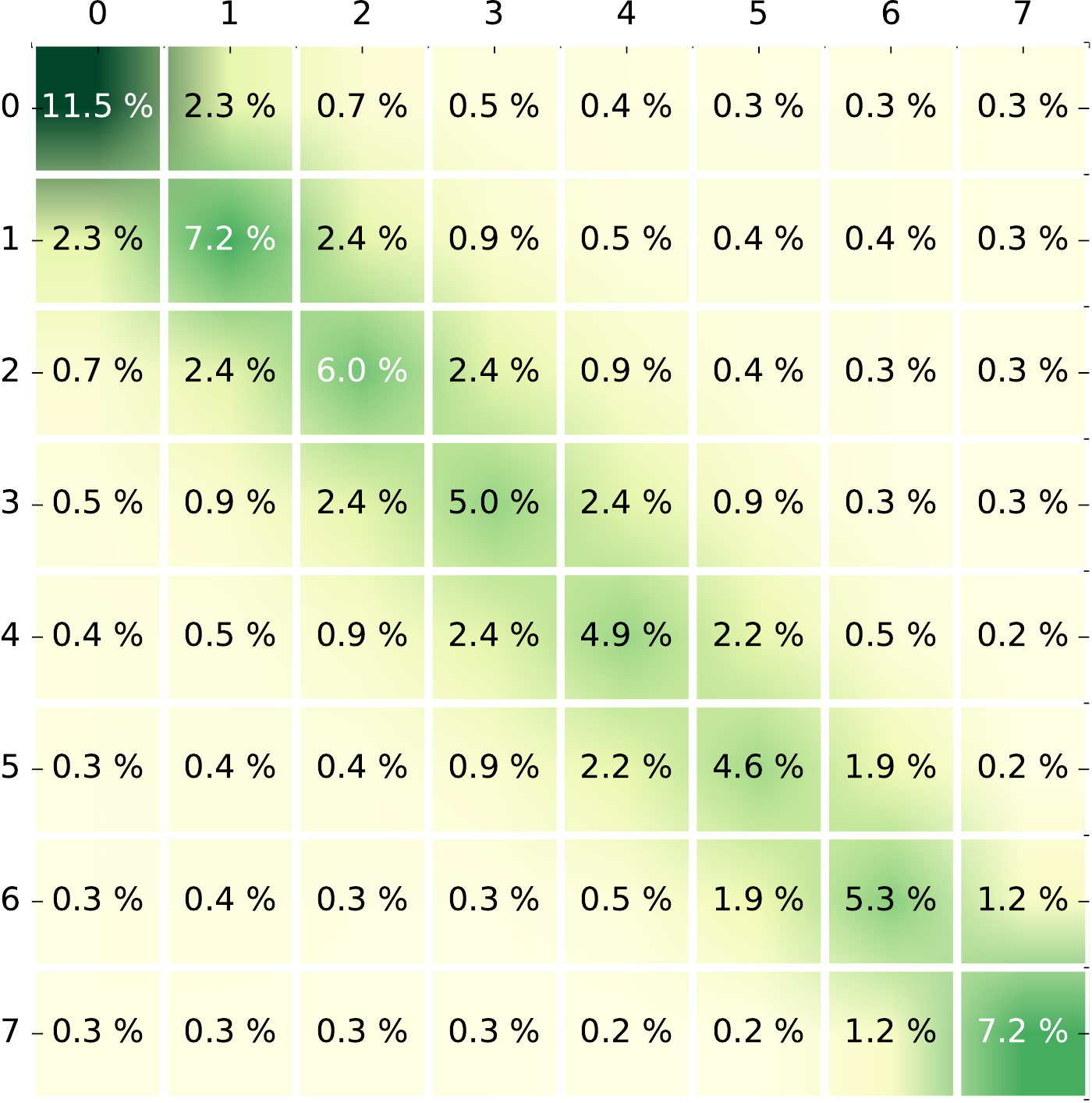}
  \label{fig:sp-amazon-01}}
  \hspace{1em}
  \subfloat[$s=0.01$]{\includegraphics[width=.3\linewidth]{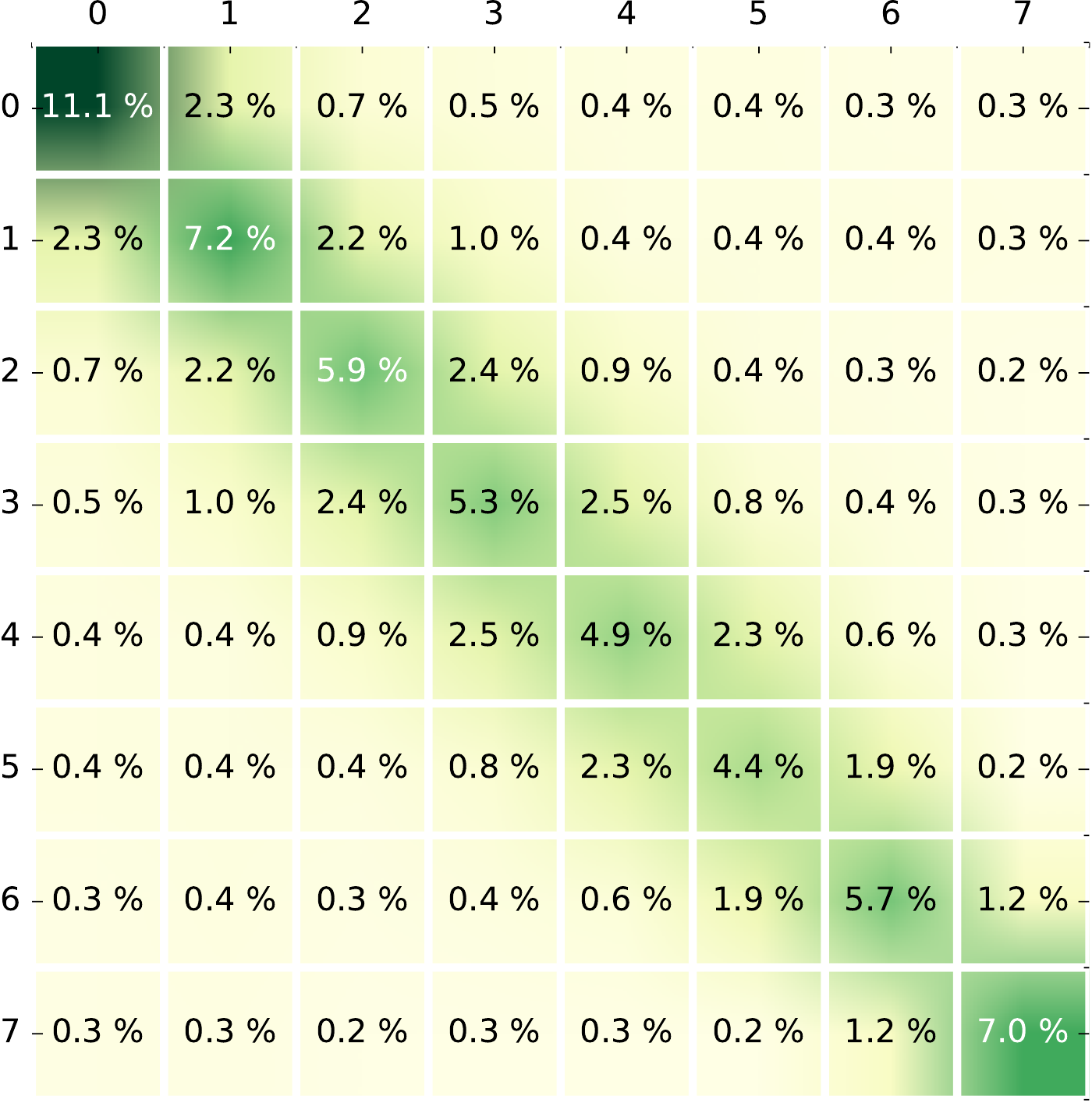}
  \label{fig:sp-amazon-001}}
  \caption{Sparsification example using the Amazon-0312 graph.}
  \label{fig:sp-ex}
\end{figure}

\Cref{fig:sp-ex} illustrates the affect of the sparsification on the Amazon-0312 graph ($\sim$0.4 million vertices and $\sim$3 million directed edges). In this figure
we plot the heat map of the Amazon-0312 graph when it is partitioned into $8\times 8$ uniform tiles for three cases;
without sparsification, sparsification factor of $s=0.1$ (i.e., keeping $10\%$ of the nonzeros) and
sparsification factor of $s=0.01$ (i.e., keeping $1\%$ of the nonzeros). We observe that when we only
keep $10\%$ of the edges, nonzero distribution of the matrix is almost the same and there are slight
changes when we keep only $1\%$ of the edges.

We can control expected relative error by automatically adjusting sparsification.
For a given partitioned matrix, if there are $Z$ nonzeros
in a tile, then the expected value is $Zs$, for the number of nonzeros after flipping, $Z'$.
$Z'$ follows the binomial distribution, $Z' \sim B(Z, s)$. The variance of the distribution of
the number of nonzeros in a tile is $Zs(1-s)$. Thus, the expected relative error of the
estimation of the nonzeros in the tile will be on the order of:

\begin{align*}
    1 - \epsilon &\leq \frac{Z'}{Zs} \leq 1 + \epsilon,\\
    \text{where}\; \epsilon &\approx \frac{\sqrt{Zs(1-s)}}{Zs} = \sqrt{\frac{1-s}{Zs}}.
\end{align*}

The above inequality implies that if a matrix has $m$ nonzeros,
then under any given partition vector that divides the matrix into $p^2$ tiles, the
maximum loaded tile will have at least $\frac{m}{p^2}$ nonzeros.
Then, the relative error will be on the order of $\sqrt{(1-s)p^2/(ms)}$.
For instance, if a matrix has $m = 11 \times 10^6$ nonzeros, the maximum loaded tile
will have not less than $\frac{11 \times 10^6}{64} \approx 1.7 \times 10^5$
nonzeros under any $8 \times 8$ ($p=8$) partitioning. If the probability $k$ we define
for flipping coins is $0.1$, then the relative error will be around $0.007$. Hence, all
the algorithms in this paper can be run on the sparsified matrix without significant change
in the quality. Let $A'$ be a sparsified version of the given matrix $A$ and $C$ be a
partition vector. The related load imbalance formula can be defined as:

\begin{align*}
    \lambda(A, C, C) = &\frac{L_{max}\{A, C, C\}}{L_{avg}\{A, C, C\}} \text{ and } 1 - \epsilon \leq \frac{L_{max}\{A', C, C\}}{L_{max}\{A, C, C\}\times s} \leq 1 + \epsilon\\
    \implies &\frac{L_{max}\{A, C, C\}}{L_{avg}\{A, C, C\}}(1 - \epsilon) \leq \frac{L_{max}\{A', C, C\}}{L_{avg}\{A, C, C\}\times s} \leq \frac{L_{max}\{A, C, C\}}{L_{avg}\{A, C, C\}}(1 + \epsilon) \\
    \implies &\lambda(A, C, C)(1 - \epsilon) \leq \frac{L_{max}\{A', C, C\}}{L_{avg}\{A, C, C\}\times s} \leq \lambda(A, C, C)(1 + \epsilon)
\end{align*}
Meaning that the load imbalance will be off on the order of $\epsilon$.
From now on, we will call $\epsilon$ as error tolerance for automatic sparsification factor
selection.

\section{Implementation Details}
\label{sec:implementation}

We implemented our algorithms using C++ standard $17$ and compile our
code-base with GCC version $9.2$. We have collected all of our implementations
in a library we named SpatiAl Rectilinear Matrix pArtitioning (SARMA). Source code
of SARMA is publicly available at \url{http://github.com/GT-TDAlab/SARMA} via a BSD-license.
C++ added support for parallel algorithms to the standard library by integrating
Intel's TBB library starting from the standard $17$.
Note that, in this work
our goal is not parallelizing the partitioning framework, to provide better performance
with the minimal work, in our code-base we simply enabled parallel execution
policy of the standard library functions and parallelized pleasingly parallel
loops wherever it is possible.

\begin{figure}[ht]
  \centering
  \subfloat[Nicol's algorithm.]{\includegraphics[width=.45\linewidth]{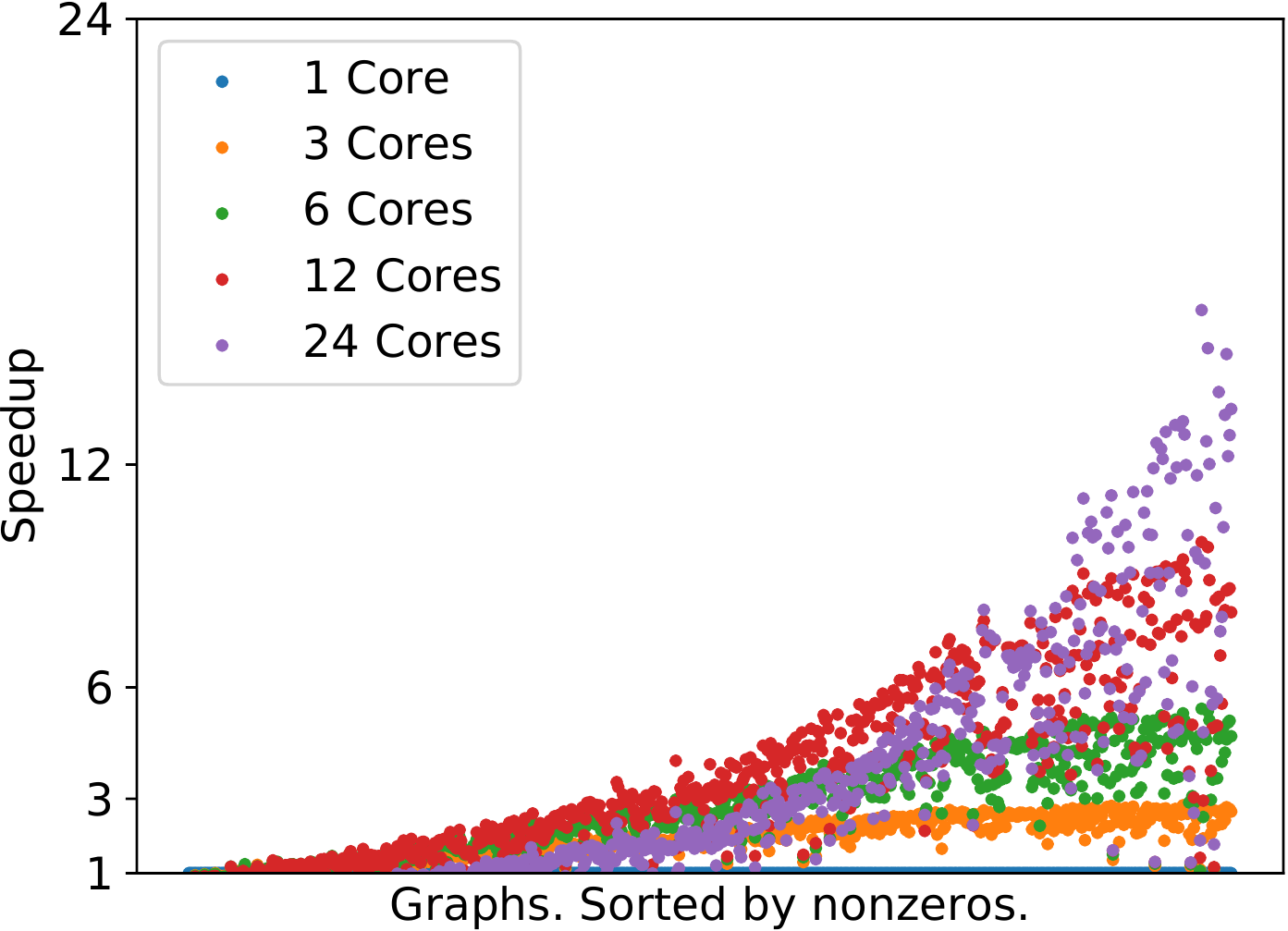}
  \label{fig:ss-nic}}
  \subfloat[\pal algorithm.]{\includegraphics[width=.45\linewidth]{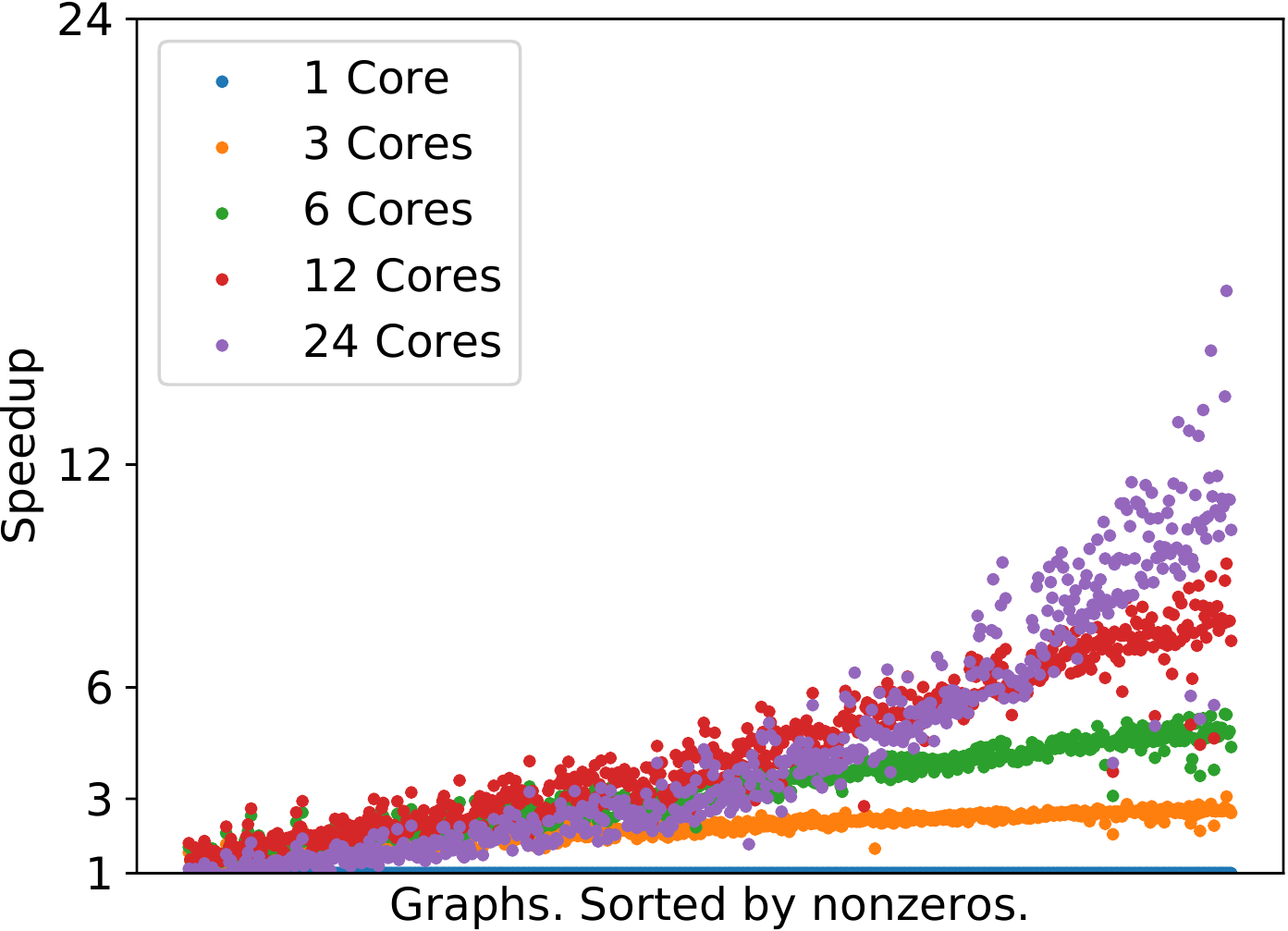}
  \label{fig:ss-pal}}
  \caption{Strong scalability experiments on \nic and \pal algorithms. Graphs are
    sorted based on their nonzeros on the x-axis and partitioned as $32\times 32$.
    Number of cores: $\{1$, $3$, $6$, $12$, $24\}$}
  \label{fig:ss}
\end{figure}

\Cref{fig:ss} presents strong scaling speedup of \nic and \pal algorithms on
$687$ graphs on an Intel architecture that have $24$ cores and no hyper-threading.
In the plot, graphs are sorted based on their number of nonzeros on the x-axis.
Adjacency matrices of graphs are partitioned into $32\times 32$ tiles.
Achieved speedup is provided on the y-axis for each graph on different cores. In this experiment, we ran each algorithm
$10$ times on $1$, $3$, $6$, $12$ and $24$ cores and report the median of the runs.
As expected, we observe that achieved speedup increases with the graph size. The
\nic algorithm achieves up-to $15$ times and \pal algorithm achieves up-to $17$ times
speedup on $24$ cores. With small graphs we observe very limited speedups because these
graphs can fit into the cache in the sequential case and parallelization does not compensate
poor cache utilization.

\section{Experimental Evaluation}
\label{sec:experimental}

We ran our experiments on a $416$-node
cluster owned by the Partnership for an Advanced Computing Environment (PACE)
of Georgia Institute of Technology equipped with $2\times 12$ cores $2.7$ GHz Intel
Xeon $6226$ CPUs, $192$ GB of RAM and at least $512$ GB of local storage.
We ran each algorithm for $4$ different cuts, $p=\{4, 8, 16, 32\}$ or
$4$ different target loads, $Z=\{m/4, m/9, m/16, m/25\}$, and
without sparsification and with $\epsilon=0.01$.
We used  the Moab scheduler along with the Torque resource manager
that runs every partitioning algorithm one-by-one on a matrix on one of the available
nodes.

We evaluated our algorithms on real-world and synthetic graphs from the
SuiteSparse Matrix Collection~\cite{Davis11-TOMS}. We excluded non-square matrices and
matrices that have less than $1$ million or more than $2$ billion nonzeros.
There were $687$ matrices satisfying these
properties (there were a total of $2,856$ matrices at the time of this experimentation).
We also chose a subset of $16$ graphs from those graphs. \Cref{table:subset} lists
those graphs that we used in some of our experiments along with the graph name,
origin/source of the graph, number of rows ($n$), number of
nonzeros ($m$) and average number of nonzeros per row ($m/n$).

\begin{table}[t]
\caption{Properties of the subset of our dataset along with their name,
origin, number of rows ($n$), number of nonzeros ($m$) and average number
of nonzeros per row ($m/n$).}
\label{table:subset}
\begin{center}
  \begin{tabular}{ |l | l | r  r  r | }

    \hline
    {\bf Matrix Name} &
        {\bf Matrix Origin} &
        $\boldsymbol{n}$ &
        $\boldsymbol{m}$ &
        $\boldsymbol{m/n}$ \\

    \hline \hline

    wb-edu &
        Web &
        $9,845,725$  &
        $57,156,537$  &
        $5.8$ \\

    road\_usa &
        Road &
        $23,947$,347  &
        $57,708,624$  &
        $2.4$ \\

    circuit5M &
        Simulation &
        $5,558,326$  &
        $59,524,291$  &
        $10.7$ \\

    soc-LiveJournal1 &
        Social &
        $4,847,571$  &
        $68,993,773$  &
        $14.2$ \\

    kron\_g500-logn20 &
        Kronecker &
        $1,048,576$  &
        $89,239,674$  &
        $85.1$ \\

    dielFilterV3real &
        Electromagnetics &
        $1,102,824$  &
        $89,306,020$  &
        $81.0$ \\

    europe\_osm &
        Road &
        $50,912,018$  &
        $108,109,320$  &
        $2.1$ \\

    hollywood-2009 &
        Movie/Actor &
        $1,139,905$  &
        $113,891,327$  &
        $99.9$ \\

    Cube\_Coup\_dt6 &
        Structural &
        $2,164,760$  &
        $124,406,070$  &
        $57.5$ \\

    kron\_g500-logn21 &
        Kronecker &
        $2,097,152$  &
        $182,082,942$  &
        $86.8$ \\

    nlpkkt160 &
        Optimization &
        $8,345,600$  &
        $225,422,112$  &
        $27.0$ \\

    com-Orkut &
        Social &
        $3,072,441$  &
        $234,370,166$  &
        $76.3$ \\

    uk-2005 &
        Web &
        $18,520,486$  &
        $298,113,762$  &
        $16.1$ \\

    stokes &
        Semiconductor &
        $11,449,533$  &
        $349,321,980$  &
        $30.5$ \\

    kmer\_A2a &
        Biological &
        $170,728$,175  &
        $360,585,172$  &
        $2.1$ \\

    twitter &
        Social &
        $41,652,230$  &
        $1,468,365,182$  &
        $35.3$ \\

    \hline
  \end{tabular}
\end{center}
\end{table}

We present some of our results using performance profiles~\cite{Dolan02-MP}.
In a performance profile plot, we show how bad a specific algorithm performs within
a factor $\theta$ of the best algorithm that can be obtained by any of the compared
algorithms in the experiment.
Hence, the higher and closer a plot is to the y-axis, the better the method is.

\subsection{Comparison with the optimal solution}
\label{ssec:optimal-comp}

\begin{figure}[ht]
  \centering
  \includegraphics[width=.6\linewidth]{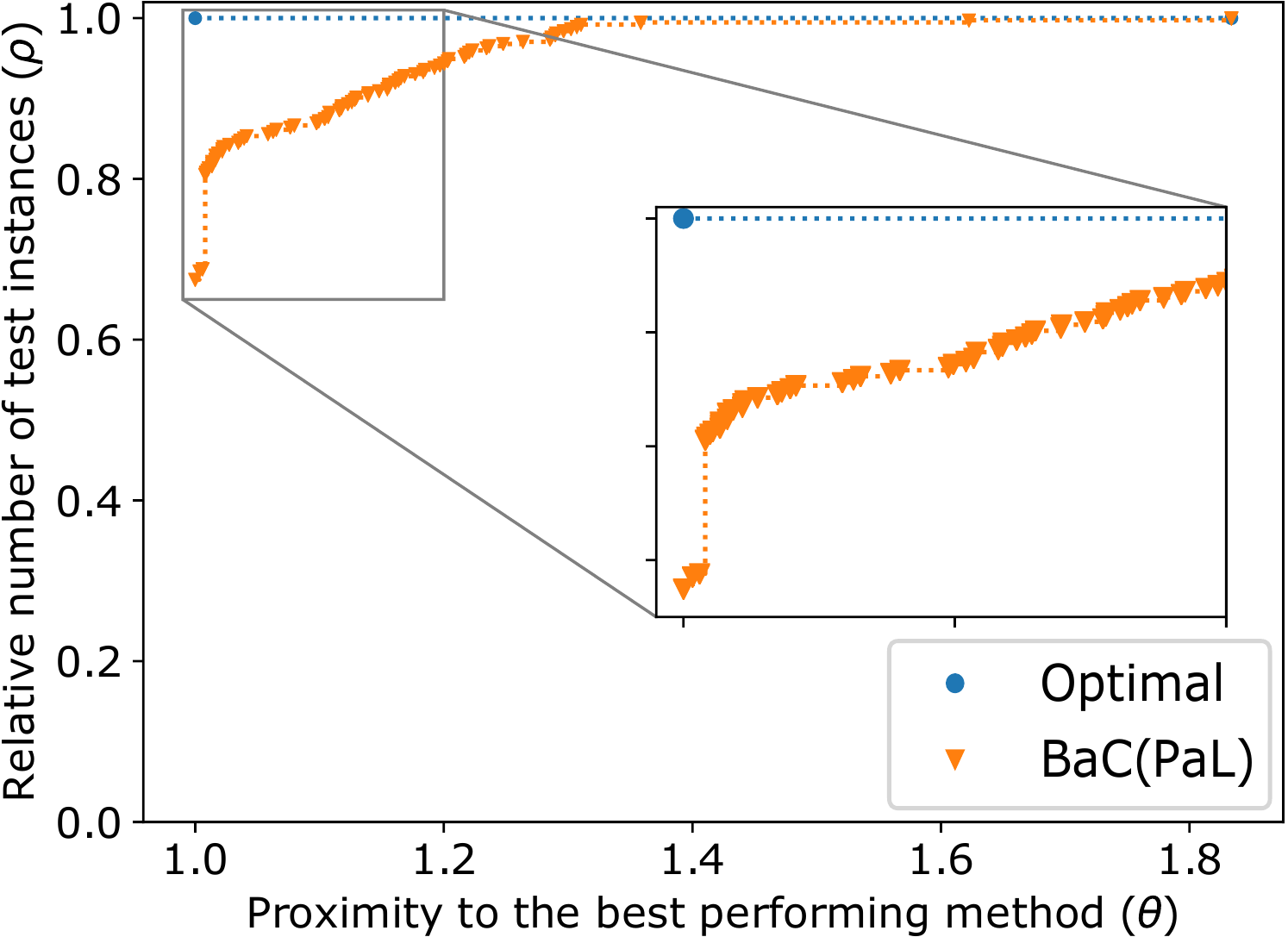}
  \caption{Comparison with the optimal solutions using a performance profile.}
  \label{fig:comp-optimal}
\end{figure}

To the best of our knowledge, this is the first work that tackles the
symmetric rectilinear partitioning problem. Hence, we do not have a fair
baseline. To understand the quality of the partitioning algorithms, in this
experiment we compare \bac(\pal) algorithm's load imbalance with the optimal solution.
We implemented our mathematical model (as shown in Section~\ref{ref:ssec:mm})
using Gurobi~\cite{gurobi}. Since finding the optimal solution
is computationally expensive, in addition to our dataset, we downloaded $375$
small graphs from SuiteSparse matrix collection~\cite{Davis11-TOMS} that have
less than $9,000$ nonzeros. We partition those graphs into $8\times8$ ($p=8$) tiles.
\Cref{fig:comp-optimal} illustrates the performance profile for the load-imbalance
between the optimal solution and the \bac(\pal) algorithm. We observe that \bac(\pal)
algorithm achieves the optimal solution on $67\%$ of the test instances and give
nearly the optimal solution on $80\%$ of the test instances. At the
worst case, the \bac(\pal) algorithm outputs at most $1.9$ times worse results
than the optimal case.

\subsection{Experiments on the sample dataset}
\label{ssec:exp:sample}
In the following experiments, we show raw load-imbalance and execution time results
of different algorithms on chosen $16$ graphs under various settings. Later, we evaluate
the effect of the sparsification on the load imbalance and the execution time.
In those experiments, we consider \uni, \nic, \rac and \bac(\pal) algorithms for the \mLI problem.
For \uni, \rac, and \bac(\pal) algorithms, we choose $p=32$ and for the \nic algorithm,
we choose $p = q = 32$. Hence, every graph is partitioned into $32 \times 32$ tiles. We ran experiments
without sparsification, with $s=1\%$, $s=0.1\%$ and $\epsilon=0.01$.
Note that $s$ is the sparsification factor and $\epsilon$ is the error tolerance
for automatic sparsification factor selection.

\begin{figure}[htb]
  \centering
  \subfloat{\includegraphics[width=.98\linewidth]{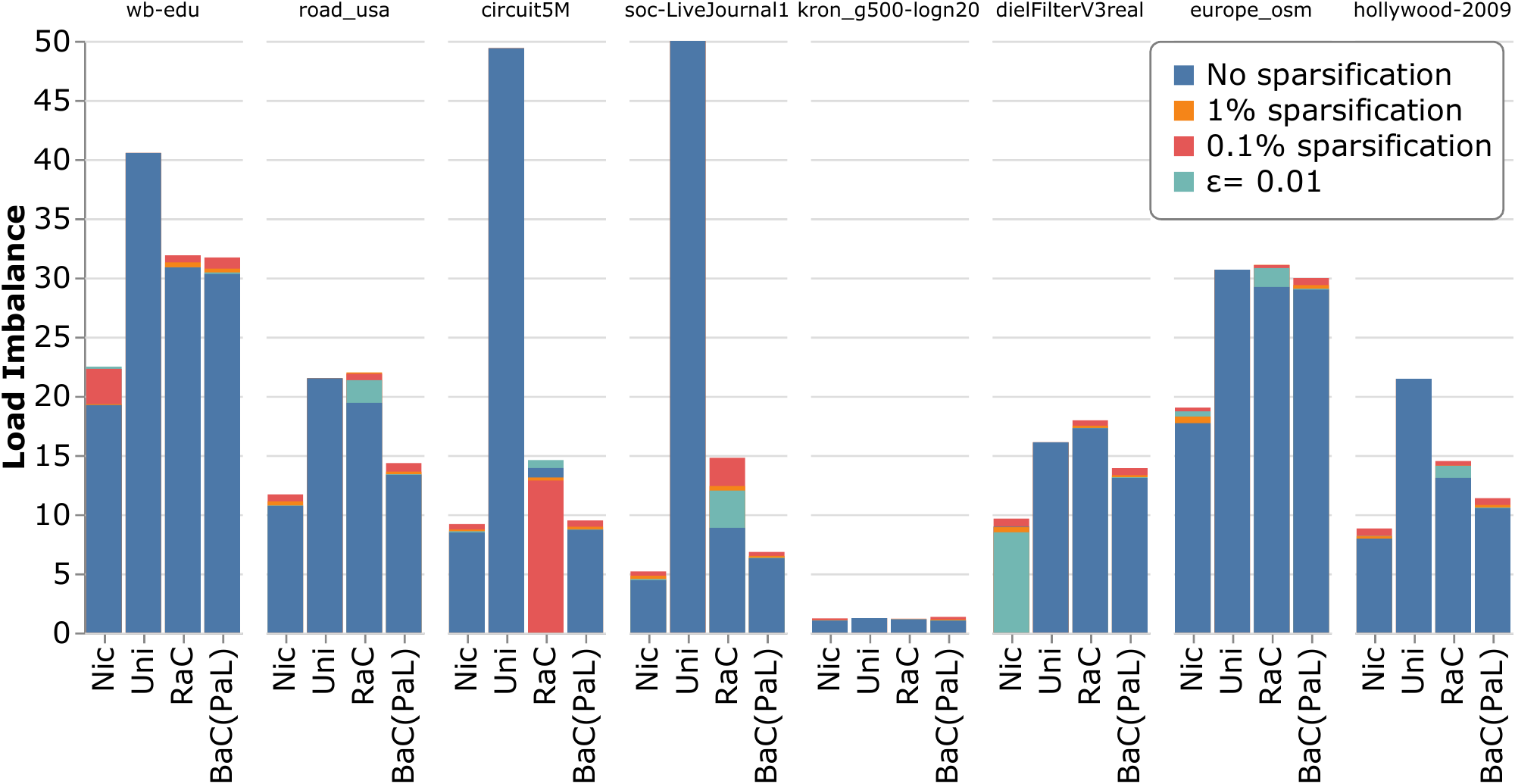}
  \label{fig:exp:li-32-1}}

  \subfloat{\includegraphics[width=.98\linewidth]{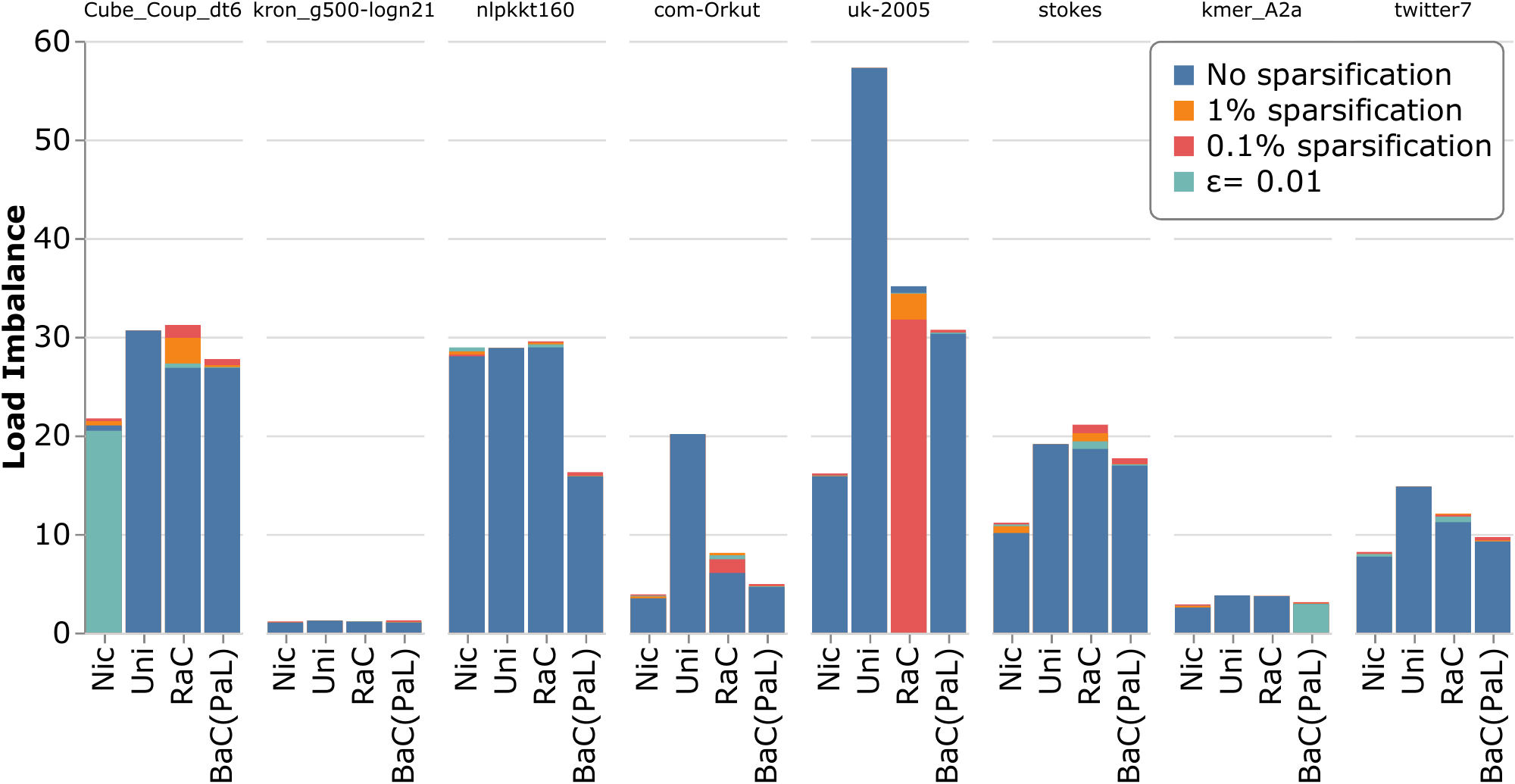}
  \label{fig:exp:li-32-2}}

  \caption{Load imbalance comparison on sample dataset.}
  \label{fig:exp:subset-li}
\end{figure}

\subsubsection{Effect of sparsification on the load imbalance}
\label{sssec:exp:li}

\Cref{fig:exp:subset-li} reports load imbalances of four different algorithms;
\uni, \nic, \rac, and \bac(\pal) on our selected $16$ graphs. In \Cref{fig:exp:subset-li},
each bar presents the load imbalance for a graph instance. Blue bars represent the load imbalance
when sparsification is off and others represent when sparsification is on. As expected,
on majority of the cases, the \nic algorithm gives the best load imbalance.
Because, symmetric rectilinear matrix partitioning is a very restricted problem and,
being able to align different partition vectors to rows and columns gives a big flexibility
to the \nic algorithm. Even with the restrictive nature, best of our symmetric partitioning algorithm
gives no worse load imbalance than $1.7\times$ with-respect-to the \nic algorithm.
\bac(\pal) algorithm gives the best performance among three symmetric
partitioning algorithms on $15$ out of $16$ cases. Since rmat graphs have a well distributed
matrix structure, we see that all algorithms give nearly optimal load imbalance on kronecker
graphs. In the worst case, the \rac algorithm gives $\approx 1.8$ times worse load imbalance
than the \bac(\pal) algorithm. As expected, the \uni algorithm performs poor on the majority of
the matrices. Especially on the matrices that have skewed distributions such as soc-LiveJournal1 and
uk-2005.
Enabling sparsification mostly affects the \rac algorithm due to
mapping of the problem from two-dimensional case to one-dimensional case and also
applying the refinement on the same direction continuously. We observe almost negligible
errors on the other algorithms (less than $0.005$). Note that, even for the \rac
algorithm, error of the load imbalance is less than $0.01$ in the majority of graphs ($11$ out of $16$).

\begin{figure}[ht]
  \centering
  \subfloat{\includegraphics[width=.35\linewidth]{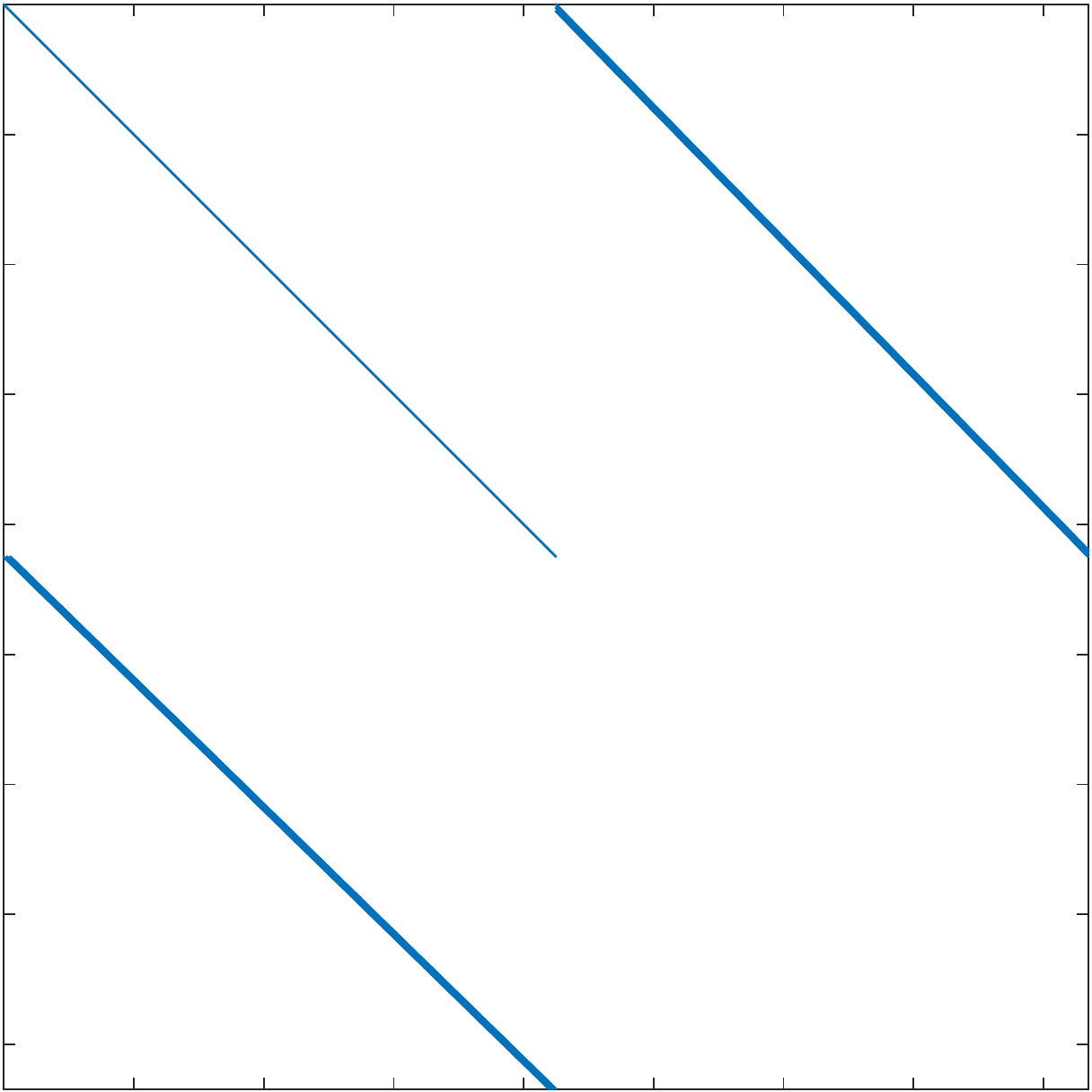}
  \label{fig:exp:nlpspy}}
  \hspace{2em}
  \subfloat{\includegraphics[width=.35\linewidth]{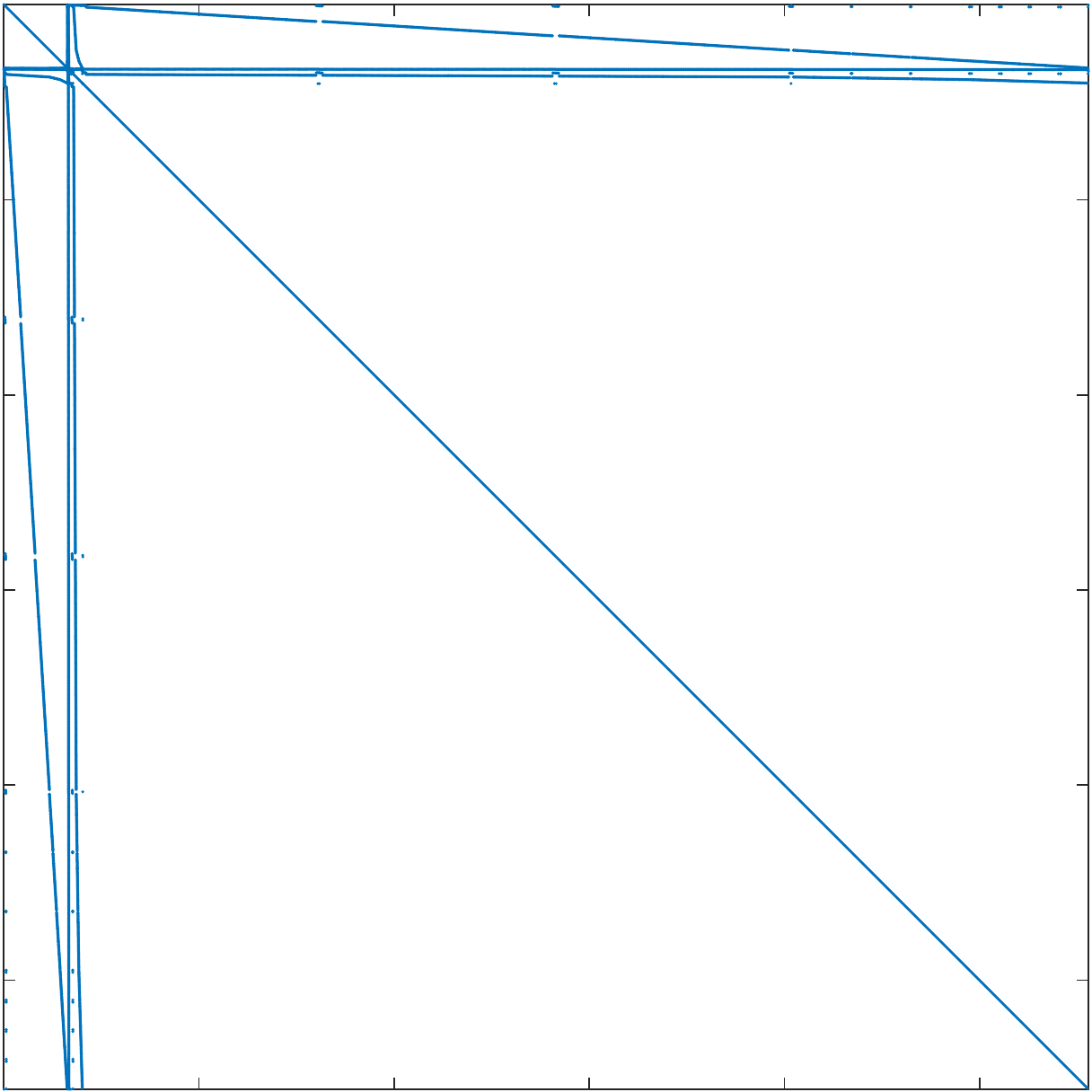}
  \label{fig:exp:circuitspy}}

  \caption{Sparsity patterns of nlpkkt160 and circuit5M matrices.}
  \label{fig:exp:mtxspy}
\end{figure}

The sparsity pattern of a matrix may play a role on the final load imbalance. We observe that
on our sample dataset the \uni partitioning gives better load imbalance than the \rac
partitioning on dielFilterV3real and nlpkkt160 matrices when there is no sparsification.
Besides, on stokes and kmer\_A2a matrices the \rac algorithm only gives slightly better load imbalance
than the \uni partitioning. Sparsity patterns of those matrices are the primary factor.
To visualize, \Cref{fig:exp:mtxspy} plots sparsity patterns of the nlpkkt160 (\Cref{fig:exp:nlpspy}) and the
circuit5M (\Cref{fig:exp:circuitspy}) matrices.
On the nlpkkt160 matrix the \uni algorithm, the \nic algorithm and the \rac algorithm outputs
similar load imbalances. As illustrated in \Cref{fig:exp:nlpspy}, the nlpkkt160 graph is really
sparse and the pattern is three regular lines. Hence, refinement algorithm outputs poor
partition vectors for both \nic and \rac. Since the pattern is regular \uni gives good load
imbalance and \bac(\pal) algorithm outperforms the other algorithm by considering more
possibilities on two-dimensional case. On the other hand, on the circuit5M matrix the \uni
algorithm gives really poor load imbalance because that matrix have regular dense regions on the first set of beginning rows and columns as illustrated in \Cref{fig:exp:circuitspy}.
Due to this dense structure on that graph \bac(\pal) and \nic algorithms gives similar results
because refinement algorithm tries to put more cuts to the beginning of the partition vectors.

\begin{figure}[ht]
  \centering
  \subfloat{\includegraphics[width=.98\linewidth]{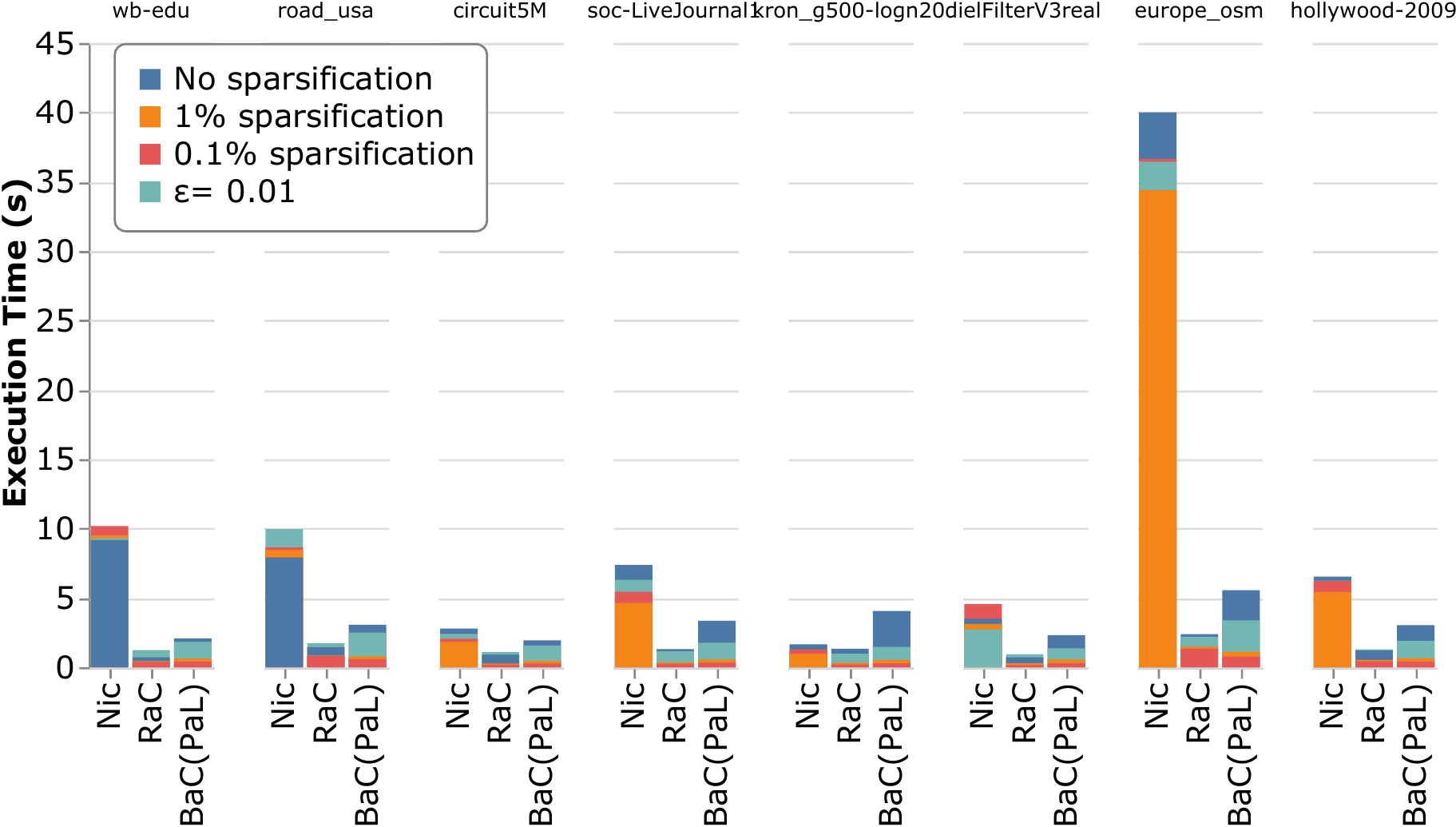}
  \label{fig:exp:et-32-1}}

  \subfloat{\includegraphics[width=.98\linewidth]{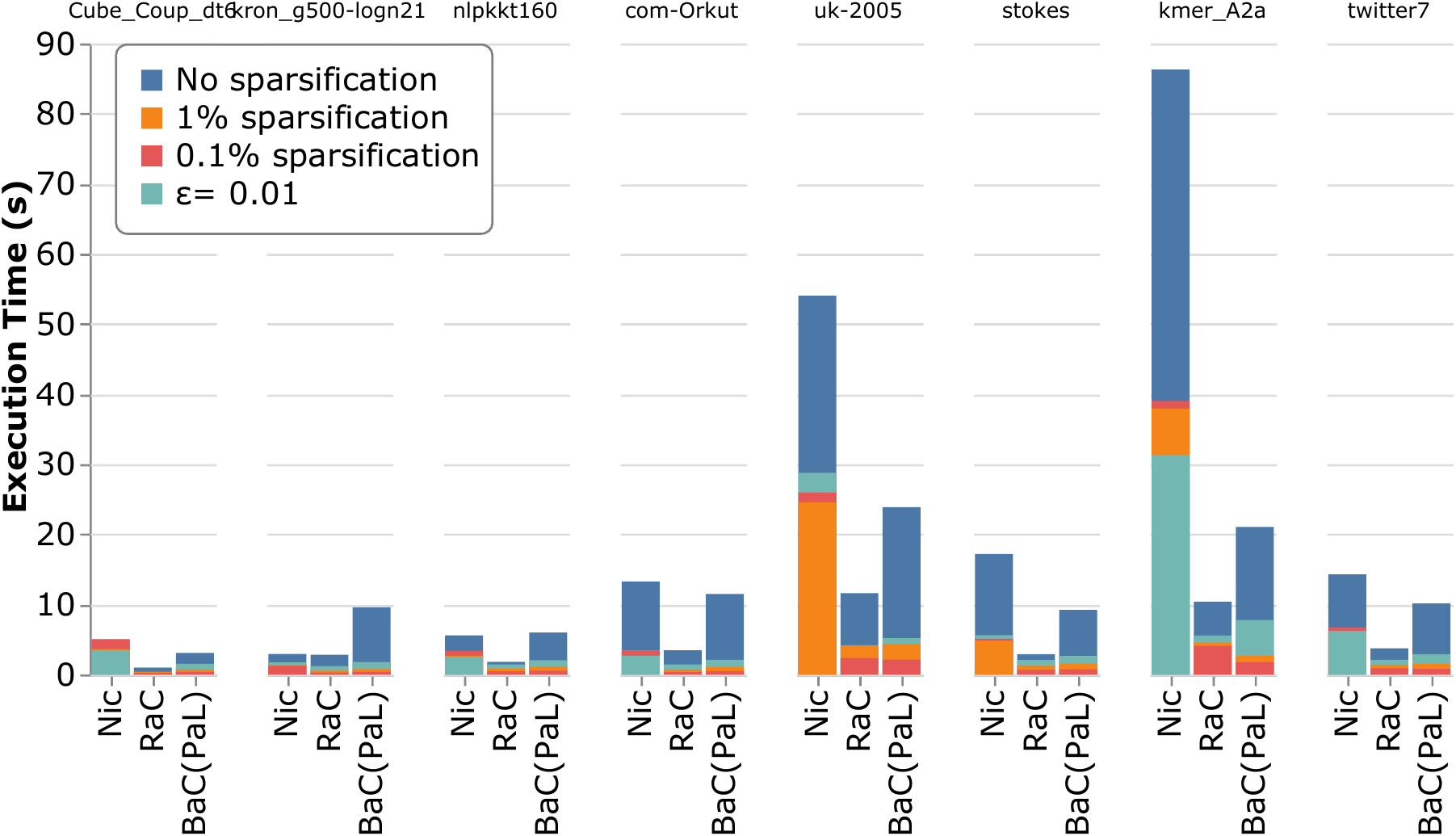}
  \label{fig:exp:et-32-2}}

  \caption{Execution time comparison on sample dataset.}
  \label{fig:exp:subset-et}
\end{figure}

\subsubsection{Effect of sparsification on execution time}
\label{sssec:exp:et}

\Cref{fig:exp:subset-et} reports execution times of three different algorithms;
\nic, \rac and \bac(\pal)
on our selected $16$ graphs. We discard the \uni algorithm from this experiment
since it can be computed in constant time.
For the \bac(\pal) algorithm, the execution time includes the generation of the
sparse-prefix-sum data structure.
In \Cref{fig:exp:subset-et}, each bar represents the execution time for a graph instance.
Blue bar represents execution time when sparsification is off and the others represent when
sparsification is on. As expected, on the majority of the test instances, the \rac algorithm
gives the best execution time, because of its lighter computational complexity. The \nic algorithm
is slower than \bac(\pal) and \rac algorithms up to $3.5\times$ and $7\times$
respectively.
We observe that, sparsification decreases the \bac(\pal) algorithm's
execution time more than $2$ times (up to $12$ times) on majority of the test instances.
The \bac(\pal) algorithm's execution time is dominated by the set-up time of the
sparse-prefix-sum data-structure. However, with sparsification, creation
cost of sparse-prefix-sum data-structure decreases significantly. Since the complexity of \nic and \rac algorithms mostly
depends on the number of rows ($n$) and number of cuts ($p$), the affect of the sparsification on
those algorithms are less significant. With sparsification, we observe decreases in their execution time
from $1.2\times$ to $2.5\times$.

Experiments above show that without loss of quality, sparsification significantly improves partitioning algorithms performance.
\subsection{Evaluation of the load imbalance}
\label{ssec:exp:li}

In this section, we evaluate the quality of the partition vectors that our proposed
algorithms output in terms of load imbalance on our complete dataset.
We run \rac and \bac(\pal) algorithms where $p=\{4,8,16,32\}$ and
we run \pal and \bal(\rac) algorithms where $Z=\{m/4, m/9, m/16, m/25\}$.
In the following experiments, we also include \uni and \bal(\uni) algorithms
as baselines.

\begin{figure}[ht]
  \centering
  \subfloat[$4\times4$]{\includegraphics[width=.45\linewidth]{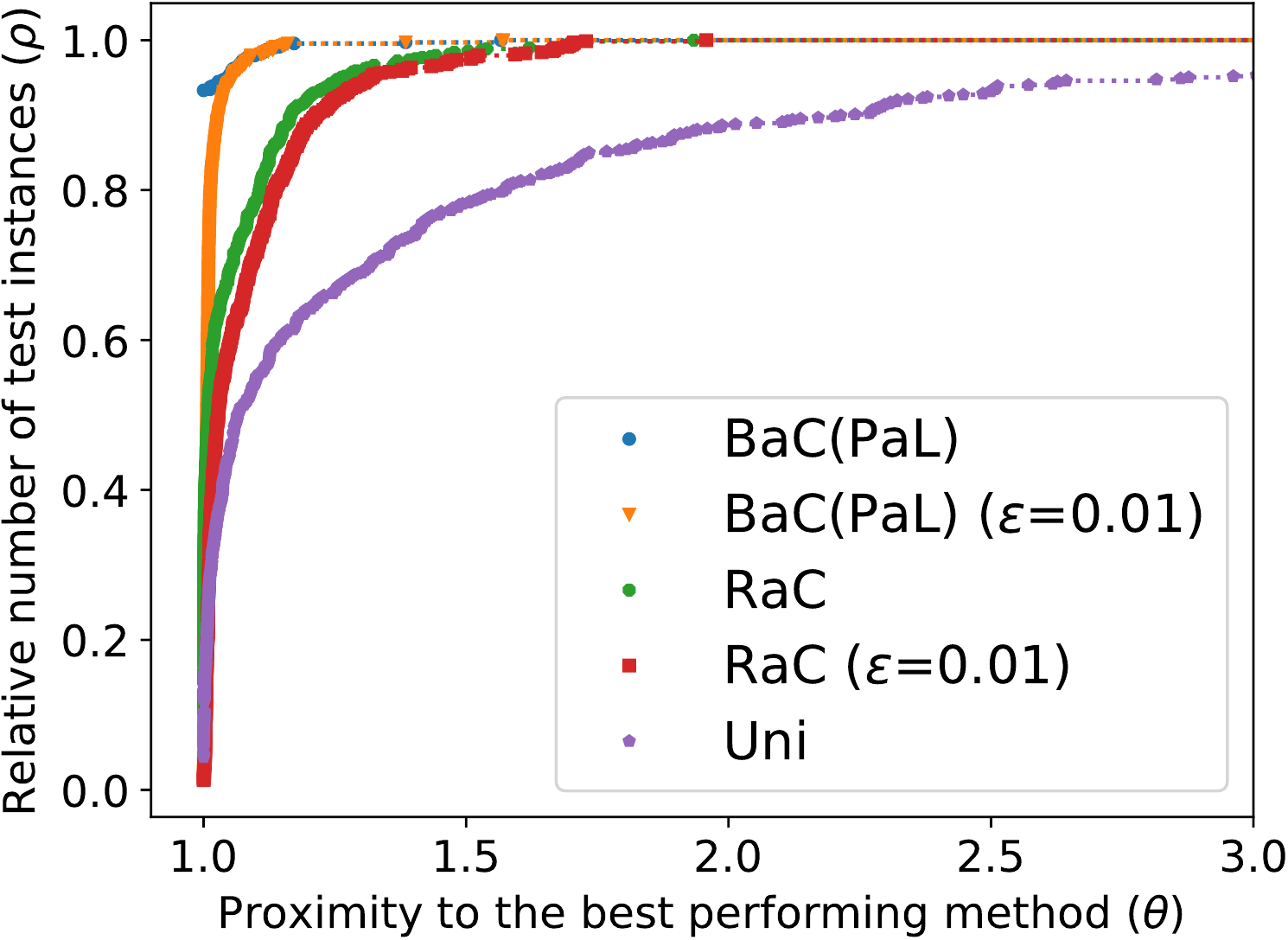}
  \label{fig:exp:comb4}}
  \hspace{1em}
  \subfloat[$8\times8$]{\includegraphics[width=.45\linewidth]{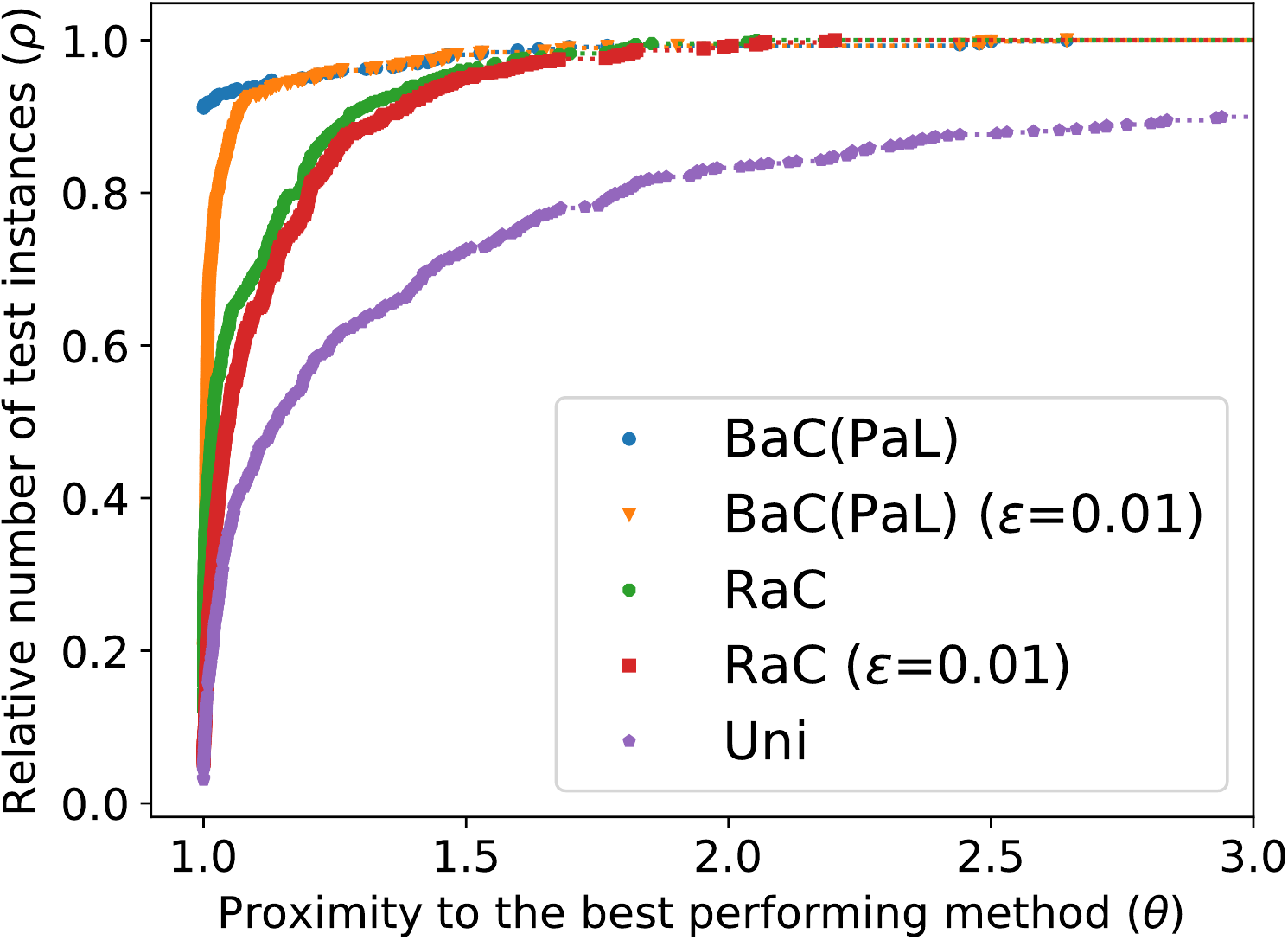}
  \label{fig:exp:comb8}}

  \subfloat[$16\times16$]{\includegraphics[width=.45\linewidth]{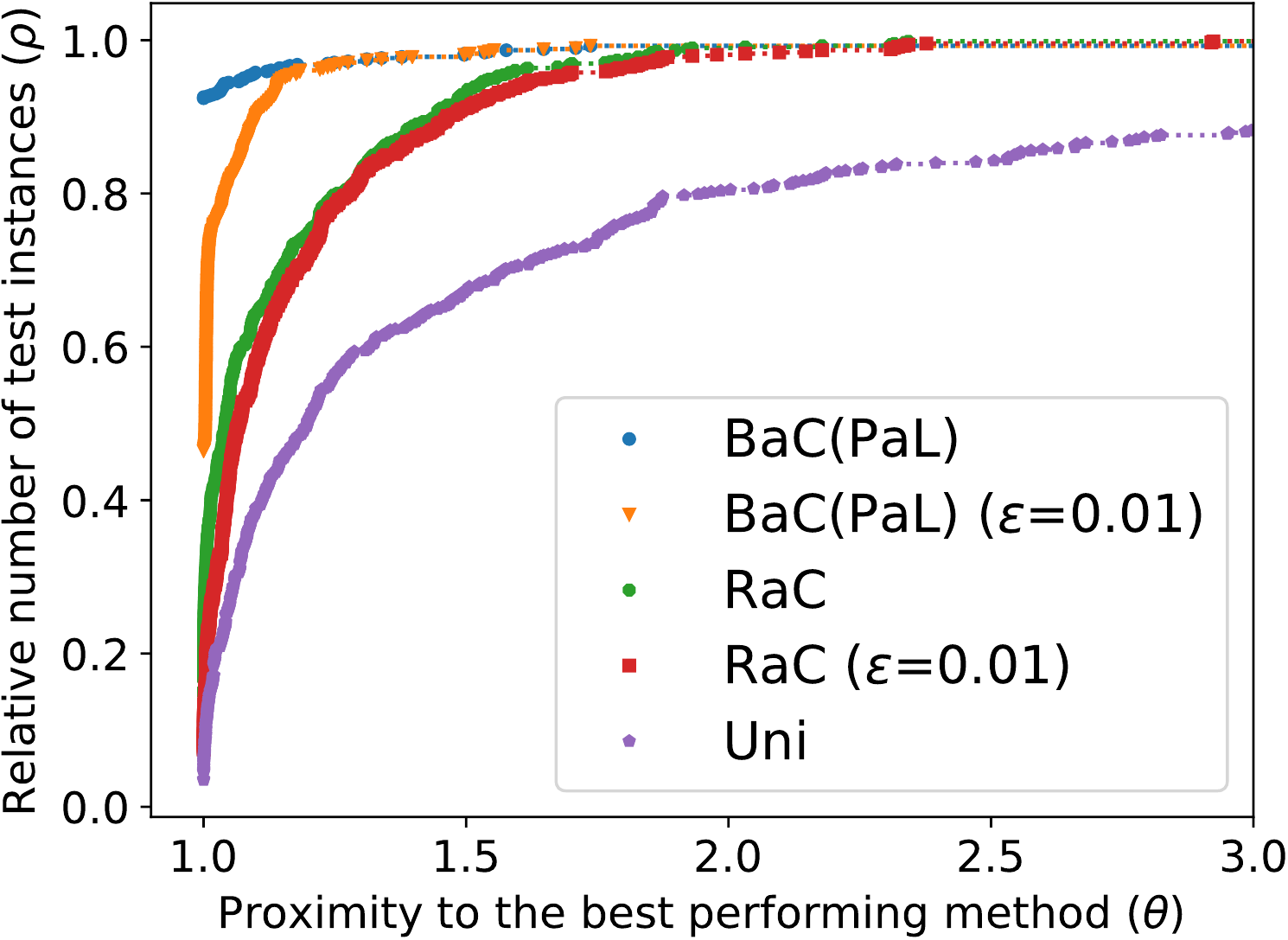}
  \label{fig:exp:comb16}}
  \hspace{1em}
  \subfloat[$32\times32$]{\includegraphics[width=.45\linewidth]{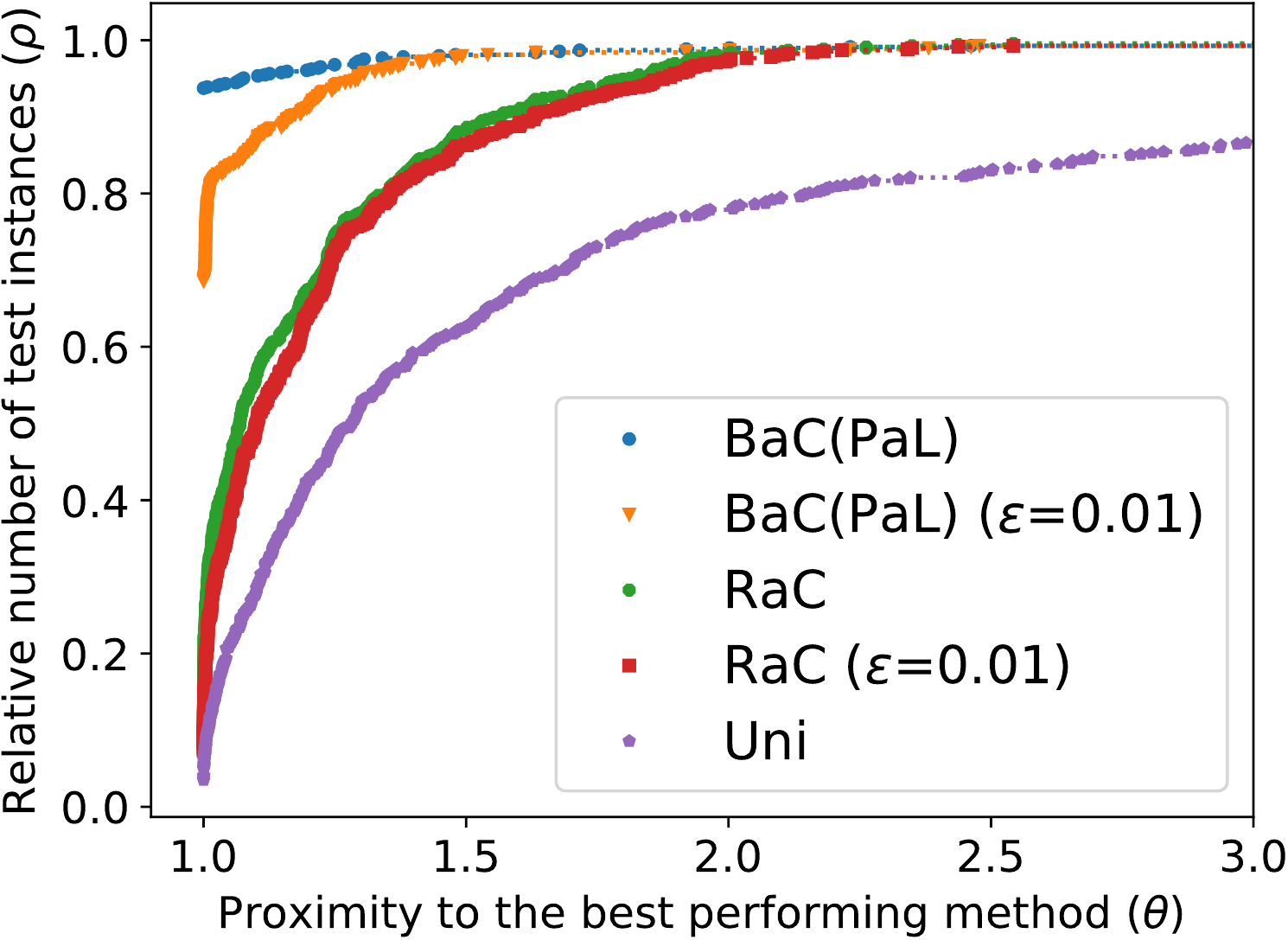}
  \label{fig:exp:comb32}}

  \caption{Load imbalance comparison using performance profiles (\mLI problem).}
  \label{fig:exp:all-li}
\end{figure}

\subsubsection{Load imbalance on the \mLI problem}
\label{ssec:exp:li:mli}

We evaluate relative load imbalance performances of \rac, \bac(\pal) and \uni
algorithms. The aim is to illustrate the efficiency of the proposed algorithms with
respect to the \uni algorithm. In this experiment, we choose $p = \{4, 8, 16, 32\}$
and we report results without sparsification and with sparsification where the error tolerance
for automatic sparsification factor selection is set to; $\epsilon=0.01$.
\Cref{fig:exp:all-li} illustrates the performance profiles of the algorithms for
different $p$ values. Note that, in the performance profiles, we plot how bad a specific algorithm
performs within a factor $\theta$ of the best algorithm. We observe that in all
test instances (\Cref{fig:exp:comb4,fig:exp:comb32})
\bac(\pal) algorithm gives the best performance.
\rac algorithm is the second-best algorithm and in the worst case, it outputs a partition
vector that gives less than $3$ times worse load-imbalance when $p=32$ with respect to
the best algorithm.
We observe that number of test instances where sparsification does not change the \bac(\pal)
algorithm's output increases
for larger $p$ values. For instance, when $p=32$, in $\approx 70\%$ of the test instances
that run on sparsified instances gives the same load imbalance as non-sparsified instances.
This ratio is $\approx 35\%$ when $p=4$. On the other hand, with sparsification when $\epsilon$ is set to $0.01$, we observe that \rac algorithm performs slightly worse. This was
expected because the \rac algorithm maps two-dimensional problem into one dimension
hence it is more error prone.

\begin{figure}[ht]
  \centering
  \subfloat[$Z=m/4$]{\includegraphics[width=.45\linewidth]{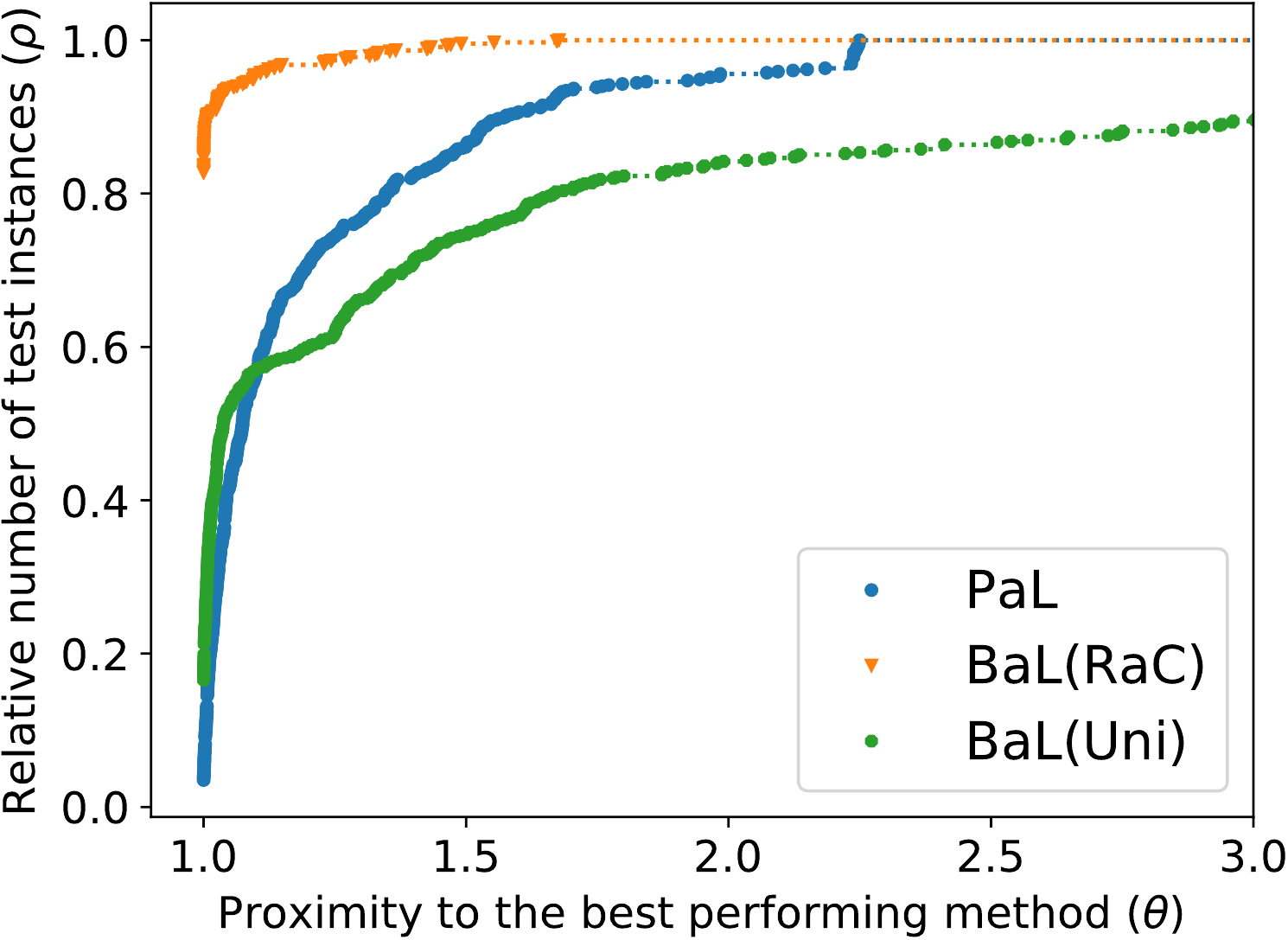}
  \label{fig:exp:mnc-li4}}
  \hspace{1em}
  \subfloat[$Z=m/9$]{\includegraphics[width=.45\linewidth]{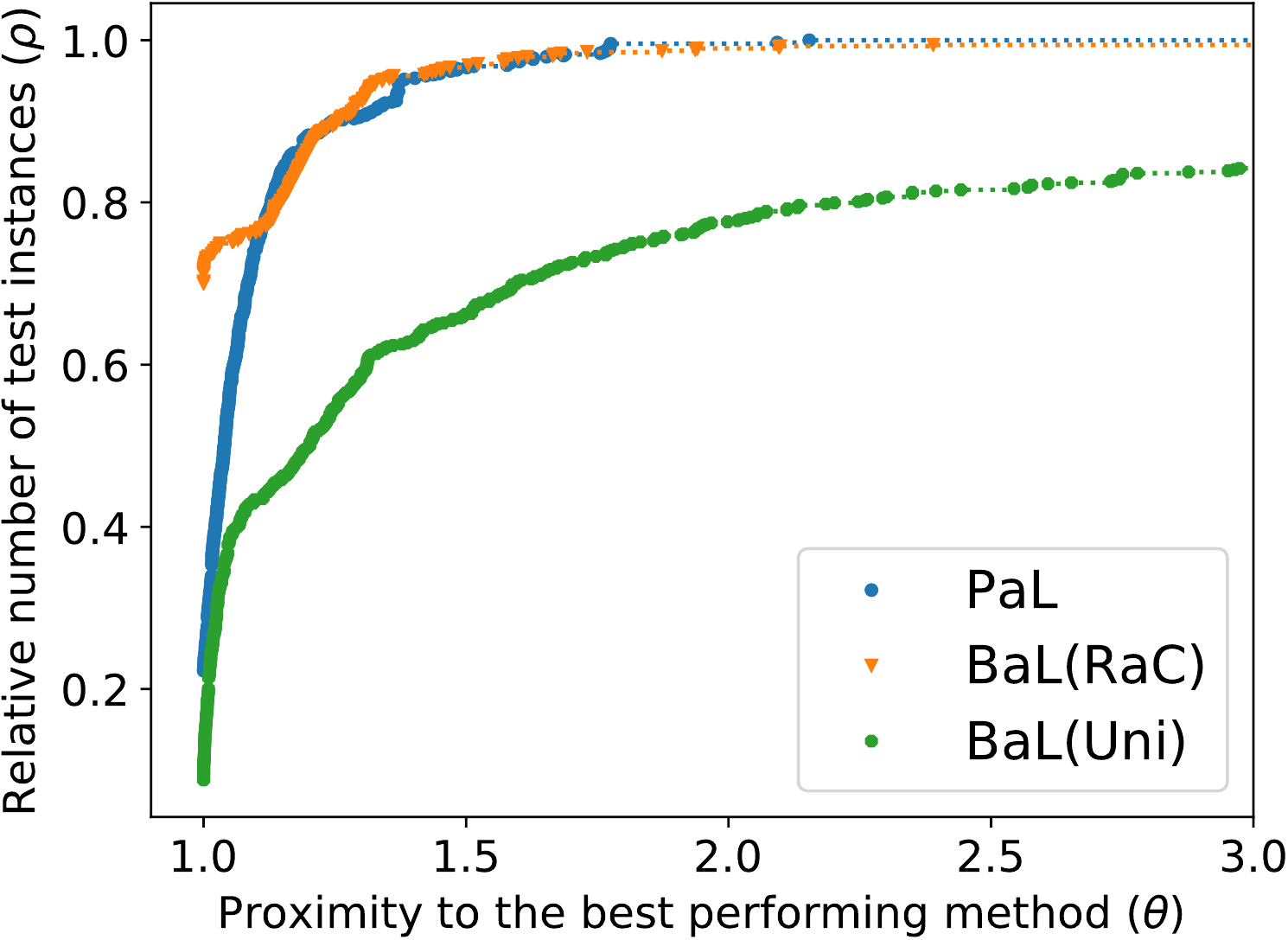}
  \label{fig:exp:mnc-li9}}

  \subfloat[$Z=m/16$]{\includegraphics[width=.45\linewidth]{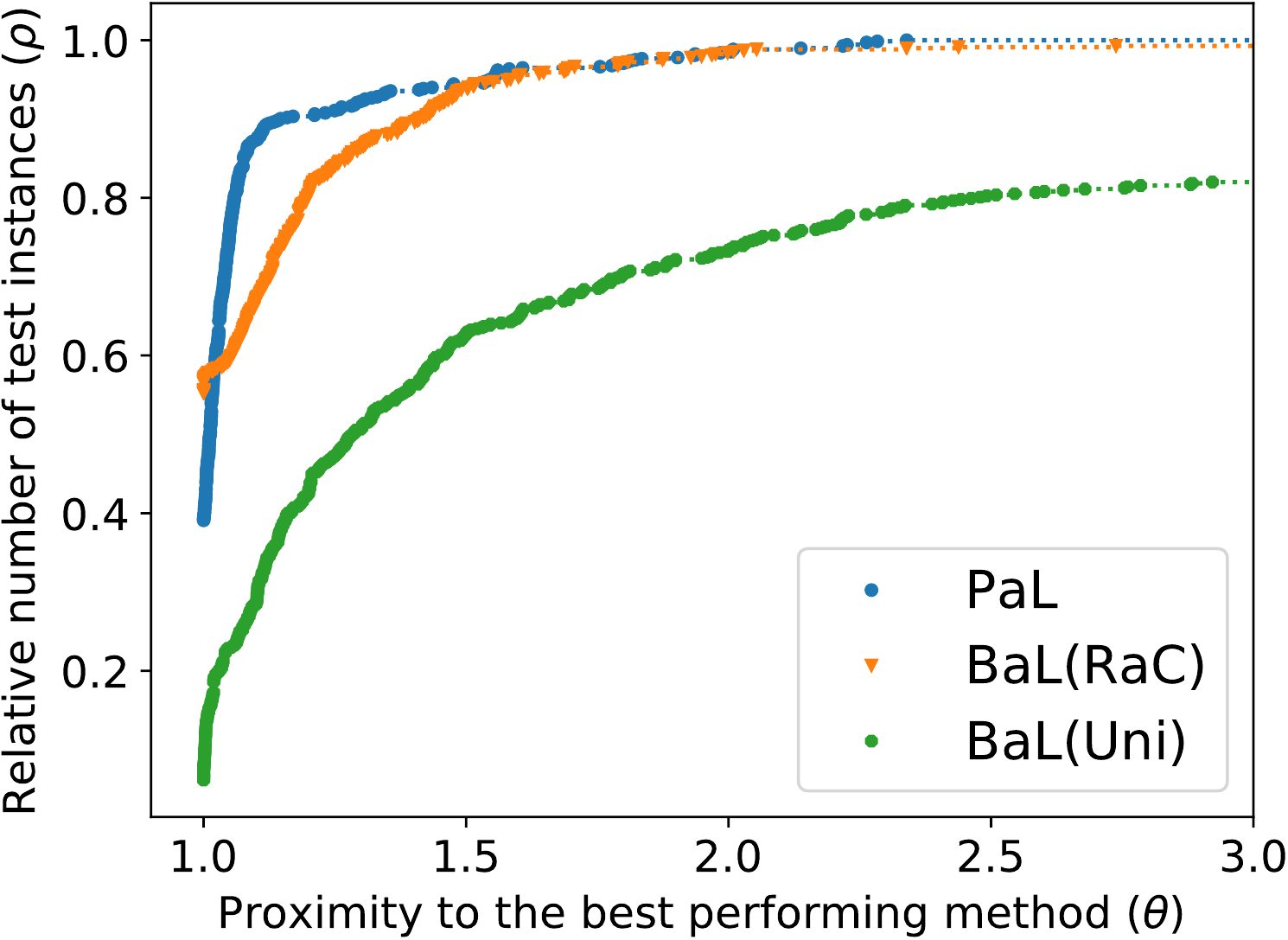}
  \label{fig:exp:mnc-li16}}
  \hspace{1em}
  \subfloat[$Z=m/25$]{\includegraphics[width=.45\linewidth]{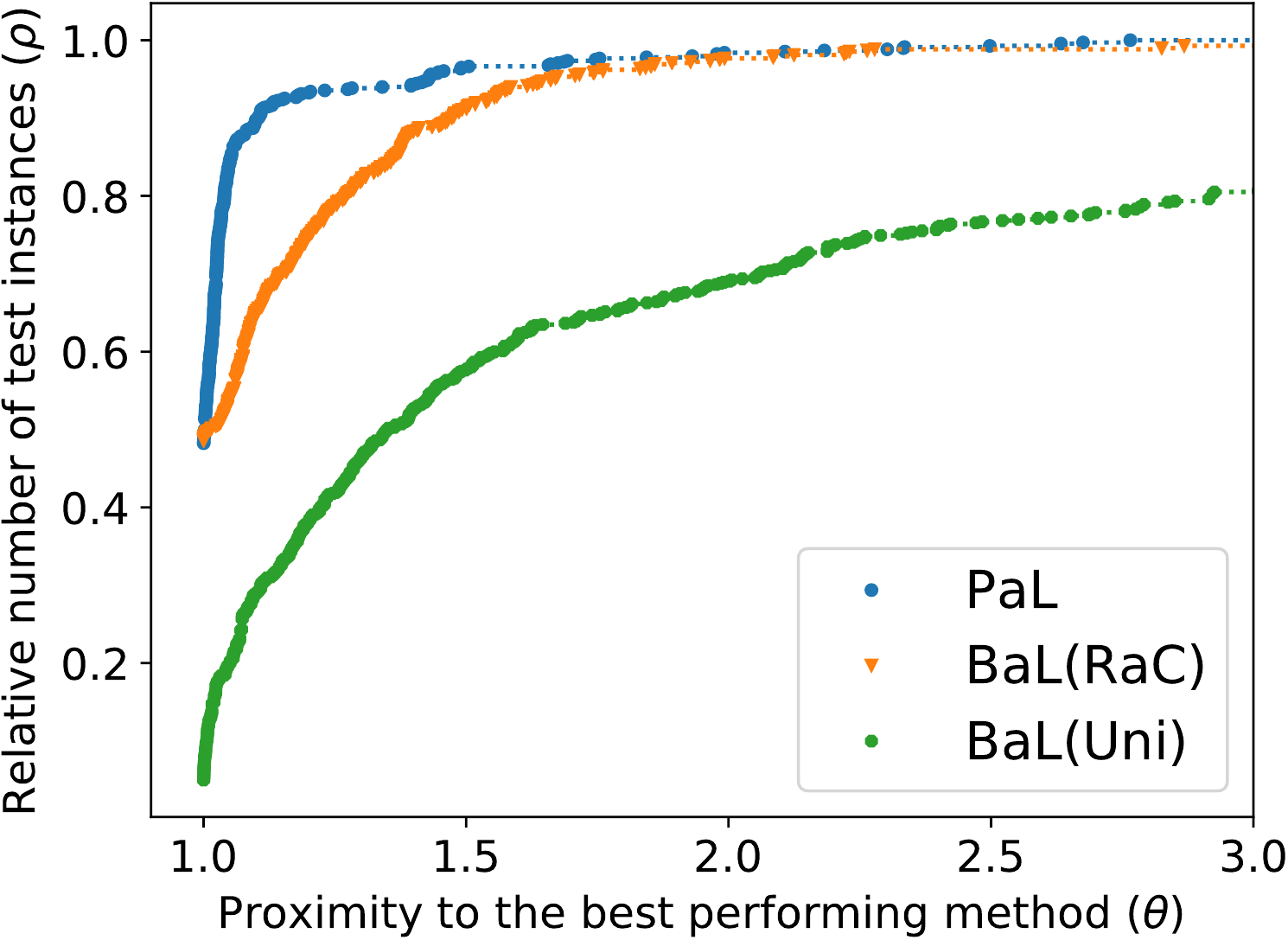}
  \label{fig:exp:mnc-li25}}

  \caption{Load imbalance comparison using performance profiles (\mNC problem).}
  \label{fig:exp:mnc:all-li}
\end{figure}

\subsubsection{Load imbalance on the \mNC problem}
\label{ssec:exp:li:mnc}
We evaluate relative load imbalance performances of \pal, \bal(\rac) and \bal(\uni)
algorithms. The aim is to illustrate the efficiency of the proposed algorithms with
respect to the \bal(\uni) algorithm. In this experiment, we choose $Z = \{m/4, m/9, m/16, m/25\}$
and we report results without sparsification because sparsification may cause bigger errors
in the \mNC problem. \Cref{fig:exp:mnc:all-li} illustrates the performance profiles of the algorithms for
different $Z$ values. We observe that when $Z$ is larger (see \Cref{fig:exp:mnc-li4,fig:exp:mnc-li9})
The \bal(\rac) algorithm performs slightly better than \pal algorithm. The \pal
algorithm outperforms for smaller $Z$ values (see \Cref{fig:exp:mnc-li16}).

\subsection{Evaluation of the partitioning time}
\label{ssec:exp:et}

In this section we evaluate the execution times of the algorithms that we proposed
for \mLI and \mNC problems on our complete dataset. Reported execution
times include sparsification time, sparse-prefix-sum data-structure
construction time, and partitioning time.
We run \rac and \bac(\pal) algorithms where $p=\{4,8,16,32\}$ and
we run \pal and \bal(\rac) algorithms where $Z=\{m/4, m/9, m/16\}$.
In the following experiments we also include \uni and \bal(\uni) algorithms
as baselines. In the following experiments we report the median of $10$ runs
for each test instance.

\begin{figure}[ht]
  \centering
  \subfloat[$4\times4$]{\includegraphics[width=.45\linewidth]{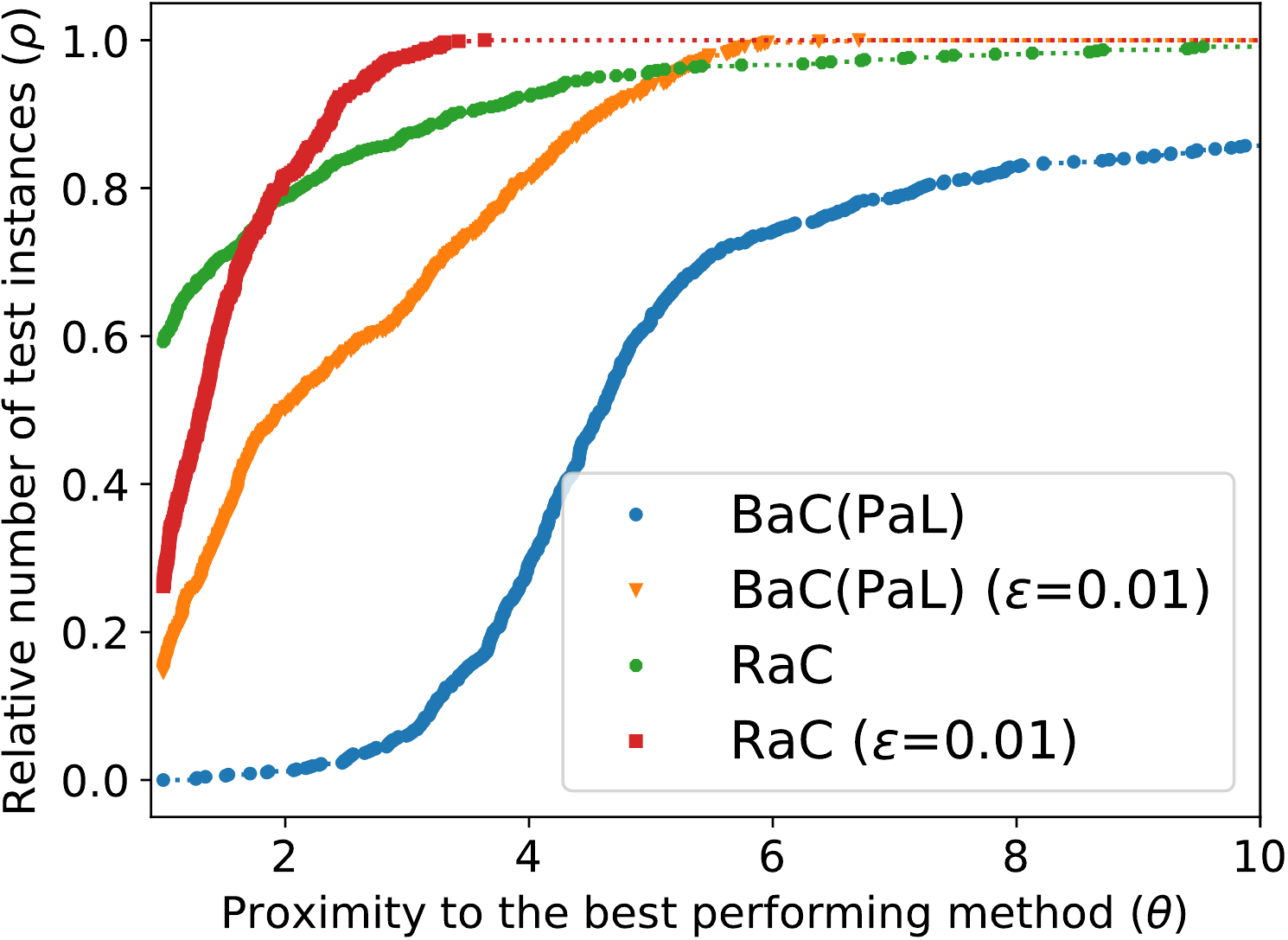}
  \label{fig:exp:comb-et4}}
  \hspace{1em}
  \subfloat[$8\times8$]{\includegraphics[width=.45\linewidth]{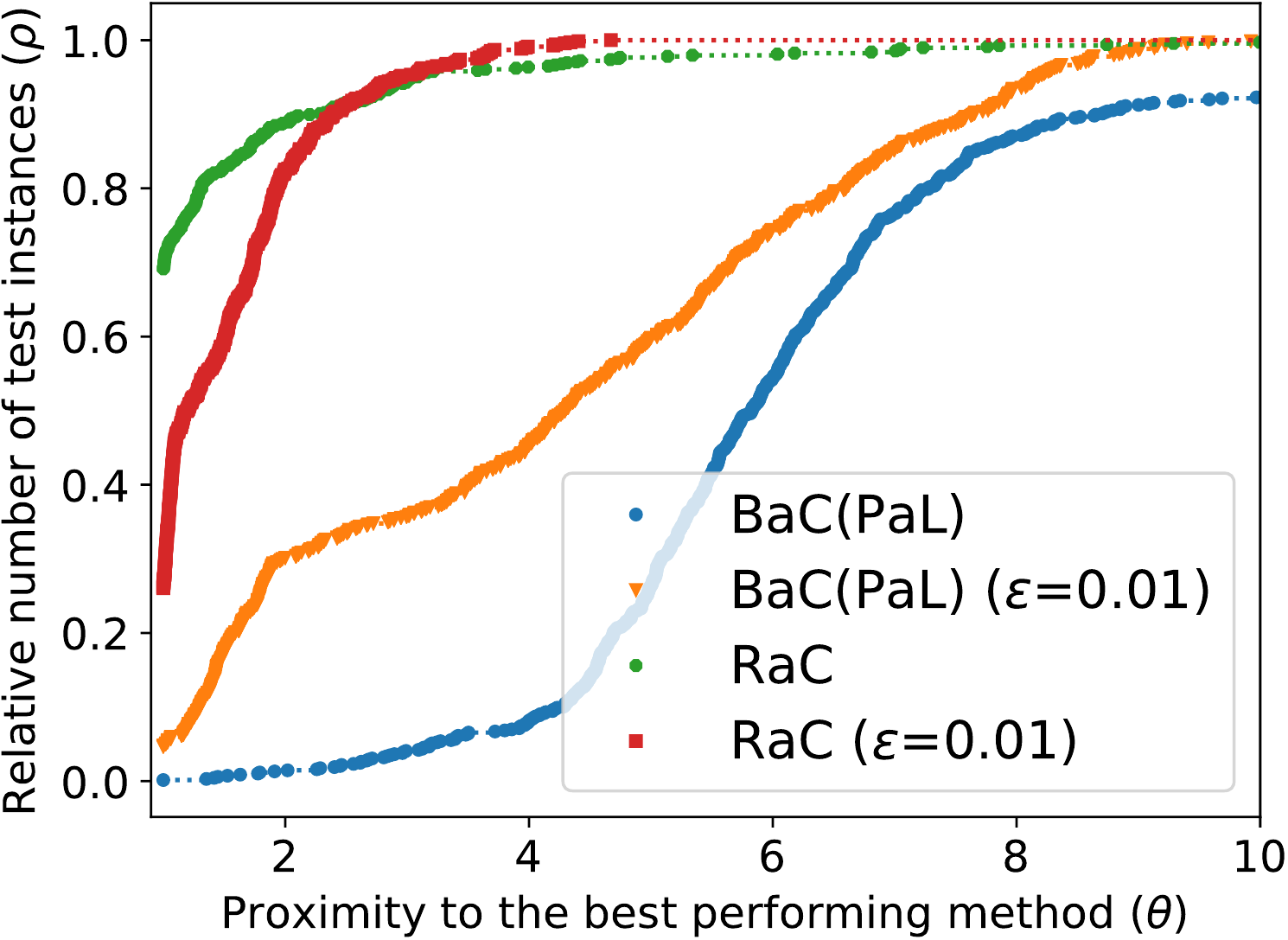}
  \label{fig:exp:comb-et8}}

  \subfloat[$16\times16$]{\includegraphics[width=.45\linewidth]{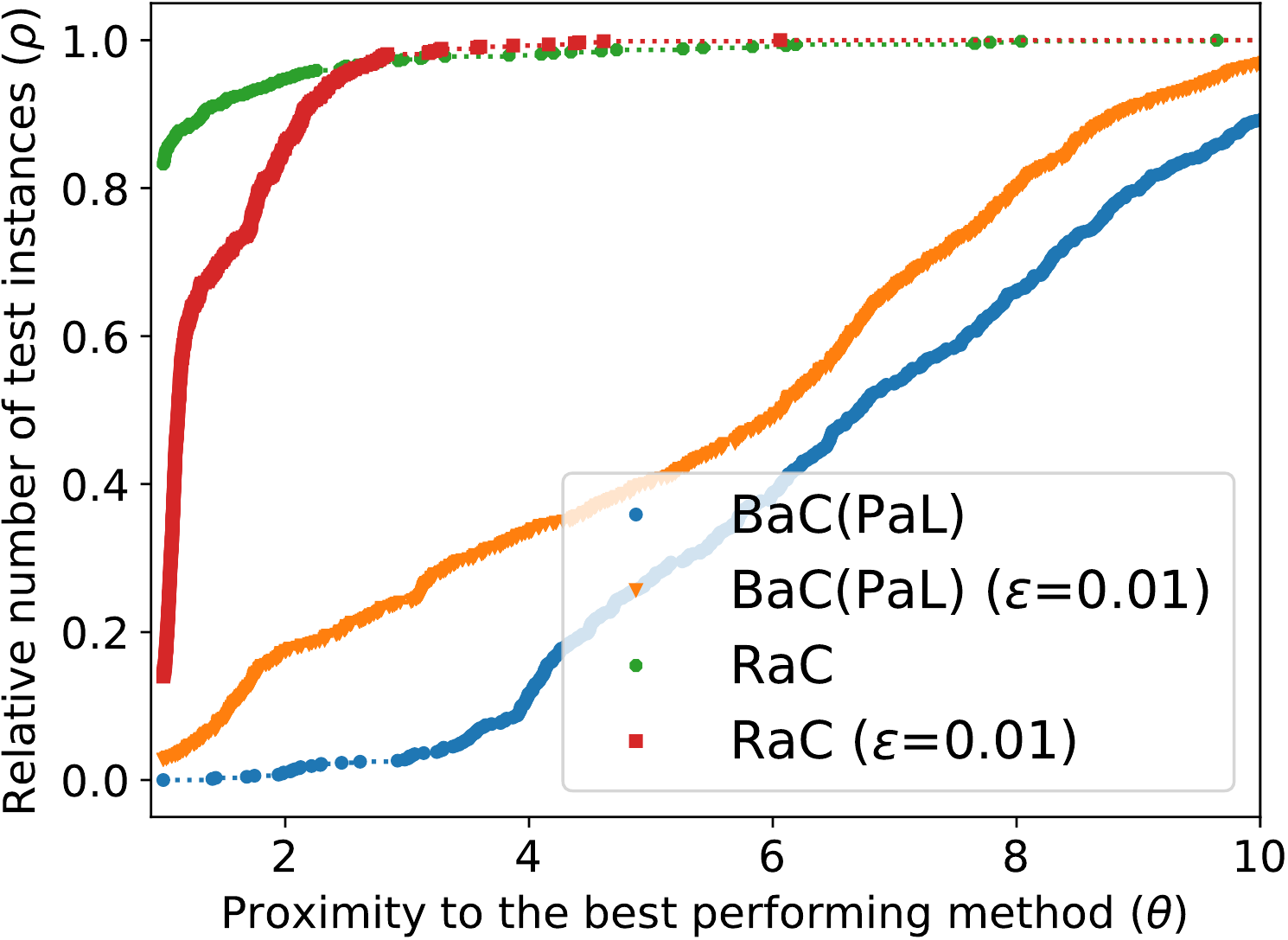}
  \label{fig:exp:comb-et16}}
  \hspace{1em}
  \subfloat[$32\times32$]{\includegraphics[width=.45\linewidth]{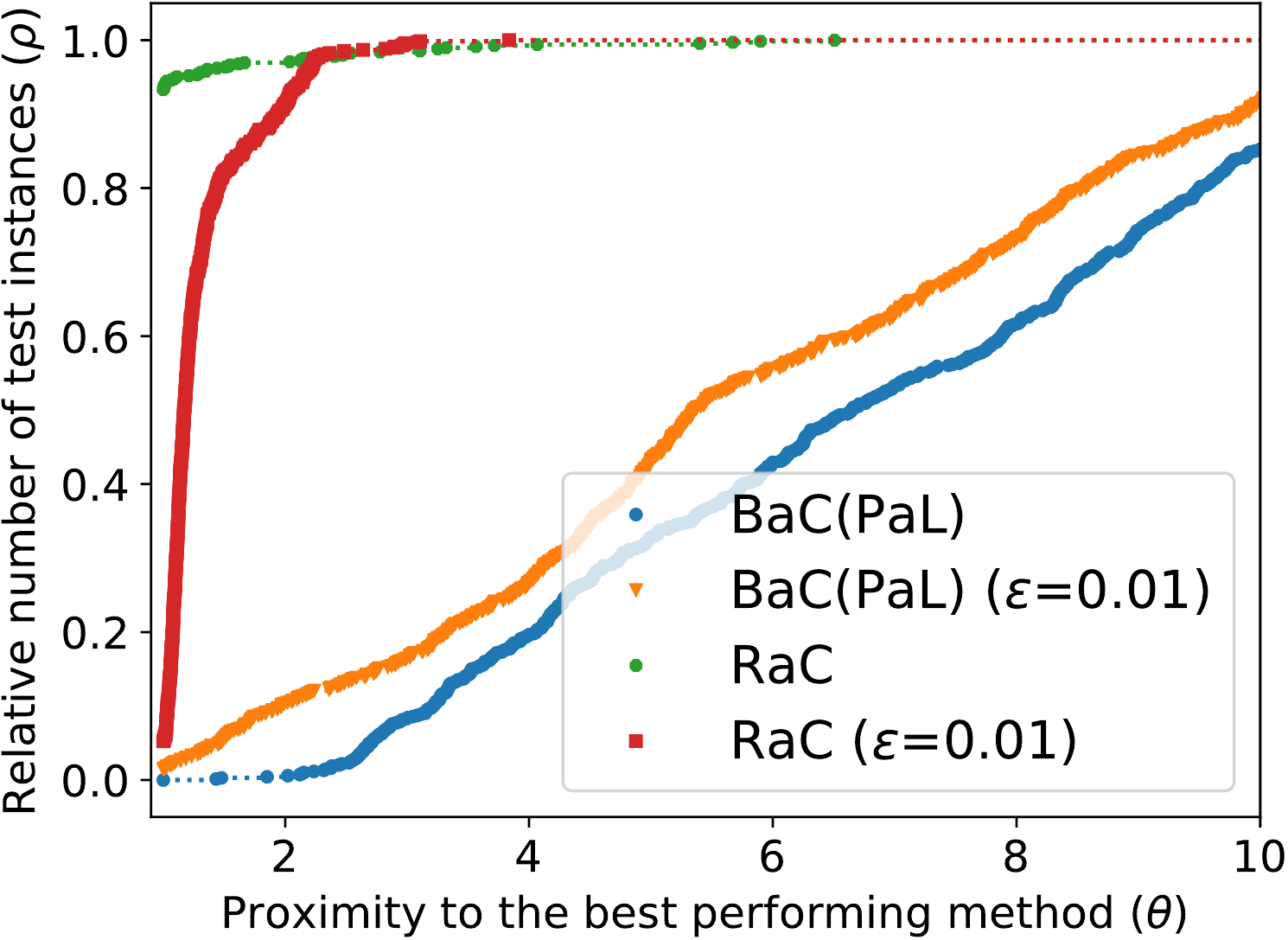}
  \label{fig:exp:comb-et32}}

  \caption{Execution time comparison using performance profiles (\mLI problem).}
  \label{fig:exp:all-et}
\end{figure}

\subsubsection{Partitioning time on the \mLI problem}
\label{ssed:exp:et:mli}

We evaluate relative executions times of \rac and \bac(\pal)
algorithms. In this experiment, we choose $p = \{4, 8, 16, 32\}$
and we report results without sparsification and with sparsification where the load imbalance
error is set to be off on the order of one percent; $\epsilon=0.01$.
\Cref{fig:exp:all-li} illustrates the performance profiles of the algorithms for
different $p$ values. We observe that in all cases (\Cref{fig:exp:comb-et4,fig:exp:comb-et32})
as expected, \rac algorithm gives the best execution time,
because of its lighter computational complexity.
The \bac(\pal) algorithm's execution time is decreases significantly when the sparsification is
on. In overall, sparsification slightly improves the \rac algorithms execution time
because the gain in the partitioning time do not compensate the sparsification time for smaller graphs .

\begin{figure}[ht]
  \centering
  \subfloat[$Z=m/4$]{\includegraphics[width=.45\linewidth]{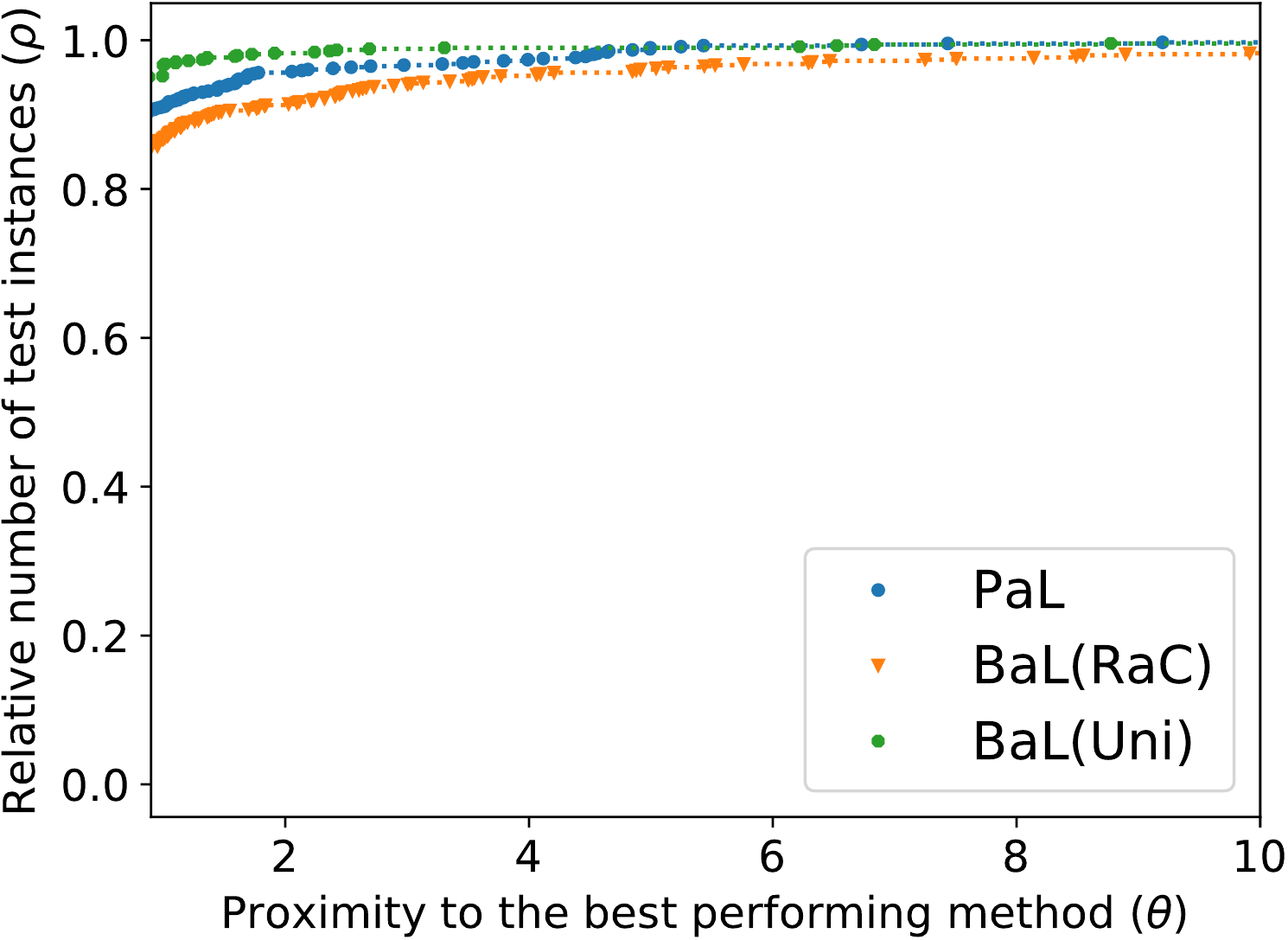}
  \label{fig:exp:mnc-et4}}
  \hspace{1em}
  \subfloat[$Z=m/9$]{\includegraphics[width=.45\linewidth]{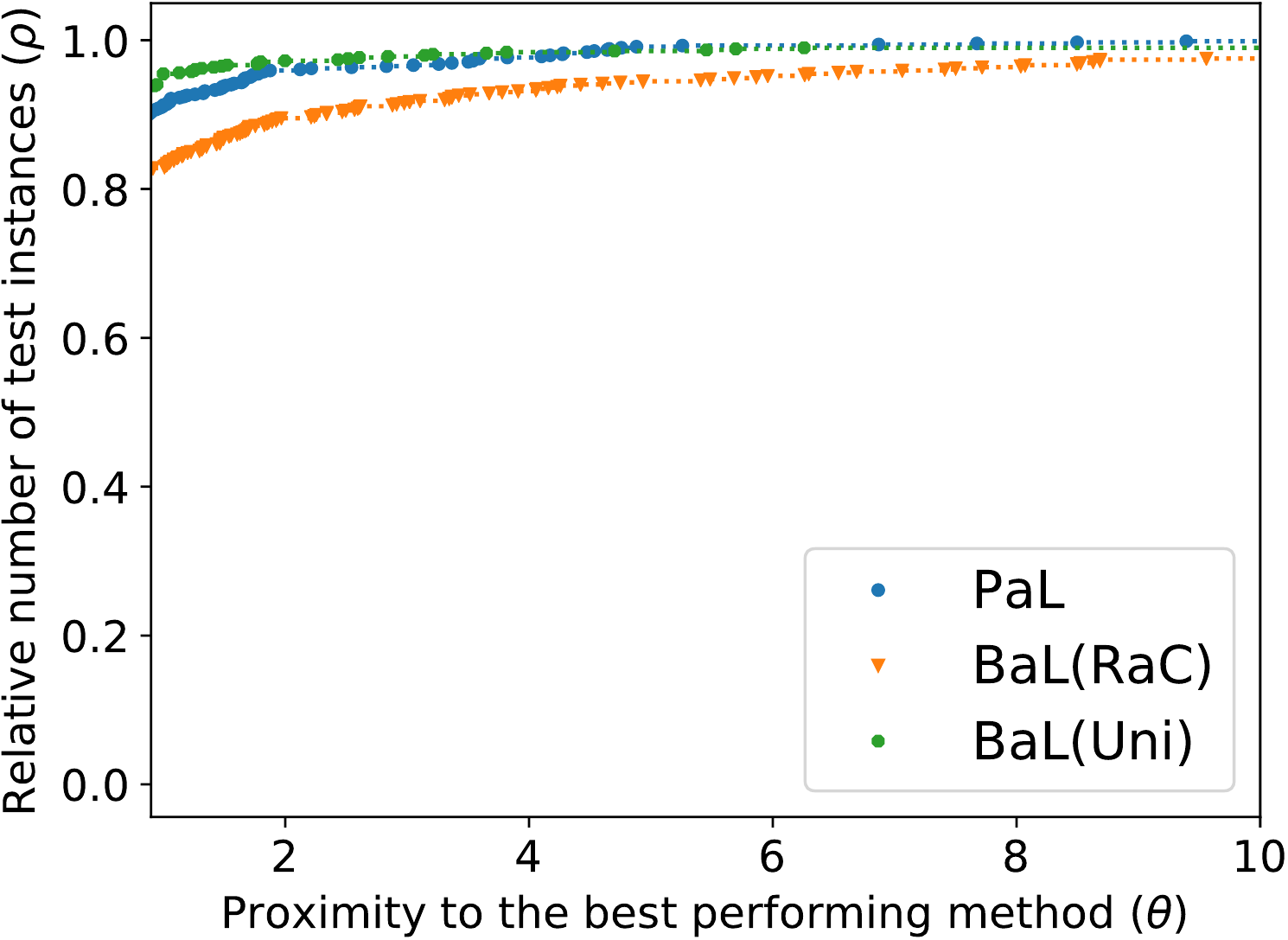}
  \label{fig:exp:mnc-et9}}

  \subfloat[$Z=m/16$]{\includegraphics[width=.45\linewidth]{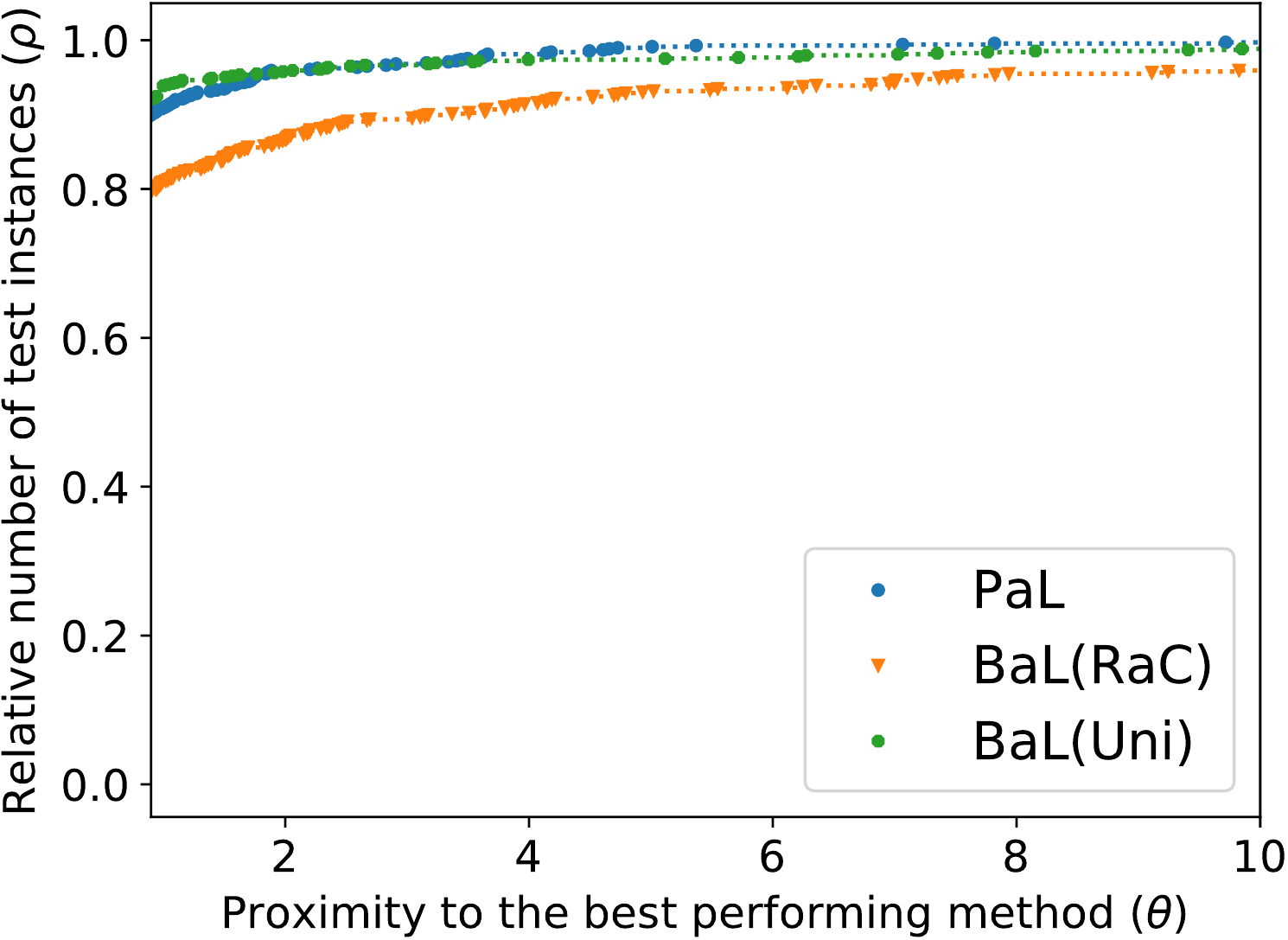}
  \label{fig:exp:mnc-et16}}
  \hspace{1em}
  \subfloat[$Z=m/25$]{\includegraphics[width=.45\linewidth]{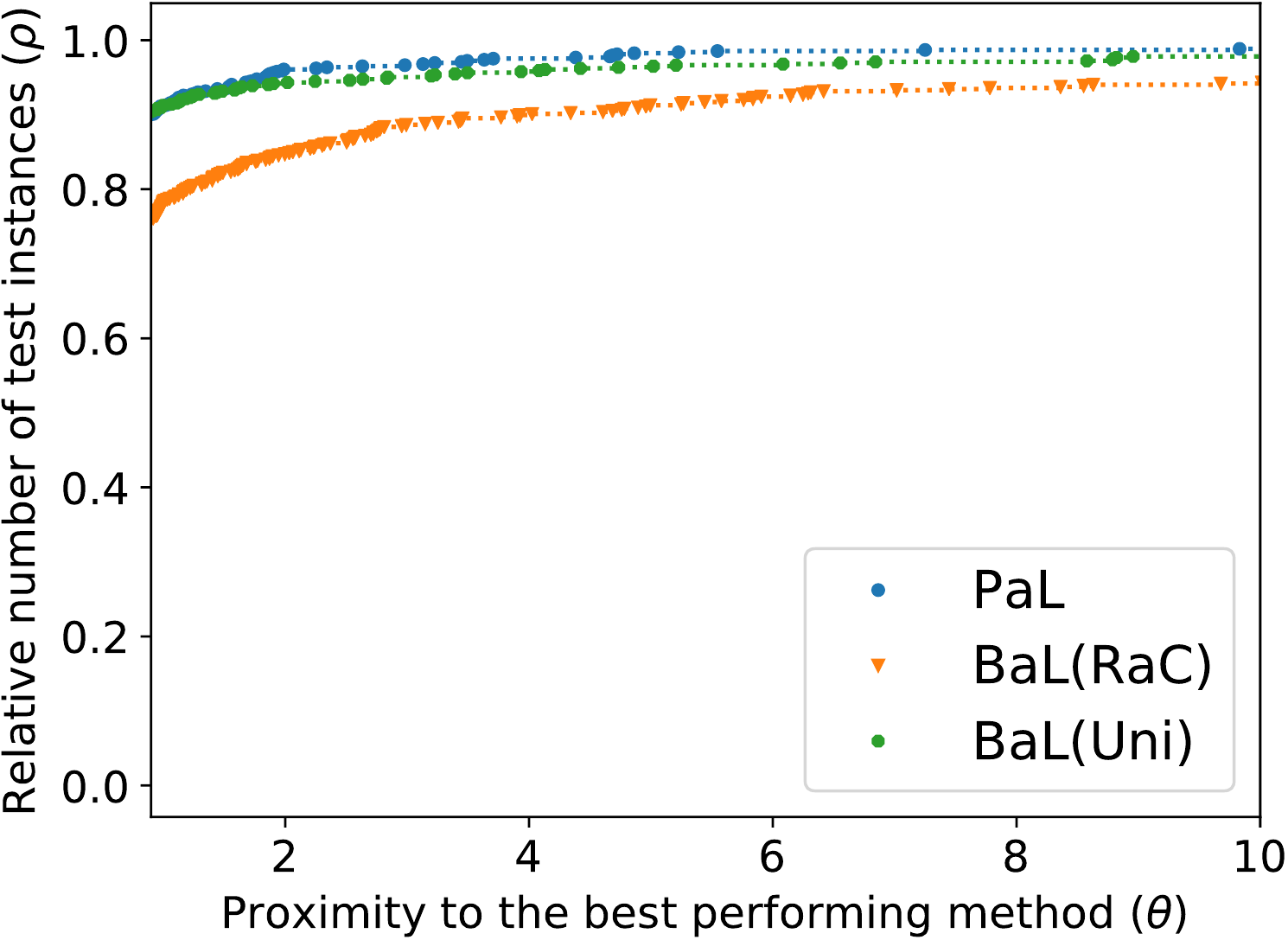}
  \label{fig:exp:mnc-et25}}

  \caption{Execution time comparison using performance profiles (\mNC problem).}
  \label{fig:exp:mnc:all-et}
\end{figure}

\subsubsection{Partitioning time on the \mNC problem}
\label{ssec:exp:et:mnc}

We evaluate relative execution time performances of \pal, \bal(\rac), and \bal(\uni)
algorithms. The aim is to illustrate the efficiency of the proposed algorithms with
respect to the \uni algorithm. In this experiment, we choose $Z = \{m/4, m/9, m/16\}$
and we report results without sparsification because sparsification may cause bigger errors
in the \mNC problem.
\Cref{fig:exp:mnc:all-et} illustrates the performance profiles of the algorithms for
different $Z$ values. We observe that, in all instances the \pal
algorithm outperforms the other algorithms. Because, both \bal(\rac) and \bal(\uni)
algorithms do many tests for different target cuts and load lookups. Hence, their computational
complexities are higher than the \pal algorithm.

\section{Conclusion}
\label{sec:conc}
In this paper, we show that the optimal solution to the symmetric rectilinear partitioning is NP-Hard, and
we propose refinement-based and probe-based heuristic algorithms to two variants of this problem.
After providing complexity analysis of the algorithms, we implement a data-structure and sparsification
strategies to reduce the complexities. Our experimental evaluation shows that our proposed algorithms
are very efficient to find good-quality solutions, such that we achieve a nearly optimal solution
on $80\%$ instances of 375 small graphs. We also open source our code at
\url{http://github.com/GT-TDAlab/SARMA} for public usage and future development.

As future work, we are working on decreasing the space requirements
of our sparse prefix sum data structure. In addition, we will also
investigate approximation techniques, and parallelization of the proposed algorithms.

\section*{Acknowledgements}
We would like to extend our gratitude to M. M{\"{u}}cahid Benlio{\u{g}}lu for his valuable
comments and feedbacks for the initial draft of this manuscript and code-base.
This work was partially supported by the NSF grant CCF-1919021.

\bibliographystyle{siamplain}
\bibliography{paper,tdalab}

\end{document}